\documentclass[preprint]{aastex}
\usepackage{subfigure}
\usepackage{lscape}

\begin{document}

\title{ $z\sim4$ H$\alpha$ Emitters in GOODS: \\
        Tracing the Dominant Mode for Growth of Galaxies }

\author{ Hyunjin Shim\altaffilmark{1},  Ranga-Ram Chary\altaffilmark{1}, 
         Mark Dickinson\altaffilmark{2}, 
         Lihwai Lin\altaffilmark{3}, Hyron Spinrad\altaffilmark{4},
         Daniel Stern\altaffilmark{5},
         Chi-Hung Yan\altaffilmark{3} }
\altaffiltext{1}{Spitzer Science Center,
California Institute of Technology, MS 220-6, Pasadena, CA 91125 }
\altaffiltext{2}{National Optical Astronomy Observatory}
\altaffiltext{3}{Institute of Astronomy \& Astrophysics, Academia Sinica, Taipei 106}
\altaffiltext{4}{Department of Astronomy, University of California at Berkeley,
Berkeley, CA 94720}
\altaffiltext{5}{Jet Propulsion Laboratory,
California Institute of Technology, Pasadena, CA 91109}

\begin{abstract}

    We present evidence for                                                  
strong H$\alpha$ emission
in galaxies with spectroscopic redshifts in the range of $3.8<z<5.0$
  over the Great Observatories Origins Deep Survey (GOODS) fields.
    Among 74 galaxies detected in the \textit{Spitzer}
  IRAC 3.6 and 4.5\,$\mu$m bands, more than 70\,\% of the galaxies show
  clear excess at 3.6\,$\mu$m compared to the expected flux density
  from stellar continuum only. 
   We provide evidence that this 3.6\,$\mu$m excess is due to H$\alpha$ emission 
  redshifted into the 3.6\,$\mu$m band, and classify these 3.6\,$\mu$m excess
  galaxies to be H$\alpha$ emitter (HAE) candidates.
   The selection of HAE candidates using an excess in broad-band filters
  is sensitive to objects whose rest-frame H$\alpha$ equivalent width
  is larger than $350\,\mbox{\AA}$.
  The H$\alpha$ inferred SFRs of the HAEs range between 
  $20$ and $500\,M_{\odot}$\,yr$^{-1}$ and is a a factor of $\sim$6 larger than SFRs inferred from the UV continuum.
   The ratio between the H$\alpha$ luminosity and UV luminosity of HAEs 
  is also on average larger than that of local starbursts. 
Possible reasons for
such strong H$\alpha$ emission in these galaxies include 
different dust extinction properties, young stellar population ages, 
  extended star formation histories, low metallicity, and a top-heavy stellar initial 
  mass function. Although the correlation between UV slope $\beta$ and 
  $L_{H\alpha}/L_{UV}$ raises the possibility that HAEs prefer a 
  dust extinction curve which is steeper in the UV, the most dominant factor 
  that results in strong H$\alpha$ emission appears to be star formation history.  
   The H$\alpha$ equivalent widths of HAEs are large despite their
  relatively old stellar population ages constrained by SED fitting, 
  suggesting that at least 60\,\% of HAEs 
produce stars at a constant rate.
   Under the assumption that the gas supply is sustained, HAEs are able to 
  produce $\gtrsim50$\,\% of the stellar mass density 
  that is encompassed in massive ($M_* >10^{11}\,M_{\odot}$) galaxies at $z\sim3$. 
   This `strong H$\alpha$ phase' of star formation plays a dominant role 
  in galaxy growth at $z\sim4$, and they are likely progenitors of massive red 
  galaxies at lower redshifts. 

\end{abstract}

\keywords{ cosmology: observation -- galaxies: evolution -- galaxies: starburst
 -- galaxies: high-redshift }

\section{Introduction}

Rest-frame near-infrared observations of 
massive, luminous galaxies at $z\sim3$ suggests
a mode of galaxy growth driven by active star formation at higher redshifts.
  Despite the discovery of galaxies with extremely large star formation
 rates (SFR) such as submm galaxies (SMGs), the number density of 
 observed SMGs significantly drops at $z>4$
 (Schinnerer et al. 2008; Capak et al. 2008; Coppin et al. 2009, Pope \& Chary 2010)
 and is insufficient
 to explain the number density of massive galaxies at lower redshifts
 (Coppin et al. 2009).
  This is either because we miss a significant fraction of 
 $z>4$ star-forming galaxies, or SFR estimates for 
 $z>4$ galaxies is uncertain. The two possibilities require 
 an understanding of the nature of star formation at high-redshifts, 
 especially how the star formation is powered and
 how well the SFR can be measured.

  The dominant mechanism that enables large SFR at high redshifts
 is unclear.
  Merger-induced star formation is the preferred mechanism expected in
 the hierarchical galaxy evolution scenario.
 Yet the observed number density of galaxies with merger-induced morphologies 
 challenges such a scenario, 
 while the values for merger rates are still
a matter of debate 
 (e.g., Ravindranath et al. 2006; Conselice \& Arnold 2009).
  As an alternative way to explain the formation of
 disks and spheroids at high redshifts, Dekel et al. (2009) suggested
 a model whereby accretion of cold gas along dark matter filaments powers star formation in galaxies. 
  Although no direct evidence for cold gas accretion has been observed
 that supports this scenario,
 the effect of cold stream feeding into massive dark matter halos 
 increases at $z>2$. Thus, the expected number density of galaxies at $z>2$ with
 large SFR would be larger than the number density of
 merger-powered star-forming galaxies at high redshifts.
  In order to assess whether merger or accretion is the main 
 mechanism for growth of stellar mass in galaxies, 
 large samples of galaxies with
well-measured SFR are necessary.

  The most popular probe of star formation history at high redshift
 is the rest-frame UV emission due to the practical reason
 that the rest-frame UV is redshifted to optical wavelengths, 
 and easily accessible at $z>3$. 
  The Lyman break galaxies 
 (LBGs; Steidel et al. 2003; Ouchi et al. 2004; Bouwens et al. 2007, 2009, 2010) 
 selected using rest-frame UV colors and the Lyman alpha emitters 
 (LAEs; Malhotra \& Rhoads 2002; Shioya et al. 2009; Taniguchi et al. 2009) 
 selected in narrow-band and grism observations are
 the best tracers of the star-forming galaxy population at $z>3$. 
  Despite their wide usage, it is not clear whether these UV-selected
 galaxy samples and UV-inferred SFRs account for
 most of the ongoing star formation, since the UV emission is 
 by nature, sensitive to dust extinction (Bouwens et al. 2009).

  While high-redshift star-forming galaxies
 do show evidence for dust attenuation
 (e.g., Adelberger \& Steidel 2000; Reddy et al. 2010),
 a precise understanding of their internal reddening properties is missing.
  The UV spectral slope $\beta$ ($f_{\lambda} \propto \lambda^{\beta}$)
 has been widely used for extinction correction, using its tight
 correlation with other extinction measures in the local universe 
 (Meurer, Heckman, \& Calzetti 1999). Several studies have 
 tested its validity at higher redshifts up to $z\sim2$ 
 (Papovich et al. 2006; Daddi et al. 2007; Reddy et al. 2010),
 yet the studies for star-forming galaxies at even higher
 redshifts suggest that this is not true: for example,  
 Siana et al. (2009) show that it may not be valid
 to use the extinction law for local starbursts
 (Calzetti et al. 2000) for young LBGs at $z\sim3$,
 since the UV slope $\beta$ overpredicts the level of
 dust extinction. Carilli et al. (2008) demonstrated that radio-derived
 SFRs for $z\sim3$ LBGs indicate UV extinction 
 smaller than estimated before. 
  Hayes et al. (2010) claimed that the extinction in Ly$\alpha$ emitters
 should be larger than that suggested by the starburst extinction law 
 (Calzetti et al. 2000). These recent studies suggest that 
 the dust extinction properties of high-redshift star-forming galaxies
 could be different from that of the local galaxies.

  This potential difference in dust extinction of high-redshift
 star-forming galaxies compared to local starbursts is constrained
 through the comparison of UV-inferred star formation to the 
 star formation from other independent indicators that is less 
 sensitive to the dust extinction (e.g. Reddy et al. 2010).  
  The optical emission lines, especially the Balmer recombination
 lines, enable a direct comparison between 
 local starbursts and high-redshift star forming galaxies. 
  Moreover, these lines are less sensitive to dust extinction compared 
 to UV emission lines or continuum measurements.
  However, it is not easy to investigate optical emission lines of
 high-redshift star-forming galaxies through spectroscopic observations
 when the lines are redshifted to the near-infrared. 
  It is impossible to measure H$\alpha$ line flux at $z>4$ ($>3\,\mu$m)
 with current facilities, 
 although there are studies measuring H$\alpha$ line flux for $z\sim2$
 galaxies through NIR spectroscopy (Erb et al. 2006, Reddy et al. 2010)
 or narrow-band imaging (Geach et al. 2008; Hayes et al. 2010). 

  In order to overcome these observational limits, we propose to use
 photometric information from high-redshift star-forming galaxies
 that reflect the presence of the line emission when the line is sufficiently strong.
  Chary, Stern, \& Eisenhardt (2005) have constrained SFR
 of a $z=6.56$ lensed galaxy using the observed increase in 
 \textit{Spitzer} 4.5\,$\mu$m flux which is likely to be due to H$\alpha$ emission.  
  Other studies of the spectral energy distributions of high-redshift
 star-forming galaxies have also suggested that strong emission lines
affect the broad-band photometry (Schaerer et al. 2009; Reddy et al. 2010). 
 Unbiased surveys for emission-line galaxies (e.g., WFC3 grism surveys,
 Atek et al. 2010) showed that there do exist galaxies with very faint
 continuum but strong H$\alpha$ emission line at $z\sim1-2$. 
 Such galaxies could be selected using the existing NIR/MIR photometric
 observations in the appropriate redshift window. This paper
 is focused on the identification and investigation of such galaxies with strong rest-frame 
 optical emission lines. 

  In this paper, we present the SFRs of $z\sim4$
 galaxies measured using the H$\alpha$ line. 
  This is the first rest-frame optical estimation of SFRs for
 spectroscopically confirmed $z\sim4$ galaxies.
  The H$\alpha$ line flux is estimated through an excess 
 in the broad-band (i.e., \textit{Spitzer} 3.6\,$\mu$m) 
 photometry over the stellar continuum.
  By comparing the stellar population model parameters
 derived by fitting the multiwavelength photometry
 with population synthesis models and H$\alpha$ SFRs,
 we address several critical issues of star formation at $z\sim4$:
 (1) whether the dust extinction corrections applied for local starbursts
 are still valid at $z\gtrsim4$; (2) how star-forming galaxies at $z\gtrsim4$ 
 are related to massive galaxies at lower redshifts; 
and (3) whether mergers or stream-fed accretion is the dominant process
 that powers the star formation at $z\gtrsim4$.  
 Throughout this paper, we use a cosmology with 
 $\Omega_{M}=0.27$, $\Omega_{\Lambda}=0.73$,
 and $H_0=71\,\mbox{km}\,\mbox{s}^{-1}\,\mbox{Mpc}^{-1}$.

\section{Sample}

\subsection{ $\mathbf{3.8<z_{spec}<5.0}$ Galaxies in GOODS }

  From the various spectroscopic observing programs over the
 Great Observatories Origins Deep Survey (GOODS) North and South field 
 (including Vanzella et al. 2005, 2006, 2008; Ando et al. 2004),
 we select galaxies with secure spectroscopic redshifts between 
 $z=3.8$ and 5.0. The redshift window of $3.8<z<5.0$ is 
 chosen to select galaxies whose redshifted H$\alpha$ emission line
 enters into the \textit{Spitzer} IRAC channel 1 ($3.6\,\mu$m) band
 (Figure \ref{fig:introA}; see Section 3 for details). 
  Provided that the signal-to-noise ($S/N$) ratios in IRAC ch1 and ch2 
 ($4.5\,\mu$m) images are high enough, the underlying stellar continuum
 and hence the level of possible ch1 excess due to the H$\alpha$
 emission line are accurately derived in these galaxies through 
 spectral energy distribution (SED) fitting. 
The IRAC ch1 photometry point is excluded in the SED fitting. 
  Furthermore, in this redshift range, strong emission lines other than 
 H$\alpha$ -- e.g., H$\beta$, [OIII] -- do not affect the fluxes 
 in other filters since these lines fall in the gap between the
 IRAC ch1 and $K$-band.
The [OII]$\lambda\lambda$3727 emission line does  
 fall in the $K$-band at $z>4.4$, yet it would not affect 
 the determination of the stellar continuum. 
 We estimated the excess in $K$-band photometry due to the 
 [OII] emission line, considering the ratio between [OII] and H$\alpha$
 line (Mouhcine et al. 2005) and the estimated H$\alpha$ luminosity 
 for $z\sim4$ galaxies in our sample. The excess in $K$-band photometry
 due to the existing [OII] line over the stellar continuum is at most 
 10\,\%. This is much smaller than the case of IRAC ch1, where the excess
 in the flux density is larger than 30\,\% (see Section 3 for details).
  This is mainly due to the difference in equivalent widths for [OII] and 
 H$\alpha$, where [OII] equivalent widths are in general less than 
 50\,$\mbox{\AA}$ while H$\alpha$ equivalent widths are even larger than 
 100\,$\mbox{\AA}$ for actively star-forming galaxies.

  Over the total area of 330 arcmin$^2$ of the GOODS-North 
 (centered on $12^h 36^m 49.4^s, 62\arcdeg 12\arcmin 58.0\arcsec$) and
 the GOODS-South ($3^h 32^m 28.0^s, -27\arcdeg 48\arcmin 30.0\arcsec$),
 the initial number of galaxies with reliable spectroscopic redshifts
 of $3.8<z<5.0$ is 124: 53 galaxies in the GOODS-North field, 
 and 71 galaxies in the GOODS-South field. The surveyed comoving volume 
 at this redshift range is $1.2\times10^6$\,Mpc$^3$.
  Among these galaxies, we apply a $S/N$ ratio cut for IRAC ch1 
 and ch2 band ($S/N_{3.6\mu m},~ S/N_{4.5\mu m}>5$) to guarantee the 
 accurate determination of underlying stellar continuum.
  We also apply an isolation criteria, removing galaxies with close 
 neighbors ($z$-band magnitude brighter than $m+2$, when $m$ is 
 the magnitude of the galaxy in consideration) within $1\arcsec$ 
 from the final sample.
  This is to ensure that their \textit{Spitzer}/IRAC photometry
 is not contaminated by neighboring galaxies due to the fact that 
 the point spread function of \textit{Spitzer}/IRAC images 
is 1.5$\arcsec$ in channel 1 and 2, and $1.8-2.0\arcsec$ in 
channel 3 and 4, respectively.
  The value of $1\arcsec$ is less than the true FWHM in IRAC images,
 which is $\sim2\arcsec$. If we restrict the allowed radius for neighbors 
 to be $2\arcsec$ to match with the FWHM, $\sim25$\,\% of galaxies 
 would be thrown out from the sample. The postage stamp images for
 sample galaxies are presented later in Section 5.1
 (Figure \ref{fig:stamp_burst} and Figure \ref{fig:stamp_cont}).
  In the end, we are left with 31 (GOODS-North) and 43 
 (GOODS-South) isolated galaxies at $3.8<z<5.0$.

\subsection{ Optical to Mid-Infrared Photometry }

In this study, we used the optical-to-MIR merged 
photometric catalog over GOODS-South and North 
to obtain multi-band photometric data 
for those objects with spectroscopic redshifts in the range of 
$3.8<z<5.0$.

  The optical bands included are the \textit{HST}/ACS F435W, F606W, 
 F775W, and F850LP (generally referred to as $B, V, i$ and $z$) bands
(Giavalisco et al. 2004).
For optical photometry, sources were detected in $z$-band images, 
and photometry was carried out through matched apertures in other
ACS bands using SExtractor (Bertin \& Arnouts 1996). 
The range of $z$-band magnitudes is $23.5-26.5$\,mag.
   Most $3.8<z<5.0$ galaxies are clear $B$-dropouts with non-detections
 in $B$-band.
   In the \textit{Spitzer}/IRAC images of 
 GOODS\footnote{The \textit{Spitzer}/IRAC observation on GOODS 
 have been obtained as a part of the \textit{Spitzer} 
 Legacy program. The details of the observation and the data description
 is on the \textit{Spitzer} Legacy website.} (Dickinson et al. 2003),
 the flux densities are measured using $4\arcsec$ diameter apertures with
 aperture corrections applied afterward. In order to check whether the high
 confusion in IRAC bands matters in the photometry, we compared
 this ``aperture'' photometry with the result of ``TFIT'' photometry
 (Laidler et al. 2007), which is obtained using the matched apertures 
defined in $z$-band. 
For bright galaxies ($S/N>10$ in $3.6\,\mu$m
 and $4.5\,\mu$m), there are no systematic differences between the 
 aperture photometry and TFIT photometry. On the other hand, TFIT 
 results in higher flux measurements compared to aperture photometry 
 for low $S/N$ galaxies since it applies a larger correction factor to account
for the contribution from outskirts
of galaxies that are below the noise threshold in the \textit{Spitzer} images. 
Since we adopted a high $S/N$ cut in 
 sample selection and constrained the sample to be 
 relatively isolated (Section 2.1), the photometric uncertainties
 in the IRAC bands due to source confusion are small.  
 Nevertheless, we included the systematic uncertainties in IRAC fluxes 
 based on the difference between the aperture and TFIT photometry
 in the estimation of flux errors by adding the square of the
 systematic uncertainties to the photometric uncertainties.
  For IRAC ch3 and 4, if the source is undetected,
 we use the $3\,\sigma$ upper limit to the flux density in the $4\arcsec$
 aperture at the location of the sources.
 
  NIR photometry is important for these galaxies since
 the Balmer break, which is a critical constraint
 on the age of the stellar population,
 falls between the $K$- and the \textit{Spitzer}/IRAC 3.6\,$\mu$m bands. 
  We derived the NIR photometry of sample galaxies by matching the
 sample catalog with the available data from NIR observing programs.
  When available, we used \textit{HST}/NICMOS $J_{110}$ and $H_{160}$ fluxes 
 (Conselice et al. 2011; Dickinson 1999).
  For galaxies in GOODS-South, we used \textit{VLT}/ISAAC $J/H/K$-band
 photometry of galaxies (Retzlaff et al. 2010). More than 90\,\% of the
 $3.8<z<5.0$ galaxies have counterparts in NIR source catalogs. 
The NIR observations in GOODS-North were carried out with
Wide-Field Near Infrared Camera (WIRCAM) on the CFHT during the years 
$2006-2009$. This includes 27.4\,hours of integration in $J$-band made by a
Taiwanese group and 31.9\,hours of integration in $K_s$-band obtained by
Hawaiian and Canadian observing programs. The WIRCAM $K_s$-band data has recently
been published by Wang et al. (2010), but here we use our own version of the
reductions (Lin, L. et al. \textit{in prep}).
The data were first pre-processed using
the SIMPLE Imaging and Mosaicking Pipeline (Wang et al. 2010), and then
combined to produce deep stacks with SCAMP and
SWARP\footnote{http://www.astromatic.net/}. The resulting 5$\sigma$
limiting magnitudes using 2$\arcsec$ diameter circular apertures reach 
$J$ = 24.8\,mag(AB) and $K_s$ = 24.35\,mag(AB).
For galaxies 
 with non-detections in the NIR, 
 we provide $3\,\sigma$ flux density upper limits
 in the $4\arcsec$
 aperture measured in the corresponding images. 

  The multi-band photometry for the 74 galaxies
at $3.8<z<5.0$ considered here, 
 including upper limits when needed, is presented in Table \ref{tab:phot}.

\section{H$\alpha$ Emitter Identification}

  The basic assumption in this photometric study of the H$\alpha$ emission
 in $z\sim4$ galaxies is that the redshifted H$\alpha$ emission line causes
 an enhancement in the broad-band flux density. Specifically, the 
 H$\alpha$ emission line is redshifted into the \textit{Spitzer}/IRAC ch1 band
 at $3.8<z<5.0$. We define such galaxies with a large rest-frame H$\alpha$ 
 equivalent width that yields an excess in IRAC ch1 photometry compared
 to the stellar continuum as ``H$\alpha$ Emitters (hereafter HAEs)''. 
 In this section, we show that the observed photometric excess is due to 
 the redshifted H$\alpha$ emission and describe the procedure used to derive 
 the H$\alpha$ line luminosity and equivalent width of HAEs.

  \subsection{IRAC Color}

\begin{figure*}
\begin{center}
 \subfigure{ \label{fig:introA} \includegraphics[width=130mm]{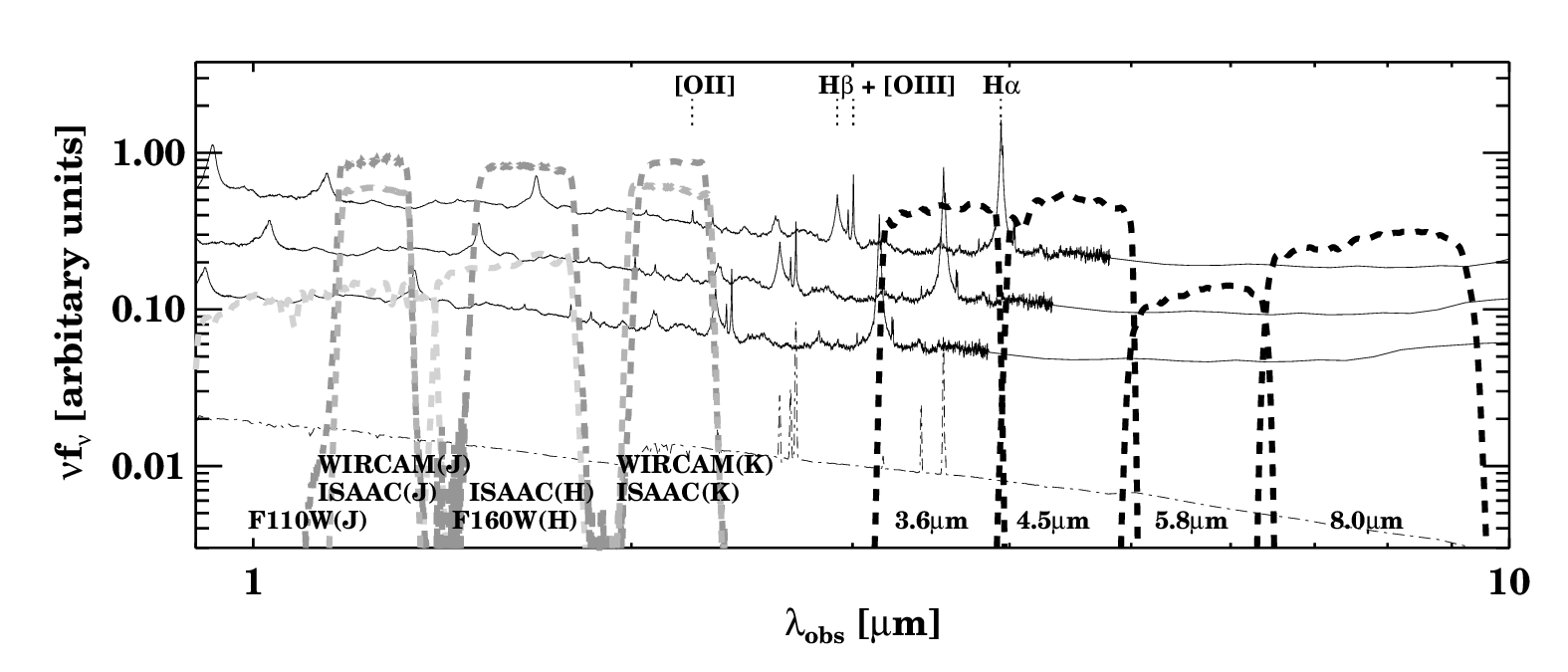} }
 \subfigure{ \label{fig:introB} \includegraphics[width=130mm]{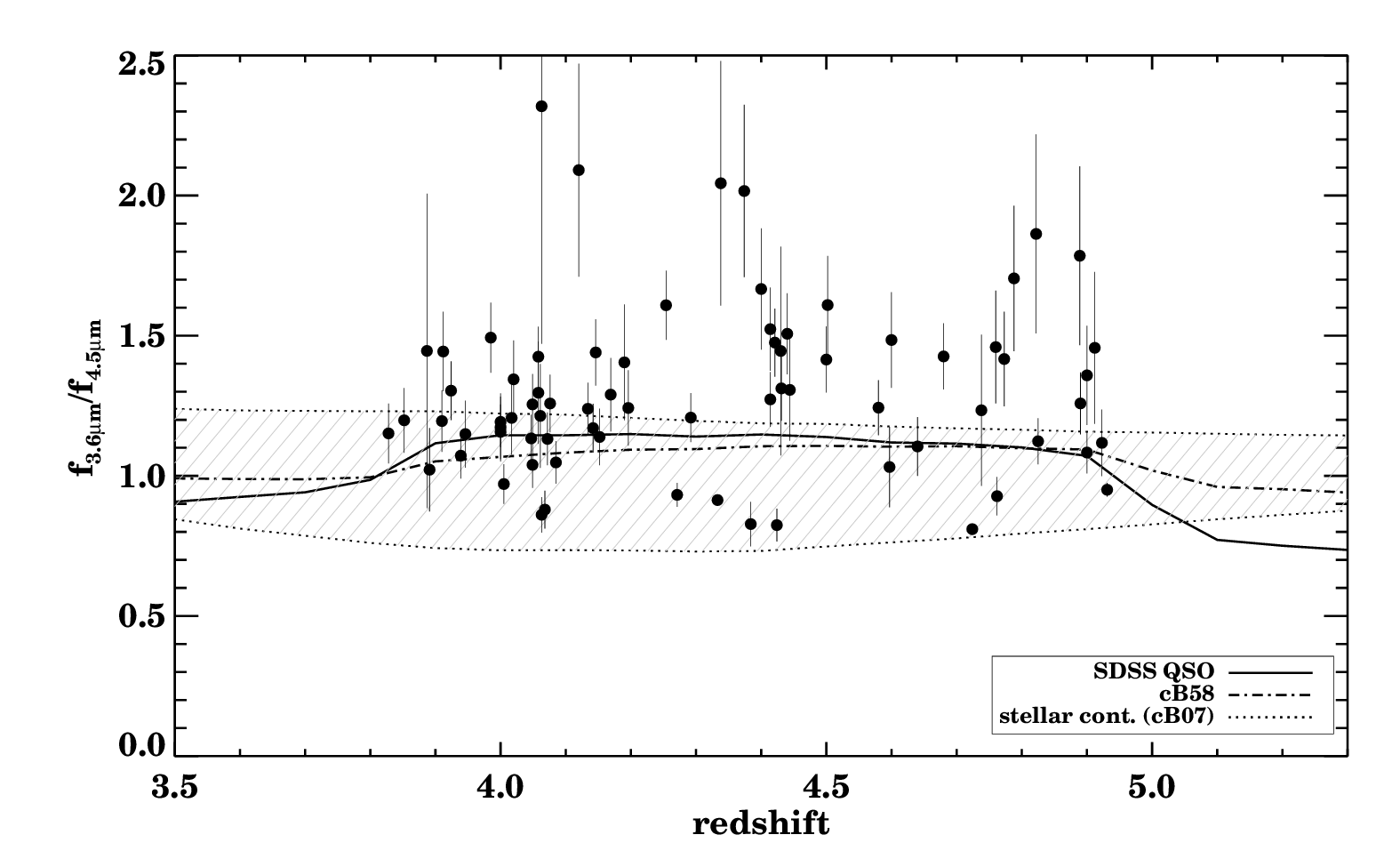} }
  \caption{
    (\textit{Top}): 
   The expected spectra of objects with emission lines at $3.8<z<5.0$.
    The \textit{solid} lines indicate the spectrum
   of the Sloan Digital Sky Survey composite quasar (Vanden Berk 2001), 
   redshifted to $z=3.8, 4.4,$ and 5.0, respectively (from bottom to top).
   The \textit{dot-dashed} line is the spectrum of the $z=2.73$ star-forming
   galaxy MS1512-cB58 (Teplitz et al. 2004) redshifted to $z=4.4$.
   The H$\alpha$ emission line enters IRAC ch1 ($3.6\,\mu$m band) at
   $z\sim3.8$, and exits ch1 at $z\sim5.0$.
    (\textit{Bottom}):
    The change of the flux ratio $f_{3.6\mu m}/f_{4.5\mu m}$
   as a function of redshift.
    The \textit{solid} line and \textit{dot-dashed} line indicate
   $f_{3.6\mu m}/f_{4.5\mu m}$ for the redshifted SDSS quasar and cB58, respectively.
    As the H$\alpha$ emission line enters 
   IRAC ch1, the ratio $f_{3.6\mu m}/f_{4.5\mu m}$ increases
   at $3.8<z<5.0$.
    The \textit{dotted} lines, and the hatched region indicate the range of 
   $f_{3.6\mu m}/f_{4.5\mu m}$ spanned by stellar population synthesis model
   templates (CB07; Bruzual 2007) 
   which are constrained by the rest-frame UV to optical SEDs of the spectroscopically 
   selected $3.8<z<5.0$ galaxies studied in this paper. The observed $f_{3.6\mu m}/f_{4.5\mu m}$
   values of our sample of galaxies at $3.8<z<5.0$ 
   are overplotted as \textit{filled circles} with error bars.
   A flux density excess at 3.6\,$\mu$m
is clearly evident. We use full SED fits 
   to the multiwavelength photometry to measure the exact contribution of 
   line emission to the broadband photometry. 
   }
\end{center}
 \end{figure*}

  The amount of excess in the \textit{Spitzer}/IRAC ch1 band 
reaches $\sim10$\,\% 
 if the galaxy observed at $3.8<z<5.0$ is 
a star-forming galaxy with H$\alpha$
 equivalent width of $100\,\mbox{\AA}$ (Figure \ref{fig:introA}).
  This is illustrated in Figure \ref{fig:introB}, 
 where the expected $f_{3.6\mu m}/f_{4.5\mu m}$ ratio of
 model galaxy templates are plotted as a function of redshift.
  The dot-dashed line is the expected $f_{3.6\mu m}/f_{4.5\mu m}$ ratio
 calculated using the template of star-forming galaxy MS1512-cB58 
 with $EW_{H\alpha}=100\,\mbox{\AA}$. The $f_{3.6\mu m}/f_{4.5\mu m}$ ratio
 increases significantly at $3.8<z<5.0$, where the H$\alpha$ line
 is redshifted into $3.6\,\mu$m band. The Sloan quasar template (solid line)
 shows a similar trend as the star-forming galaxy.

  Figure \ref{fig:introB} also reveals if the increase in $f_{3.6\mu m}/f_{4.5\mu m}$
 can be explained using changes to the properties of the stellar population only,
 by showing the bluest and the reddest $f_{3.6\mu m}/f_{4.5\mu m}$ colors
 possible from the stellar continuum for
 Charlot \& Bruzual 2007 (hereafter CB07; Bruzual 2007) population synthesis model
 spanning a range of starburst age, star formation history, and dust extinction. 
  The range of parameters used in stellar population synthesis model
is constrained by the rest-frame UV to optical colors 
 of our sample galaxies:
 ages between 3\,Myr and $\sim1000$\,Myr old,
 $E(B-V)$ between 0.0 and $\sim0.4$.
   The overplotted data points in Figure \ref{fig:introB} are the 
 observed $f_{3.6\mu m}/f_{4.5\mu m}$ ratio for our sample galaxies
 at $3.8<z<5.0$. Among 74 galaxies, 63 sources show $f_{3.6\mu m}/f_{4.5\mu m}$
 ratios larger than 1, and 58 sources show $f_{3.6\mu m}/f_{4.5\mu m}$ ratios
 larger (i.e., bluer ch1$-$ch2 color) than that of MS1512-cB58.
  The large $f_{3.6\mu m}/f_{4.5\mu m}$ ratios that are observed could be 
 produced in many cases using younger stellar populations
and less dust extinction.
 Yet, when the measured broad-band flux is larger than $\sim50-150$\,\% 
 of the expected stellar continuum
 (e.g., Chary, Stern, \& Eisenhardt 2005; Zackrisson, Bergvall, \& Leiter 2008), 
 it is impossible to reproduce the observed excess by changing the
 stellar continuum only. This requires the inclusion of strong emission lines
 to fully explain the observed broad-band photometry excess. 
We fit the spectral energy distribution 
 of each galaxy to constrain the age of the stellar population 
 and dust extinction and, in addition, measure the contribution of the
 H$\alpha$ emission line to the broad-band flux density.

  \subsection{Spectral Energy Distribution Fitting}

  Although the blue IRAC color (high $f_{3.6\mu m}/f_{4.5\mu m}$) 
 is a sign that suggests the presence of strong H$\alpha$ emission in 
 these galaxies, the color alone is insufficient to identify HAEs 
 since the effect of stellar population age and dust extinction 
 to the broad-band color is not constrained. In order to determine 
 the underlying stellar continuum in IRAC ch1, and to disentangle 
 the relative contributions from stellar radiation and from line emission 
 to the ch1-band flux, we fit the multi-band photometry of all $3.8<z<5.0$
 galaxies using CB07 stellar population synthesis model.

  We used the galaxy models of fixed metallicity $0.2\,Z_{\odot}$,
 considering the relatively low metallicity of high-redshift 
 star-forming galaxies (e.g., Pettini et al. 2002).
Salpeter initial mass function (Salpeter 1955) is used to generate galaxy templates. 
  Star formation histories of the model galaxies are varied using 
 different values for the exponentially decaying star formation time scale, 
 i.e., $\tau =$ [ 1, 10, 20, 30, 40, 50, 60, 70, 80, 90, 100, 200, 300, 400,
 500, 600, 700, 800, 900, 1000 ] Myr and $\tau=10^5$ Myr 
 (constant star formation).  The ages of model galaxies are varied to
 span the range between 1 Myr and the age of the universe at the given 
 redshift. For internal reddening of the model galaxy templates, we used 
 two forms of extinction law: the extinction law for local 
 starbursts (SB; Calzetti et al. 2000) and the extinction law for 
the Small Magellanic Cloud (SMC; Pr\'evot et al. 1984). The SMC extinction law
 is expected to describe low-metallicity galaxies well, which results in 
 more rapidly increasing extinction with decreasing wavelength in the UV. 
  The internal dust extinction is allowed to vary between
 $E(B-V)= 0$ and $E(B-V)= 0.6$ for both extinction laws.  
  The best-fit template is determined through $\chi^2$ minimization
with $\chi^2$ is defined as follows:

\begin{equation}
 \chi^2 = \displaystyle\sum_{\rm filter} \frac{ (f_{\rm obs} - f_{\rm mod})^2 }{ \sigma_{\rm obs} ^2 } 
\end{equation}

 \noindent The model flux $f_{\rm mod}$ is calculated by convolving 
 the filter response curve to the galaxy template, after applying
 the intergalactic medium (IGM) attenuation (Madau 1995).

  Because we wish to evaluate any potential flux excess in 
 \textit{Spitzer}/IRAC ch1 (3.6\,$\mu$m) from an emission line, 
 we initially exclude the ch1 photometry in the SED fitting.
 The photometry data points used in the fits are \textit{HST}/ACS 
 $B, V, i,$ and $z$-band flux,
 NIR flux (\textit{HST}/NICMOS F110W and F160W when available;
 \textit{VLT}/ISAAC $J, H,$ and $K$-band for GOODS-South;
 \textit{CFHT}/WIRCAM $J$ and $K$-band for GOODS-North) and
 \textit{Spitzer}/IRAC ch2, ch3 (5.8\,$\mu$m), and ch4 (8.0\,$\mu$m) flux.
  The photometry of each galaxy used in SED fitting is presented in 
 Table \ref{tab:phot}, as described in Section 2.2. 
  We also list the output parameters from SED fitting in 
 Table \ref{tab:params}: stellar mass, $E(B-V)$, stellar population age, 
 $e$-folding time scale of star formation $\tau$,
and UV spectral slope $\beta$.
  These parameters are derived independently for both SB and SMC 
 extinction laws. We derived the UV-slope $\beta$ by fitting
 the best-fit model spectra at UV wavelengths ($1500-2500\,\mbox{\AA}$) 
 using a power-law ($f_{\lambda} \propto \lambda^\beta$).  

  The templates used in the SED fitting (CB07; Bruzual 2007) 
 do not incorporate a contribution from nebular emission. 
 As we mentioned in Section 2.1, most of the strong nebular emission lines
 [OII], H$\beta$ and [OIII] fall in the gap between the filters
 at this redshift range  and thus do not affect the broad-band photometry at other wavelengths
 or have a negligible contribution to the broad-band flux density, typically 
 of order $<10$\,\%.      
  The role of nebular continuum on the other hand, begins to be
 significant at the 20\,\% level only longward of $4.5\,\mu$m in the observed frame
 for young, star-forming galaxies (Zackrisson, Bergvall, \& Leitet 2008).
  Therefore, we conclude that our SED fits to the photometry
 with the exception of the IRAC ch1 flux density
 is a reasonable approach to determine the
 stellar population properties of $3.8<z_{spec}<5.0$ galaxies, with
 negligibly small effects from nebular continuum emission and line
 emission other than H$\alpha$.

  Since a number of studies claim the significance of an old stellar 
 component in high-redshift galaxies, especially for red and massive galaxies
 (e.g., Muzzin et al. 2009), we tested a two-component
 stellar population model fits for these galaxies.
The two-component model is made up of the sum of a 
young stellar population and an old stellar population.
We used a
5\,Myr-old stellar population forming stars 
at a constant rate of 30\,$M_{\odot}$ yr$^{-1}$
as the young component, and a single burst, passively evolving stellar population 
with $z_f=20$ as the old component. 
The SED fitting procedure was the same as the case of 
single component SED fitting, except the stellar mass ratio between the two component is 
a free parameter in the two-component SED fitting.
  Most of the galaxies prefer the single-component fits
 while only $\sim10$\,\% of the galaxies show lower $\chi^2$ values for 
two-component fitting. The galaxies that prefer the two-component
 fitting have relatively large $\chi^2$ values compared to other objects, i.e. 
 $\chi^2>5$, suggesting biases in the photometry of the sources themselves.
  Therefore, we only use the results from single-component fitting 
 in the following analysis.

  In Figure 2, 
 we present the best-fit SED fitting results for all 74 galaxies at 
 $3.8 < z < 5.0$. The filled squares are the observed photometry points
 given in Table \ref{tab:phot}, and the overplotted lines are the best-fit
 SED templates. The best-fit parameters -- dust extinction,  stellar population 
 age -- are shown along with the $\chi^2$ value. 
  In addition to the comparison between the best-fit SED and  
 the multi-band photometry data points, a ``residual plot''
 showing the differences between the observed flux density
 and the model flux density, as a function of wavelength, is
included for each object. 
  The $y$-axis in the residual plot indicates
 $ ( f_{\rm obs} - f_{\rm mod} )/\sigma_{\rm obs}  $,
 i.e. the discrepancy between the model flux density
 and the observed flux density divided by the observed flux density error. 

  Clearly, a large fraction of our sample galaxies show 
 a significant excess in \textit{Spitzer}/IRAC ch1
 (3.6\,$\mu$m; Figure 2), 
 while no excess is observed in other filters. 
  To quantify this excess in a particular filter, for instance in
 \textit{Spitzer}/IRAC ch1, we define a significance factor $S$ as follows: 

\begin{equation}
  S_{\rm ch1} = \frac{ \Delta_{\rm ch1}}{ \displaystyle \frac{1}{N} \sqrt{\displaystyle \sum_{i}^{\rm filter} \Delta_{i}^2 } }
\end{equation}

 \noindent In this equation, $\Delta$ is defined as
 $ \Delta = (f_{\rm obs} - f_{\rm mod})/\sigma_{\rm obs} $.
 The denominator indicates the mean of the residuals in filters
 other than the IRAC ch1 -- this is basically an indicator showing
 the quality of the SED fitting. Therefore, the factor $S_{\rm ch1}$ represents
 the significance of the ch1 excess compared to typical residuals
 at other wavelengths from the SED fitting.

  We show the distribution of the significance factor $S_{\rm ch1}$ 
 in Figure \ref{fig:sfactor_hist}. The distribution of factor $S$
 in other filters ($V$, $J$, $K$, and IRAC ch2) are also presented 
 for reference.
In all but one galaxy, S15920, 
the factor $S_{\rm ch1}$ is larger than 0. 
  This differs 
from the other filters,
 in which the distribution of $S$ is close to a Gaussian distribution
 with mean value of $S=0$ -- confirming that the SED fitting 
is reasonable. The mean value of $S_{\rm ch1}$ larger than 0 
 is derived by excluding IRAC ch1 photometry 
during the SED fitting (solid line in Figure \ref{fig:sfactor_hist});
as described above. 
However even if ch1 photometry is included in 
the SED fitting (dashed line), the result is the same. 
The $S$ for other filters is distributed around $S=0$ while the $S_{\rm ch1}$
is distributed around $\langle S_{\rm ch1} \rangle > 0$,
although the mean value
 $\langle S_{\rm ch1} \rangle$ decreases by a factor of two compared
 to fits which exclude the ch1 photometry.
  The $S_{\rm ch2}$ values tend to
be negative when 
ch1 photometry is included in the SED fitting,
 since the stellar continuum adjusted to match the high ch1 flux
 overpredicts the ch2 flux.
  Therefore, it is clear that there exists an excess in 3.6\,$\mu$m photometry
 for $3.8<z<5.0$ galaxies. 
  Among 74 galaxies, 
 64 galaxies have `reasonable' SED fitting results, 
 i.e., $\chi^2$ less than 5. All 64 of these galaxies have 
 $S_{\rm ch1}$ greater than 0; thus, we call these 64 galaxies 
 with positive residual in IRAC ch1 ``HAE candidates''. 
We classify the subset of 47 galaxies with $S_{\rm ch1} >10$
as \textit{bona-fide} HAEs. Thus, at least $\sim65$\,\% (47/74) of 
the spectroscopically confirmed galaxies at $3.8<z<5.0$ are found to be HAEs.
 
  The most plausible explanation for the 3.6\,$\mu$m excess
is the addition of 
H$\alpha$ emission line to the stellar continuum.
  Other possibilities for this excess include the photometric errors
 produced by uncertainties in the zero point calibration,
 the aperture correction, and/or the color correction.
  The uncertainties in zero point calibration in
 IRAC 3.6\,$\mu$m and 4.5\,$\mu$m are less than 2\,\% 
according to the \textit{Spitzer} Data Users' 
 Manual\footnote{http://ssc.spitzer.caltech.edu/dataanalysistools/cookbook/}.
  The uncertainties in aperture correction for 
 IRAC 3.6\,$\mu$m and 4.5\,$\mu$m are less than 5\,\%.  
  Moreover, the point-spread-function of IRAC 3.6\,$\mu$m and 4.5\,$\mu$m
are comparable to each other; thus,
the aperture correction would be unlikely to cause a systematic
 flux density increase in IRAC 3.6\,$\mu$m compared to IRAC 4.5\,$\mu$m. 
   Finally, the color correction is less than 10\,\% for 
 black-body spectra of any temperature
as well as for most of the galaxy SEDs. 
  Considering all these photometric uncertainties that might affect
the 3.6\,$\mu$m photometry, we conclude that the effects are negligible and
 3.6\,$\mu$m excess should be interpreted as the contribution of 
 H$\alpha$ emission from these galaxies.

 \subsection{Estimation of H$\alpha$ Line Flux and Equivalent Width}

  The excess in the 3.6\,$\mu$m photometry compared to the determined 
 stellar continuum from SED fitting is converted into an H$\alpha$ line flux.
  We added a Gaussian emission line to the best-fit stellar continuum as a 
 proxy for H$\alpha$ emission line, and then increased the amplitude
 of the Gaussian to determine the H$\alpha$ line flux that reproduces 
 the observed excess in IRAC 3.6\,$\mu$m flux density. 
 The width of the Gaussian does not affect the derived line flux,
since the line flux is an integral over wavelength.
  The H$\alpha$ luminosity derived in this way is the sum of H$\alpha$
 and [NII] doublet at 6583, 6548\,$\mbox{\AA}$. In order to correct
 the derived $L(H\alpha)$ for [NII] contamination, we 
assume the relation [NII]$\lambda\lambda$ 6583/H$\alpha = 0.3$ 
 and [NII]$\lambda\lambda$6583/[NII]$\lambda\lambda$6548 $=3$
 (Gallego et al. 1997).
This ratio is obviously metallicity-dependent, yet since
we do not have strong evidence constraining the metallicity of 
HAEs, we use the conventional [NII]/H$\alpha$ ratio for [NII] correction.
 The luminosity, the equivalent width, 
 and the flux of H$\alpha$ line used throughout this paper are all
 corrected for [NII], using this correction: 
 $f_c (H\alpha) = 0.71\times f(H\alpha + [NII])$.

  We present the H$\alpha$ luminosity and the equivalent width of 
 all (74) galaxies at $3.8<z<5.0$ in Table \ref{tab:params}. 
 One object (S15920) has $S_{\rm ch1} < 0$ thus its H$\alpha$ luminosity 
 could not be measured. Among these 74, 64 galaxies are considered 
 to be galaxies with reasonable SED fitting and 47 are considered as 
significant (\textit{bona-fide}) HAEs as noted in Section 3.2. 
  The derived $L(H\alpha)$ of all $3.8<z<5.0$ galaxies ranges between
 $10^{42.4} < L_{H\alpha} [\mbox{erg}~ \mbox{s}^{-1}] < 10^{43.8}$,
 corresponding to SFRs of
 $20-500\,M_{\odot}$ yr$^{-1}$ assuming the SFR calibration 
 of Kennicutt (1998a). Figure \ref{fig:prop}a shows the luminosity
 distribution of $z\sim4$ HAE candidates
 (64 galaxies, with reliable SED fitting and $L(H\alpha)$ estimates)
 and the H$\alpha$-inferred SFRs
 as a function of stellar mass which is derived from the SED fitting.   
  The HAE candidates are found to be
 very actively star-forming galaxies: 
most have H$\alpha$ emission that is substantially 
 stronger than the characteristic luminosity $L_{*,~H\alpha}$ for galaxies at lower redshifts 
 (e.g., $10^{42.83\pm0.13}$ erg s$^{-1}$ at $z=2.23$, $10^{42.46\pm0.19}$ erg s$^{-1}$ at $z=1.4$;
 Geach et al. 2008; Shim et al. 2009).

  In Table \ref{tab:params}, the uncertainties in $L_{H\alpha}$ are
 presented as well.
  The major uncertainties in the derived $L_{H\alpha}$ come from:
 (1) uncertainties in the photometry at 3.6\,$\mu$m, and 
 (2) uncertainties in the determination of stellar continuum level
 at 3.6\,$\mu$m. Since our sample galaxies have relatively high $S/N$
 in 3.6\,$\mu$m (mostly $>10$), the dominant origin of
 the uncertainties is the latter.  
We derive the uncertainties using a Monte Carlo analysis. 
For each galaxy, one hundred sets of multi-band photometry were
generated by simulating a Gaussian distribution of flux densities 
around the measured flux densities. That means, the photometry sets 
are [ $S_B^i$, $S_V^i$, $S_I^i$, ... $S_{\rm ch4}^i$ ] with $S_{\rm filter}^i$
being drawn from a Gaussian distribution of flux density values at the given filter 
which reproduces the photometric uncertainty at that filter. 
 We then re-derive the best-fit stellar continuum through the SED fitting 
 as before, and re-estimate the value of $L_{H\alpha}$.
  From the distribution of the one hundred sets of the derived $L_{H\alpha}$,
 we take the standard deviation as the uncertainty in the H$\alpha$ luminosity. 
 The H$\alpha$ luminosity uncertainties in Table \ref{tab:params} 
 are derived following this procedure.

  \subsubsection{Validation of Photometric Estimates}

  We tested the validity of our $EW(H\alpha)$ estimates
 from the broad-band photometric excess using galaxies at
 lower redshifts that have spectroscopic measurements of $EW(H\alpha)$.
  The galaxies used in this verification are
 15 galaxies at $z\sim2$ (Erb et al. 2006) with available multi-band photometry at
 $U_n$, $G$, $R$, $J$, $K$, \textit{Spitzer}/IRAC 3.6\,$\mu$m and
 4.5\,$\mu$m-band (Reddy et al. 2006).
  These galaxies are vigorous star-forming galaxies selected in the rest-frame UV.
  The $EW(H\alpha)$s of these galaxies are measured to be
 $70-300\,\mbox{\AA}$, which are comparable with those of our HAE candidates at $z\sim4$.
   At $2.1<z<2.5$, the H$\alpha$ emission line is redshifted
to $\sim2.2\,\mu$m, causing the flux density excess to be in the $K$-band.

  Figure \ref{fig:method} shows the comparison between
 spectroscopically measured EW ($EW_{\rm spec}$) and SED-fitting
 derived EW ($EW_{\rm phot}$).
  We used the same method applied to the $z\sim4$ HAE candidates
 to derive $EW_{\rm phot}$, except using $K$-band excess instead of IRAC ch1-band.
 The $EW_{\rm phot}$ reproduces the $EW_{\rm spec}$ relatively well, within
 a scatter of $\sim0.25$ dex. There is no significant systematic
 difference between the two EWs.
This validation ensures the reliability
 of the derived $EW(H\alpha)$ of the $z\sim4$ HAE candidates.
 Almost all of the $z\sim4$ HAEs have $EW_{\rm phot}$ larger than
 $300\,\mbox{\AA}$, which is the maximum value in the $z\sim2$ galaxy sample (Erb et al. 2006).
The relative accuracy of the H$\alpha$ line flux measurement
 from broad-band photometry would be better for stronger lines.
 Therefore, the consistency between $EW_{\rm phot}$ and $EW_{\rm spec}$,
 even for weak H$\alpha$ lines ($<300\,\mbox{\AA}$) at $z\sim2$,
 supports that this photometric method works
for strong lines in $z\sim4$ HAEs.

  There are two objects that show large discrepancies
between $EW_{\rm spec}$ and $EW_{\rm phot}$ (one at $EW_{\rm spec} = 90\,\mbox{\AA}$
 and the other at $266\,\mbox{\AA}$). Their $J-K$ colors are bluer
 than the color
 expected by the star-forming galaxy template (see the inset plot of
 Figure \ref{fig:method}).
  It appears that the EW discrepancies of these sources are due
 to large photometric uncertainties in the $J$ and $K$ bands.

\subsection{Selection Bias for HAEs from Broad-band Photometric Surveys }

  The HAE candidates are galaxies with large $EW(H\alpha)$,
 i.e., $EW(H\alpha)=140-1700\,\mbox{\AA}$ (Table \ref{tab:params}; Figure \ref{fig:prop}b). 
 Since the EW is 
the ratio between the line luminosity
 and the continuum luminosity 
at the wavelength of the emission line,
photometric uncertainties in IRAC ch1 place a limit on the detectable
 $EW(H\alpha)$. We calculated the minimum $EW(H\alpha)$ that
 could be detected using the criterion that 
 an HAE candidate should show IRAC ch1 excess at least three times 
 larger than the uncertainties, including both photometric uncertainty 
 and stellar continuum uncertainty. The $EW(H\alpha)$ limit varies
 between $70-350\,\mbox{\AA}$ for different galaxies. This limit is 
 relatively large compared to the observational EW limit of 
 narrow-band surveys (e.g., Geach et al. 2008; 50\,$\mbox{\AA}$),
 yet comparable to the limit for low-resolution grism surveys
 (e.g., Shim et al. 2009; 150\,$\mbox{\AA}$). 
  This shows that HAE candidates selected using broad-band photometric
 excess are biased towards strong H$\alpha$ emitters, 
comparable to the strongest emission-line galaxies selected in grism surveys. 

 If the star formation in a galaxy is instantaneous,
 then such strong H$\alpha$ emission
 ($EW_{H\alpha}\gtrsim100\,\mbox{\AA}$) is short-lived,
 lasting for only $\sim5$ Myr after the starburst (Leitherer \& Heckman 1995).
  On the other hand, if the star formation is extended
 (i.e., continuous star formation),
 this phase of large H$\alpha$ equivalent width lasts longer. 
  Considering only
 instantaneous star formation for galaxies, our redshift window of
 $3.8<z<5.0$ spans $\sim470$\,Myr of cosmic time,
 implying the chance for selecting such HAEs to be only $\sim1$\,\%.
   The observed fraction of HAEs among spectroscopically selected galaxies
 at $3.8<z<5.0$ is more than an order of magnitude larger than that (47/124), 
which suggests
 that the star formation timescale of $z\sim4$ galaxies appears
 to be extended, not instantaneous.
Note however that
due to the fact that the target selection
 for spectroscopic observation is mostly based on the UV colors
 (i.e., dropout selection),
it is also probable that our sample is biased towards young galaxies
 with large ongoing SFRs.

  \section{Origin of Strong H$\alpha$ Emission}

  We have demonstrated that a significant fraction of $z\sim4$ galaxies show 
 strong H$\alpha$ emission. The H$\alpha$-derived SFR and the UV-derived SFR 
 using the relation in Kennicutt (1998a) 
suggest a median ratio of $\langle SFR(H\alpha)/SFR(UV) \rangle \sim6.1$,
although with a large scatter of 4.9. 
The H$\alpha$-to-UV luminosity ratio of $z\sim4$ HAEs 
 is on average larger than that of local starbursts (Figure \ref{fig:lumratio}). 
  What could be the origin of such strong H$\alpha$ emission in these galaxies? 
  In this section, we present several possibilities
to explain the strong H$\alpha$ emission in HAEs.

 \subsection{The Effect of Dust Extinction}

  The estimation of the internal reddening for high-redshift galaxies is
generally based on the UV spectral slope $\beta$, following the work
 on local starbursts (e.g., Meurer, Heckman, \& Calzetti 1999).
 The method requires the unverified assumption of the similarity
 in the intrinsic SEDs of local and high-redshift star-forming galaxies
 as well as a similarity in the dust obscuration properties. 
  The IR luminosity to UV luminosity ratio of $z\sim2-3$ star-forming galaxies 
 (e.g., Siana et al. 2009; Reddy et al. 2010) have suggested 
 possible discrepancies between dust extinction laws for high-redshift 
 star-forming galaxies and local star-forming galaxies. 
  The ratio between the rest-frame optical emission line and
 the rest-frame UV continuum provides an independent test for the validity of 
the assumed similarity between low- and high-redshift star-forming galaxies.

  The derived H$\alpha$ line-to-UV continuum ratio 
(hereafter line-to-continuum ratio) is compared with UV spectral slope
$\beta$ in Figure \ref{fig:lumratio}.
 The line-to-continuum ratio is not to be confused with an equivalent width 
since the line flux and continuum are measured at different wavelengths.
   The number of HAE candidates with reliable SED fitting results 
 ($\chi^2<5$) is 64, including one object with an X-ray counterpart (N23308). 
 This object lies out of the range plotted in Figure \ref{fig:lumratio}.
  Though there exists a considerable
 scatter, the $\beta$ and the line-to-continuum ratio correlate with
a Spearman's coefficient ($\rho_s$) of 0.52 which
is significant at the level of $>70$\,\%.  
  Also shown is the $\beta$ vs. line-to-continuum ratios for
 local starburst galaxies, which is taken from Meurer, Heckman, \& Calzetti 
 (1999; originally observed by Storchi-Bergman et al. 1995, McQuade et al. 1995). 
 The $z\sim4$ HAE candidates lie in clearly different regions
of $\beta$ vs. line-to-continuum space compared to local starbursts. 
 This is quantitatively supported by a Kolmogorov-Smirnov test, 
 which shows that the probability of the two groups
having the same correlation between the $\beta$ 
 and line-to-continuum ratio is less than 1\,\%. 
 
  In order to assess the origin of this difference, 
 we overplot several model tracks on the
 observed data points in Figure \ref{fig:lumratio}a.
 The model tracks are reddened assuming different extinction laws:
the starburst extinction law (Calzetti et al. 2000),
the SMC extinction law (Pr\'evot 1984), and the LMC extinction law
 (Fitzpatrick 1986) without the 2175\,$\mbox{\AA}$ graphite feature. 
When each extinction
 law is applied, different factors are used for the line and the
 continuum, following the statement in Calzetti (2001) 
 that the stellar continuum suffers roughly
 half of the dust reddening suffered by the ionized gas due to the mixed
 dust geometry. Thus, the observed $F_{H\alpha}$ is modulated using
 $E(B-V)_{\rm gas}$ at the H$\alpha$ wavelength ($6563\,\mbox{\AA}$),
defined by $E(B-V)_{\rm star} = 0.44 E(B-V)_{\rm gas}$\footnote{There 
exist controversy on the value of the factor
 $E(B-V)_{\rm star}/E(B-V)_{\rm gas}$, from 0.44 (Calzetti 2001) to $\sim1$ (Reddy et al. 2010).}.
 
  The high line-to-continuum ratio is reproduced by young galaxies
 with long star formation time scales.
 With the same $e$-folding time
 $\tau$ for star formation, there is a factor of 3 difference in 
 line-to-continuum ratio between galaxies of 1 Myr age and 10 Myrs age. 
 For the same age of 100 Myrs old, there is a factor of 2.5 difference 
in line-to-continuum ratio between instantaneous starburst galaxies and
 continuously star-forming galaxies. 
  In addition, different extinction laws produce differences in 
 line-to-continuum ratio.
The SMC/LMC extinction laws have slightly steeper
$\beta$ vs. line-to-continuum relations than the SB extinction law.
The slope is 0.24 and 0.27 for SMC and LMC extinction laws, respectively, 
and is 0.18 for the SB extinction law. 
 Figure \ref{fig:lumratio}b compares
the $\beta$ vs. line-to-continuum ratio for local starbursts 
 and HAE candidates with those expected by the SMC and SB extinction laws. 
 The slope in $\beta$ vs. line-to-continuum ratio for $z\sim4$ HAE 
 candidates is $0.27\pm0.07$.
   This indicates that the steeper shape of the extinction curve with
 decreasing wavelength that is found in the SMC/LMC
 describes the high line-to-continuum ratio observed in $z\sim4$ HAE
 candidates marginally better.
Thus, it is clear that HAE candidates prefer
 the SMC extinction law than the SB extinction law while the slope 
 uncertainty is greater than the slope difference between the SMC and
 SB extinction law. We note that the relation between line-to-continuum ratios
 and $\beta$ of local starbursts is also closer to SMC extinction law than
 SB extinction law, which is already mentioned in Meurer, Heckman, \& Calzetti (1999)
 yet the reason is not clearly understood.

 \subsection{The Effect of Age}

 Other dominant factors that affect H$\alpha$ equivalent width
are the stellar population age 
and the star formation timescale.
 For galaxies with brief bursts of star formation, the
 $EW(H\alpha)$ drops to 1\,\% of the initial value
 (ranging between $1600-3200\,\mbox{\AA}$ depending on the metallicity)
 after $\sim10$ Myrs (Leitherer et al. 1999).
 For galaxies with constant
 star formation, $EW(H\alpha)$ decreases to 10\,\% of the initial 
 value (ranging between $1800-3200\,\mbox{\AA}$ depending on the metallicity)
 after $\sim100$ Myrs.

  Figure \ref{fig:ew_age} shows a comparison between the derived
 H$\alpha$ equivalent width and the age of the stellar population compared to 
 population synthesis models of the equivalent width for different 
 star formation histories. 
  If the derived ages are assumed to be reliable, 
we find that even though the observed $EW(H\alpha)$ of $140-1700\,\mbox{\AA}$ 
 can be easily reproduced by young and continuously star-forming galaxies,
 the stellar population in the $z\sim4$ galaxies are quite heterogeneous; 
 the minority are bursty and younger than $\sim10$\,Myrs old while
 the majority are extended with an $e$-folding timescale of star formation 
 which is comparable to the Hubble time at their redshift. 
  A comparison with the $z\sim2$ galaxies which are shown in the plot reveals
 that a larger fraction of $z\sim4$ galaxies (40\,\%) show burst-like star formation histories
 compared to the $z\sim2$ galaxies (7\,\%).
We note that the $z\sim2$ galaxies were fit with extended star-formation 
histories (Erb et al. 2006) which tends to result in 
larger ages, and a smaller burst fraction. However, when we fit the $z\sim4$ HAEs 
with only extended star formation histories, we still get a larger fraction of galaxies 
with burst-like star formation histories at $z\sim4$.

 \subsection{The Effect of IMF and Metallicity}

   Another factor that affects the $EW(H\alpha)$ is
 the stellar initial mass function (IMF). A top-heavy IMF implies
 more early-type stars that dominate the Lyman continuum and thereby H$\alpha$ production. 
 By changing the power-law slope for the IMF ($N(M)\propto M^{-\alpha}$)
 from 2.3 to 2.0, 1.7, and 1.5, there is 3\,\%, 8\,\%, and 16\,\% increase 
 in the $EW(H\alpha)$ respectively 
 when the stellar population age is the same and the star formation is 
 described as a single burst (Starburst99; Leitherer et al. 1999).
   If the star formation is assumed to be continuous with a constant
 rate of $10\,M_{\odot}$ yr$^{-1}$, the change of power-law slope from 
 2.3 to 2.0, 1.7, 1.5, and 1.3 results in an increase of the $EW(H\alpha)$
 of 47\,\%, 110\,\%, 250\,\%, and 290\,\% at the stellar population age of 100\,Myrs old. 

  Metallicity also affects $EW(H\alpha)$, especially when the 
 stellar population age gets older compared to star-forming time scale. 
 For example, the $EW(H\alpha)$ ratios of $0.02\,Z_{\odot}$ and $0.2\,Z_{\odot}$ galaxy
 to $Z_{\odot}$ galaxy are 1.18 and 1.12 when the galaxy is 1\,Myr old
 after a single burst.
 The ratios increase to be 1.9 and 1.6 when the galaxy is 10\,Myrs old.

  It is difficult to assess whether the top-heavy IMF and/or metal-poor
 metallicity is the origin of large $EW(H\alpha)$ for the $z\sim4$ HAEs.
  Figure \ref{fig:ew_age} shows that the $EW(H\alpha)$ for galaxies 
 at two different redshifts ($z\sim2$ and $z\sim4$) are consistent 
 when the stellar population is less than 100\,Myrs old. Yet above 100\,Myrs old,
 the $EW(H\alpha)$ is clearly larger for $z\sim4$ galaxies compared to 
 $z\sim2$ galaxies with the same stellar population age.
If the IMF and metallicity
 were the dominant factors that affected $EW(H\alpha)$, this comparison 
 suggests that the metallicity is lower and the IMF is more top-heavy
 at $z\sim4$ compared to at $z\sim2$, at least for the systems with the largest derived $EW(H\alpha)$.

 \subsection{AGN Contamination}

  A fourth possibility is contribution to the H$\alpha$ emission
from an active galactic nucleus (AGN). 
  We matched our HAE candidates to the deep X-ray imaging catalog
 from \textit{Chandra} (Alexander et al. 2003; Luo et al. 2008) 
 to identify possible AGN.
 The sensitivity limit of the X-ray images are
 $\sim1.9\times10^{-17}$ and $\sim1.3\times10^{-16}$ erg cm$^{-2}$ s$^{-1}$
 for the 0.5--2.0 and 2.0--8.0 keV bands in GOODS-South (Luo et al. 2008).
The corresponding limits in GOODS-North are 
 $\sim2.5\times10^{-17}$ and $\sim1.4\times10^{-16}$ erg cm$^{-2}$ s$^{-1}$
(Alexander et al. 2003).  

  Two sources in GOODS-South (S14602, S23763)
 and two sources in GOODS-North (N12074, N23308) are 
 identified as known X-ray sources using the matching radius of 
 $3\arcsec$. 
 The X-ray luminosities of the matched objects are
 $5.3\times10^{42}-2.1\times10^{43}$\,erg\,s$^{-1}$
at rest-frame energies of 2.7-10.8\,keV at median redshift of $z\sim4.4$.
  After the matching, we checked the optical and
 X-ray images to ensure that X-ray emission is truly from
 our sources, not from neighboring objects. 
  We calculated
 the X-ray to optical luminosity ratios for these objects
 using the X-ray luminosity at rest-frame 2\,keV and the optical 
 luminosity at rest-frame 2500\,$\mbox{\AA}$. The resultant 
 $F_{X}/F_{\rm opt}$ ratios for four X-ray detected objects are 
 $0.038-0.073$. This is consistent with previously known 
 $z>4$ quasars detected at X-ray energies (Kaspi, Brandt, \& Schneider 2000).
  Therefore, these four objects are expected to be powered by AGN. 
  The relatively poor SED fitting results ($\chi^2>5$) of these
 objects, except for one object (N23308), also support the idea that
 these are AGNs. 
Note that N12074 is likely to have erroneous spectroscopic redshifts 
(see caption in Table \ref{tab:phot} and Figure 2), and
  S14602 is detected at 850\,$\mu$m (Coppin et al. 2009), 
 which suggests it may harbor an AGN but be starburst-dominated.
 In SED fitting, this object does not show photometric excess at 3.6\,$\mu$m.

We stacked the objects 
that are not directly detected in the X-ray images 
following the procedure of Alexander et al.
 (2003). The stacking yields non-detections in both fields, i.e., 
 $S_{\rm 0.5-2\,keV} \lesssim 3\,\sigma$ flux limits of
 $\sim3.0\times10^{-18}$ and $\sim2.0\times10^{-17}$
 erg cm$^{-2}$ s$^{-1}$ for GOODS-South and 
 $\sim5.5\times10^{-18}$ and $\sim3.7\times10^{-17}$
 erg cm$^{-2}$ s$^{-1}$ for GOODS-North, respectively. 
 At $z\sim4$, this upper limit in X-ray flux corresponds to 
 a luminosity limit of 
$<(5.3-9.5)\times10^{41}$\,erg\,s$^{-1}$ at 
 rest-frame energies of $2.5-10$\,keV, and 
 $<(3.4-6.5)\times10^{42}$\,erg\,s$^{-1}$ at rest-frame energies of 
 $10-40$\,keV. We calculated the $F_{X}/F_{\rm opt}$ ratios 
 using the same method as above, and the result is 
 $F_{X}/F_{\rm opt}\lesssim0.011$ at $z\sim4$. This value is 
 smaller by a factor of $>3$ compared to the objects with individual
 detection in X-ray. Assuming that the $F_{X}/F_{\rm opt}$ ratio is not
 dependent on the optical luminosity, as suggested in previous studies
 (Kaspi, Brandt, \& Schneider 2000; Brandt et al. 2004), this low $F_{X}/F_{\rm opt}$
ratio for our sources rule out the possibility of strong AGN contamination
 in sources 
not directly detected at X-ray energies.
  According to the number of matched objects to the X-ray catalog
 and the non-detection in the X-ray stacked images, the AGN fraction 
 among HAE candidates is $\sim5$\,\%.  
  Thus, AGN can be ruled out as the origin of strong H$\alpha$ emission
 in these galaxies.

\section{ Star Formation and Mass Assembly Since $z\sim4$ }

 As described in Section 3.4, HAE candidates have
H$\alpha$ equivalent width of 
$140-1700\,\mbox{\AA}$, indicating large
 current SFRs compared to that
 derived from the UV continuum. 
 In this section, we investigate the dominant mode of star formation
in these HAEs, and the implications for the build up of massive
 galaxies at $z\sim2-3$.

   \subsection{ Extended vs. Bursty Star Formation }

 Since the most dominant factors that drive large $EW(H\alpha)$ are
 stellar population age and star formation history 
 (Leitherer \& Heckman 1995; see Section 4.2), we investigate
 the correlation between age and $EW(H\alpha)$ for the 64 HAE 
 candidates (Figure \ref{fig:ew_age}). 
  The overplotted lines in Figure \ref{fig:ew_age} are the expected
$EW(H\alpha)$ vs. age tracks as a function of star formation history,
 metallicity, and IMF (Starburst99; Leitherer et al. 1999).
Other than the one likely AGN (asterisk; see Section 4.4), all 63
 objects lie over the tracks suggested by the models:
 24 ($\sim40$\,\%) are consistent with the instantaneous burst models, 
and the remaining 
 39 ($\sim60$\,\%) are consistent with the continuous star formation models.
This is based on the stellar population age derived using SED fitting
assuming a fixed metallicity of 0.2\,$Z_{\odot}$ for galaxies. 
If these galaxies are more metal-rich, their stellar population age 
would be even lower. The metallicity cannot be verified with current data and so we 
adopt the low metallicity assumption that is common for high-redshift
galaxies. 
  According to the model tracks, the large $EW(H\alpha)$ implies 
one or more of the following for the HAEs: (1) the galaxy is young, (2) the galaxy is more likely to 
 form stars continuously than instantaneously,
 (3) the galaxy is relatively metal-poor, 
and/or (4) the IMF is top-heavy. 

We use the derived age 
to divide the HAEs into two groups: 
24 galaxies that prefer instantaneous star formation (ages\,$<30$\,Myr),
   39 galaxies that prefer continuous star formation (ages\,$>30$\,Myr). 
We find that
 60\,\% of HAE candidates prefer `extended' star formation rather than
 `bursty' star formation, indicating that more than half of the
 $z\sim4$ galaxies are forming stars at a relatively constant rate.
  In order to investigate the factors that yield the different
 star formation histories among $z\sim4$ HAE candidates, we compare
 the stellar masses and morphologies of HAE candidates
 in the age vs. $EW(H\alpha)$ plot. Figure \ref{fig:ew_age_smass}
 shows the age vs. $EW(H\alpha)$ relation as a function of stellar
 mass, with the symbol size proportional to the stellar mass.
There is a factor of 2 difference between the 
 mean stellar mass of the two populations:
 $\mbox{log}\,M_*\,(M_{\odot}) = 9.85\pm0.36$ for continuous star forming galaxies and
 $\mbox{log}\,M_*\,(M_{\odot}) = 9.49\pm0.44$ for instantaneous burst galaxies.
 The mean stellar mass for galaxies with continuous star formation 
 is slightly larger than that of galaxies with instantaneous bursts, 
 but it should be noted that the factor of 2 difference is within the 
 stellar mass uncertainty inferred by SED fitting itself. 
   Note that there are several galaxies with extremely large 
 $EW(H\alpha)$ which is only reproduced with an extremely metal-poor 
stellar population 
and/or a top-heavy IMF. 
For example, S1478 has a 
large stellar mass ($M_* = 10^{11} M_{\odot}$)
in addition to a large $EW(H\alpha)$ and an old stellar population age. 
 It is plausible that such galaxies may harbor AGNs, whose X-ray luminosity is not large
or is heavily obscured. 

 As shown in Figure \ref{fig:stamp_burst} and \ref{fig:stamp_cont}, 
galaxies with `extended' and `bursty' star formation are not
distinguished in terms of morphologies. 
We visually divide mergers
 and non-mergers in the two groups: 
  while 13 of the 24 (54\,\%) instantaneous burst galaxies are apparent 
 merging/interacting systems, 19 of the 39 (49\,\%) continuously star-forming
 galaxies are classified as merging/interacting 
systems\footnote{We have verified the classification of merging and 
non-interacting systems using the CANDELS WFC3 data in GOODS-S and 
find the fraction of systems in those two categories to be 
consistent with that presented here.}.
  The fraction of mergers is roughly half in both cases, 
 which shows that morphology alone, especially in the rest-frame UV,
 is not enough to describe or
 represent the star formation mode in galaxies.

   Figure \ref{fig:ew_age} also shows that the mode of star formation 
 in $z\sim4$ HAE candidates is different compared to that of
 star-forming galaxies at lower redshifts. 
  The lensed $z=2.72$ Lyman break galaxy MS1512-cB58 
is well described by the 
instantaneous burst model
 ($\sim10$\,Myr old, Siana et al. 2008;
 $EW(H\alpha)\sim100\,\mbox{\AA}$, Teplitz et al. 2004). 
  This supports evidence showing that star formation
 in sub-$L^*$ galaxies at $z\sim3$ is bursty. 
  On the other hand, $z\sim2$ star-forming galaxies
 selected in the UV (Erb et al. 2006) clearly occupy the region 
 sampled by the continuous star formation models.
  As mentioned in Section 4.2, the fraction of galaxies with 
 extended star formation increases from 60\,\% at $z\sim4$ to 93\,\% at $z\sim2$.
  The difference partly results from the fact that the stellar population ages of 
 $z\sim2$ galaxies are basically derived using constant star formation history 
 models (i.e., $\tau=\infty$; Erb et al. 2006).
  The choice of $\tau$ does affect the derived stellar population ages, 
 by increasing the derived age if $\tau$ increases. 
  We test whether the use of constant star formation history models would 
 change the `young' ages ($<30$\,Myr) of $\sim40$\,\% of $z\sim4$ HAEs;
we find that most of the galaxies would still be fitted with ages less than 50\,Myr
even for $\tau=\infty$. 
 Therefore, despite the difference in the star formation history 
 of the stellar population models used in SED fitting, 
 the fraction of continuously star forming galaxies appears to
have significantly increased
 by $z\sim2$ compared to at $z\sim4$.

 \subsection{Number Density of HAEs and Massive Galaxies}

  The HAE candidates with star formation histories 
best fit by 
a continuous starburst model have a median stellar age of 
80\,Myr, and a median stellar mass of
$7.1\times10^9\,M_{\odot}$.
  Therefore, their past average SFR is $\sim90\,M_{\odot}$\,yr$^{-1}$,
 while the current SFR is observed to be in the range of 
 $30-600\,M_{\odot}$\,yr$^{-1}$.
The current SFRs are comparable or even larger than the past value. 
Assuming that the extended star-formation timescale is due to a steady supply of cold gas, 
if the HAEs continue to form stars at the measured rate, 
the HAEs are likely progenitors of massive galaxies at $z=2-3$.

  As we detect 39 robust HAE candidates having extended star formation
 time scales (e.g., best fit with continuous star formation), 
the lower limit  
on the number density of galaxies that could evolve into massive galaxies at $z=2-3$ is
 $3.3\times10^{-5}$\,Mpc$^{-3}$.
  This is more than 190 times the number density of $z\sim4$ 
 sub-millimeter galaxies (SMGs) discovered so far (Coppin et al. 2009)
 while the SMGs are 
considered likely progenitors
 for massive galaxies due to their large SFRs. 
  The stellar mass density produced by HAEs is calculated 
 by dividing the integral of their SFRs by the survey volume,  
 i.e. $ (1/V) 	\displaystyle \sum^{\rm galaxy}_{i}\int^{t_{z_0}}_0\,SFR_{i}(t)\,dt$. 
  The resultant stellar mass density the HAEs can contribute
 by $z_0=2$ and $z_0=3$ is $1.1\times10^7\,M_{\odot}$\,Mpc$^{-3}$
 and $4.4\times10^6\,M_{\odot}$\,Mpc$^{-3}$ respectively, when $SFR(t)$ is 
 fixed to the current observed SFR of each galaxy. Note that this is a
 lower limit produced by active star forming galaxies at
 $z\sim4$, since our selection of HAE candidates is incomplete
 due to the limitations of spectroscopic sample incompleteness,
IRAC sensitivity limit, and source photometry contamination. 
 
  The estimated mass density produced by HAE candidates is 
 $15-20$\,\% of the average global stellar mass density at $z=2$
 and $z=3$ (Figure \ref{fig:mass_z}). 
The estimated stellar mass of individual HAEs is larger than $10^{11}\,M_{\odot}$
at $z=2$ and $5\times10^{10}\,M_{\odot}$ at z=3. Therefore, 
when compared with the stellar 
 mass density of massive ($M_* >10^{11}\,M_{\odot}$) galaxies only, 
 the values are $\sim80$\,\% and $\sim50$\,\% of the stellar mass 
 density at $z=2$ and $z=3$ respectively. 
  This result suggests that $z\sim4$ HAE candidates may produce
 at least $50-80$\,\% of massive galaxies at $z=2-3$. 
  At $z=3$, this is $\sim5$ times higher
than the value provided
 by $z\sim4$ SMGs (10\,\%, Coppin et al. 2009), which are characterized as 
 an ultraluminous star formation phase for a short duty cycle of 
 $\sim100$\,Myrs. 
   Considering that our HAE selection is still incomplete, 
HAEs can likely account for most of the massive galaxies at $z=3$. 
 That is, most of the massive galaxies at $z=3$ 
apparently are formed through
steady, extended star formation rather than through 
the violent, burst-like star formation frequently reported in mergers.

\subsection{Star Formation Rates vs. Stellar Mass}

The SFRs of HAE candidates are correlated with their stellar masses
 (Figure \ref{fig:sfr_smass}),
 similar to the tight correlation between the SFR and the stellar mass 
 of star-forming galaxies at lower redshifts 
 (Elbaz et al. 2007; Noeske et al. 2007; Daddi et al. 2007). 
The correlation between SFR vs. stellar mass of $z\sim4$ HAEs 
is significant with a Pearson correlation coefficient of $r=0.77$.
 Considering the stellar mass is roughly proportional to the halo mass, 
 and to the gas mass, the tight SFR-$M_*$ correlation 
 is another reflection of Schmidt-Kennicutt law (Schmidt 1959; Kennicutt 1998b)
 that connects gas density and SFR density.

The SFR-$M_*$ correlation
evolves as a function of redshift, with the same 
 slope.  This may indicate a difference in star formation
 efficiency between galaxies at different redshifts. 
  The big difference between $z\sim4$ HAE candidates and star-forming
 galaxies at other redshifts is the comparison with submm galaxies
at similar redshifts. At $z\sim2$, submm galaxies are found to be      
 significant outliers in the SFR-$M_*$ relation (Daddi et al. 2009),
 with SFR a factor of 10 larger than normal star-forming
 galaxies. The discrepancy implies that star formation efficiency 
 (or gas fraction) is higher in submm galaxies compared to 
 normal star-forming galaxies at that epoch, which is very likely 
 considering submm galaxies are wet mergers. 
  On the other hand, known $z\sim4$
 SMGs are located close to the extension of SFR-$M_*$ relation of $z\sim4$
 HAE candidates. We interpret this as the high star formation 
 efficiency (or high gas density) of $z\sim4$ HAE candidates. The morphologies of 
 HAE candidates are not dominated by merging systems; 
thus, other mechanisms are required to explain the high star formation
 efficiency and large SFR for HAE candidates. 
The cold accretion flow scenario can explain the observed 
 number density of HAE candidates that produce SFR larger than
 200\,$M_{\odot}$ yr$^{-1}$ (Dekel et al. 2009). Yet the
 contribution and the importance of merger-induced star formation
 at this redshift range is difficult to constrain and it
 is difficult to classify merging systems and non-merging systems 
based on UV morphologies alone.

\section{Predictions for Future Observations}

The expected H$\alpha$ line flux of HAEs is 
$10^{-17}-10^{-16}$ erg s$^{-1}$ cm$^{-2}$, 
which indicate the HAEs presented in this paper are 
potential targets for James Webb Space Telescope to confirm 
and measure the strength of H$\alpha$ emission at $z>4$ for the first time. 
With the expected line flux and the observed 
4.5\,$\mu$m continuum level of $23-25$\,mag(AB), 
we expect to get $S/N$ of $\sim10$ for the emission line
and $S/N\sim3$ for the continuum
with an exposure of $1200-1800$ seconds 
using the G395($R\sim1000$)/F290LP grating/filter setting of 
\textit{JWST}/NIRSPEC\footnote{http://jwstetc.stsci.edu/etc/input/nirspec/spectroscopic/}. 

  We have discussed different scenarios for the origin of 
 such strong H$\alpha$ emission in the previous sections.
 If the large H$\alpha$ EWs are due to dust obscuration,
 we calculate the range of the  
 intrinsic SFR of the HAEs is
 $20-500\,M_{\odot}$\,yr$^{-1}$, corresponding to an infrared luminosities
 of $\lesssim2\times10^{12}\,L_{\odot}$.
  We calculated the expected far-infrared/millimeter fluxes
 for the HAE candidates using the derived $SFR(H\alpha)$,
 and checked whether these objects are detectable in 
 future surveys at long wavelengths.
  The $L_{IR}$ of the HAE candidates derived from $SFR(H\alpha)$
 ranges from 
 $1.2\times10^{11}\,L_{\odot}$ to $2.3\times10^{13}\,L_{\odot}$. 
 This is consistent with the $L_{IR}$ calculated using the 
 difference between the best-fit template and its unattenuated 
 form, with a scatter of $\sim0.5$ dex. 

  Figure \ref{fig:sed_ir} shows the IR templates 
 that represent $L_{IR}$ of each object at 
its observed redshift.
  The IR templates are from Chary \& Pope (2010). 
 The observed optical-to-MIR fluxes are overplotted as circles
 and lines that connects the circles. 
  At wavelengths longward of 70\,$\mu$m, the HAE candidates 
 show a wide range of flux densities. 
The limits for future space and ground-based
 missions are indicated as horizontal shaded region:
 blue for the Herschel/Photodetector Array Camera and Spectrometer
 limits (PACS; GOODS-Herschel program), 
 green for the Herschel/Spectral and Photometric Imaging REceiver
 limits (SPIRE; GOODS-Herschel), and 
 red for the expected sensitivities of
the Atacama Large Millimeter/Sub-millimeter Array (ALMA).  
  Assuming that these IR SEDs are valid for HAE candidates, 
 we expect to detect $\sim10$\,\% of HAE candidates using GOODS-Herschel data, 
primarily by SPIRE. Our candidates are relatively
 free of source confusion, at least on scales of a few arcseconds, due to 
 their selection in IRAC images.
However, the SPIRE beam size of $15-20\arcsec$ 
 is still too large to avoid source confusion. 
 Thus the flux density uncertainties due to source confusion 
will be significant.
To definitively distinguish between extinction and stellar age effects,
 we will need to await the start of ALMA which can detect
 the thermally reprocessed far-infrared emission from these objects. 
Spectroscopy with ALMA will also measure the ratio between the 158\,$\mu$m [CII] line
and $L_{IR}$ in these objects which
can be used to discriminate between brief mergers and 
temporally extended cold-flow driven star-formation. 
The former scenario would show ratios that are almost an order of magnitude 
lower than the latter.

\section{ Summary }

  We have studied the multiwavelength properties of a sample of 74 galaxies
 that have spectroscopic redshifts in the range $3.8<z<5.0$ over 330\,arcmin$^2$
 of the GOODS-North and South fields. 
  The stellar continuum of these objects is well
 determined through SED fits to the multi-band photometry
 from the optical to the MIR ($B$, $V$, $i$, $z$, $F110W$, $F160W$, $J$, $H$, $K$,
 3.6\,$\mu$m, 4.5\,$\mu$m, 5.8\,$\mu$m, and 8.0\,$\mu$m).  
Their stellar population ages, star formation histories, extinction laws,
 $E(B-V)$, and stellar masses are determined through this SED fitting.
  We demonstrate that the majority of galaxies show excess in the
 IRAC 3.6\,$\mu$m band relative to the expected stellar continuum,
 which is likely due to the redshifted H$\alpha$
 emission line. 
  We define the significance factor $S$ of the 3.6\,$\mu$m excess
 compared to the average SED fitting residuals in other filters, 
 and used this factor to select H$\alpha$ emitting galaxies.
 H$\alpha$ emitters (HAEs) are defined as galaxies with $S_{3.6\,\mu m}>10$.
 The HAE selection using this excess in a broad-band filter
 identifies galaxies with large H$\alpha$ equivalent widths and high SFR. The
equivalent width ranges between $140-1700\,\mbox{\AA}$ while SFRs are in the ranges
between $20-500\,M_{\odot}$ yr$^{-1}$.
  The derived H$\alpha$ line luminosity ranges between
 $2.5\times10^{42}\,\mbox{erg s}^{-1}$ and $6.3\times10^{43}\,\mbox{erg s}^{-1} $.
  The z$\sim4$ galaxy population appears to show a factor of $\sim2-3$
stronger H$\alpha$ emission
than $z\sim1$ and $z\sim2$ galaxies.

The factors
 that affect H$\alpha$ emission include: dust extinction, stellar population 
age, star formation history, IMF, and galaxy metallicity.
The flux ratio 
$F_{H\alpha}/F_{UV}$ is an extinction indicator independent of UV spectral 
slope $\beta$. The relationship between $\beta$ and $F_{H\alpha}/F_{UV}$ 
depends on the form of the dust extinction law, with $F_{H\alpha}/F_{UV}$
ratio being higher for the SMC extinction law than for the starburst extinction
 law when the UV slope $\beta$ is fixed. The large H$\alpha$ fluxes, 
and large $F_{H\alpha}/F_{UV}$ ratio of $z\sim4$ HAEs imply that $z\sim4$ HAE 
 galaxies prefer the SMC extinction law rather than the SB extinction law. 
A lower metallicity stellar population and a 
more top-heavy IMF compared to lower redshift (e.g., $z\sim2$)
galaxies are expected to be another possible reason for strong H$\alpha$ emission. 
However, the most dominant driver for strong H$\alpha$ emission appears
to be the temporally extended star formation history of these HAEs 
based on their location in the stellar age vs. $EW(H\alpha)$ phase-space.

  At least 60\,\% of HAEs
 are classified as continuously star-forming galaxies
suggesting that this phase is more common than the bursty, short-duration star formation 
that occurs through
mergers. If the HAEs continues to form stars constantly with the observed
SFR, HAEs would evolve into 
 massive galaxies ($>10^{11} M_{\odot}$) at $z\sim2-3$.
 While the number density of $z\sim4$ SMGs discovered to date is not
 sufficient to  explain the number density of massive galaxies 
 at $z\sim2-3$, HAEs can account for $\gtrsim50$\,\%
 of the stellar mass density in massive galaxies at $z\sim2-3$
indicating that the strong H$\alpha$ phase plays a dominant role 
in the growth of galaxies at high redshift. 
  We believe that these luminous HAEs are ideal targets for future observations with
JWST, which would enable their H$\alpha$ emission to be spectroscopically confirmed,
and with ALMA which would place strong constraints on the nature of dust extinction as well as
the dominant physical mechanism powering star-formation at $z>4$.

\acknowledgements
 We thank the entire GOODS team for their effort in compiling and
 reducing different components of the data. 
We thank Rychard Bouwens and Avishai Dekel for helpful comments and discussions and the anonymous referee for a
thorough referee report.
 This work is based, in part, on observations made with the Spitzer
 Space Telescope, which is operated by the Jet Propulsion Laboratory,
 California Institute of Technology under a contract with NASA.
 Part of the NIR data are
 based on observations obtained with WIRCam, a joint project of
 CFHT, Taiwan, Korea, Canada, France, at the Canada-France-Hawaii
 Telescope (CFHT) which is operated by the National Research Council
 (NRC) of Canada, the Institute National des Sciences de l'Univers of the
 Centre National de la Recherche Scientifique of France, and the
 University of Hawaii.

\clearpage

\begin{figure*}
\centering
\subfigure{\label{fig:seds} \includegraphics[width=135mm]{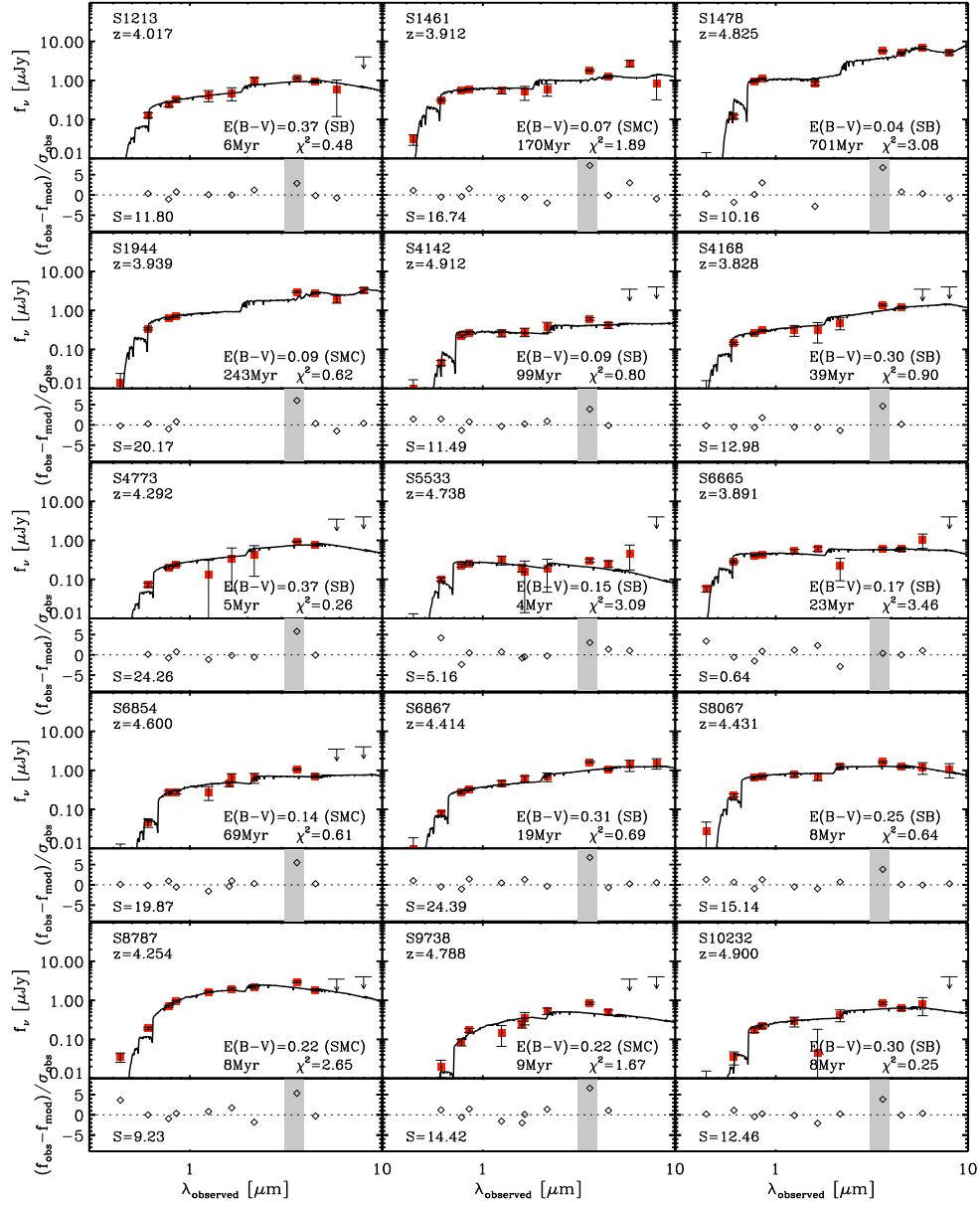} }
 \caption{
  \small{ The SEDs of galaxies at $3.8<z_{spec}<5.0$ over the GOODS fields
  (first 43 from GOODS-South, next 31 from GOODS-North).
  The \textit{filled squares} indicate \textit{HST}/ACS $B$, $V$, $i$,
  and $z$-band photometry, $J$, $H$, and $K$-band photometry for objects
  selected in GOODS-South, $J$ and $K$-band photometry for objects
  in GOODS-North, and \textit{Spitzer}/IRAC four band photometry.
   For some objects, \textit{HST}/NICMOS F110W/F160W photometry points
  are available. For objects that are not detected in NIR bands or
  \textit{Spitzer}/IRAC channel 3/4, we use the 3\,$\sigma$ flux density upper limits
  marked as arrows. Overplotted solid lines are the
  best-fit galaxy spectral templates (CB07; Bruzual 2007).
   The best-fit galaxy age, $E(B-V)$, and the extinction law is
  specified in addition to the lowest $\chi^2$ value.
  Below each SED, we show the fitting residuals as a function of wavelength.
   The $y$-axes indicate the residuals divided by the observational
  flux uncertainty (i.e., $ ( f_{\rm obs} - f_{\rm mod} ) / \sigma_{\rm obs} $). 
  The significance factor for the 3.6\,$\mu$m band flux density excess,
  $S$ (see text for definition) is also provided. For some objects with
  $ ( f_{\rm obs} - f_{\rm mod} ) / \sigma_{\rm obs} $ at 3.6\,$\mu$m 
  exceeding the $y$-axis range, we use
a triangle in the
  residual plot and also indicate the value. Two of the objects, S15920 and N12074, appear to
show multiwavelength SED that are inconsistent with their quoted redshift and likely have erroneous
spectroscopic redshifts. 
   }  }
\end{figure*}
\addtocounter{figure}{-1}

\begin{figure*}
\centering
\subfigure{ \includegraphics[width=135mm]{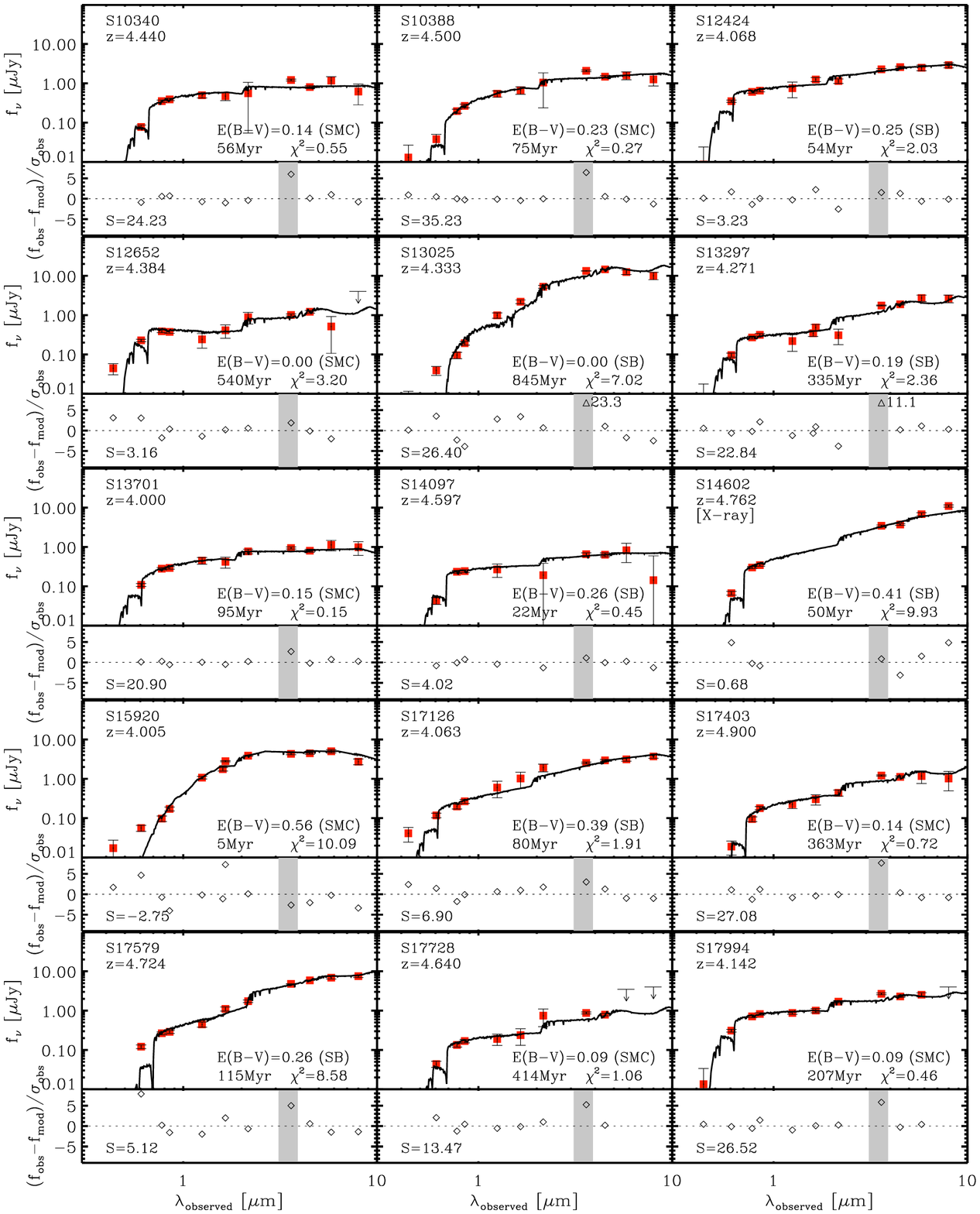}}
\caption{ \textit{Continued.}}
\end{figure*}
\addtocounter{figure}{-1}

\begin{figure*}
\addtocounter{subfigure}{1}
\centering
\subfigure{ \includegraphics[width=135mm]{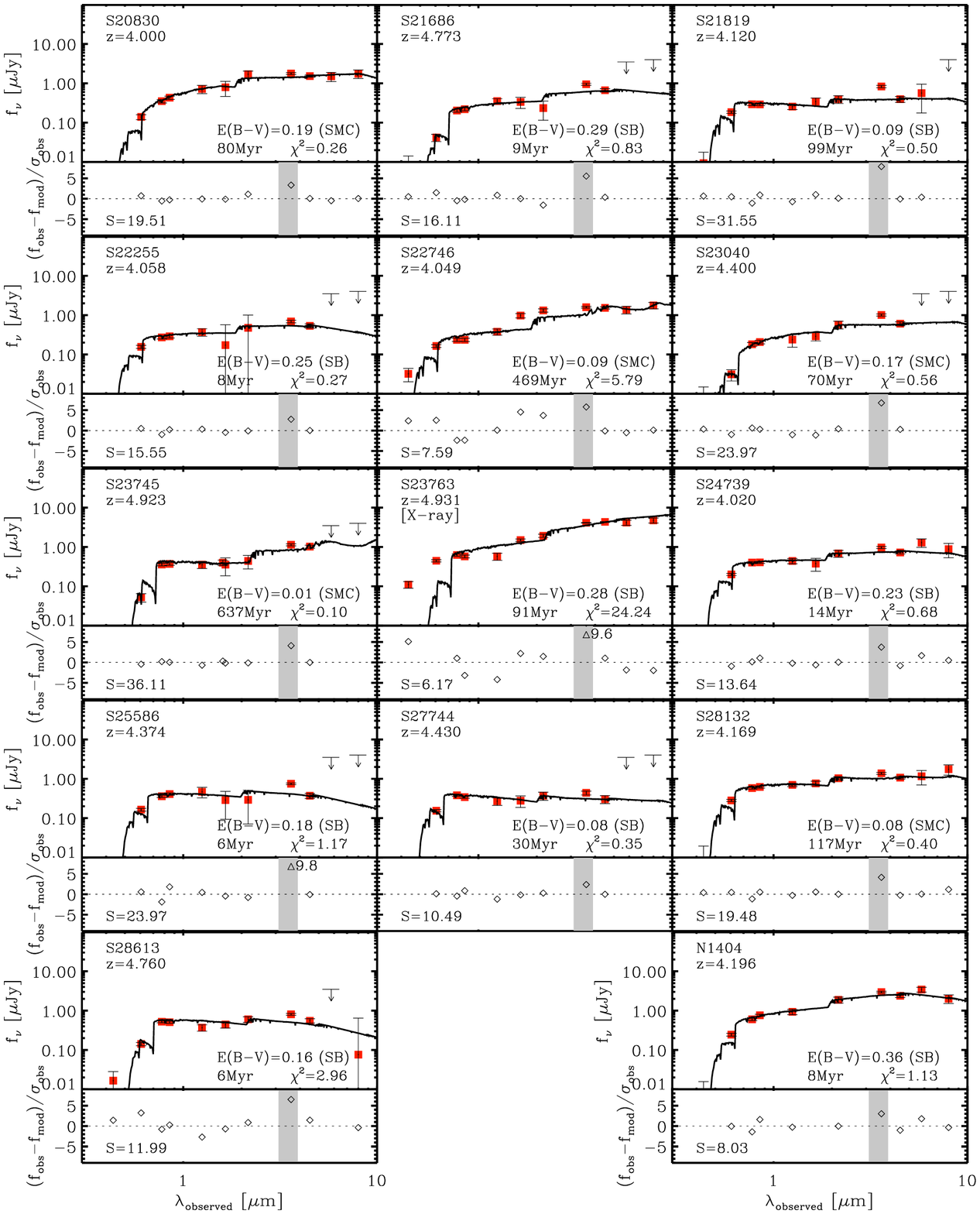}}
\caption{ \textit{Continued.}}
\end{figure*}
\addtocounter{figure}{-1}

\begin{figure*}
\addtocounter{subfigure}{1}
\centering
\subfigure{ \includegraphics[width=135mm]{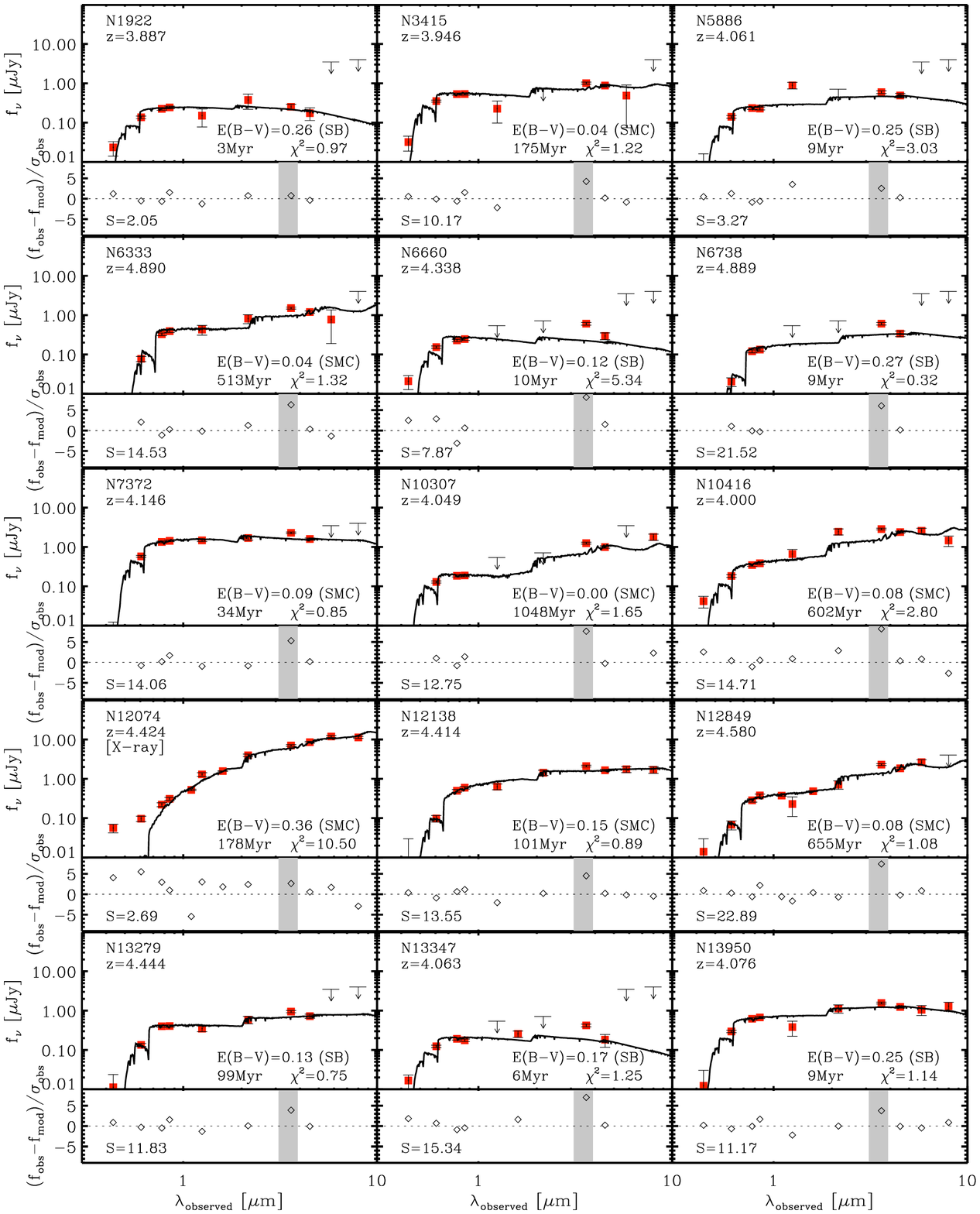}}
\caption{ \textit{Continued.}}
\end{figure*}
\addtocounter{figure}{-1}

\begin{figure*}
\addtocounter{subfigure}{1}
\centering
\subfigure{ \includegraphics[width=135mm]{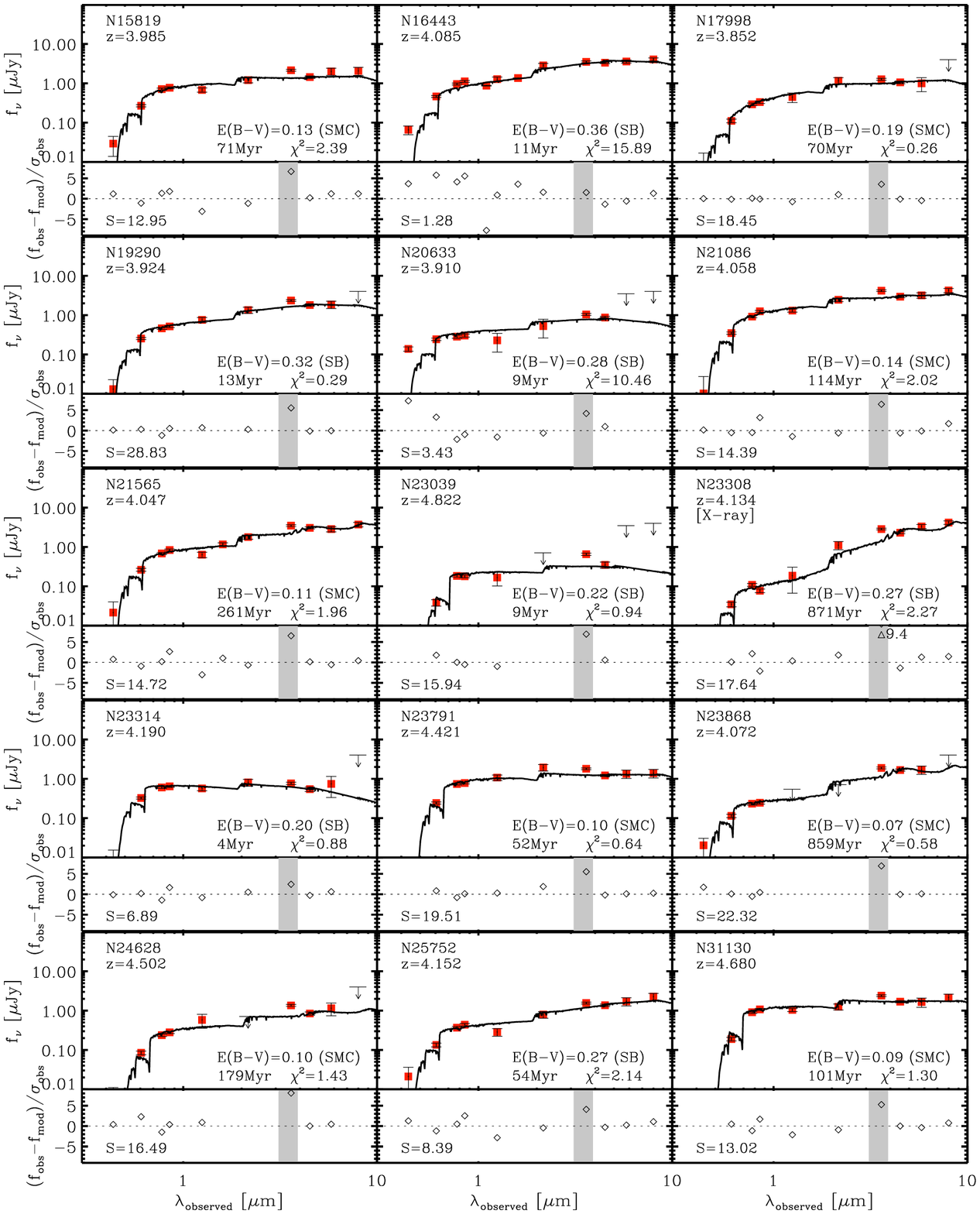}}
\caption{ \textit{Continued.}}
\end{figure*}

 \begin{figure*}
   \plotone{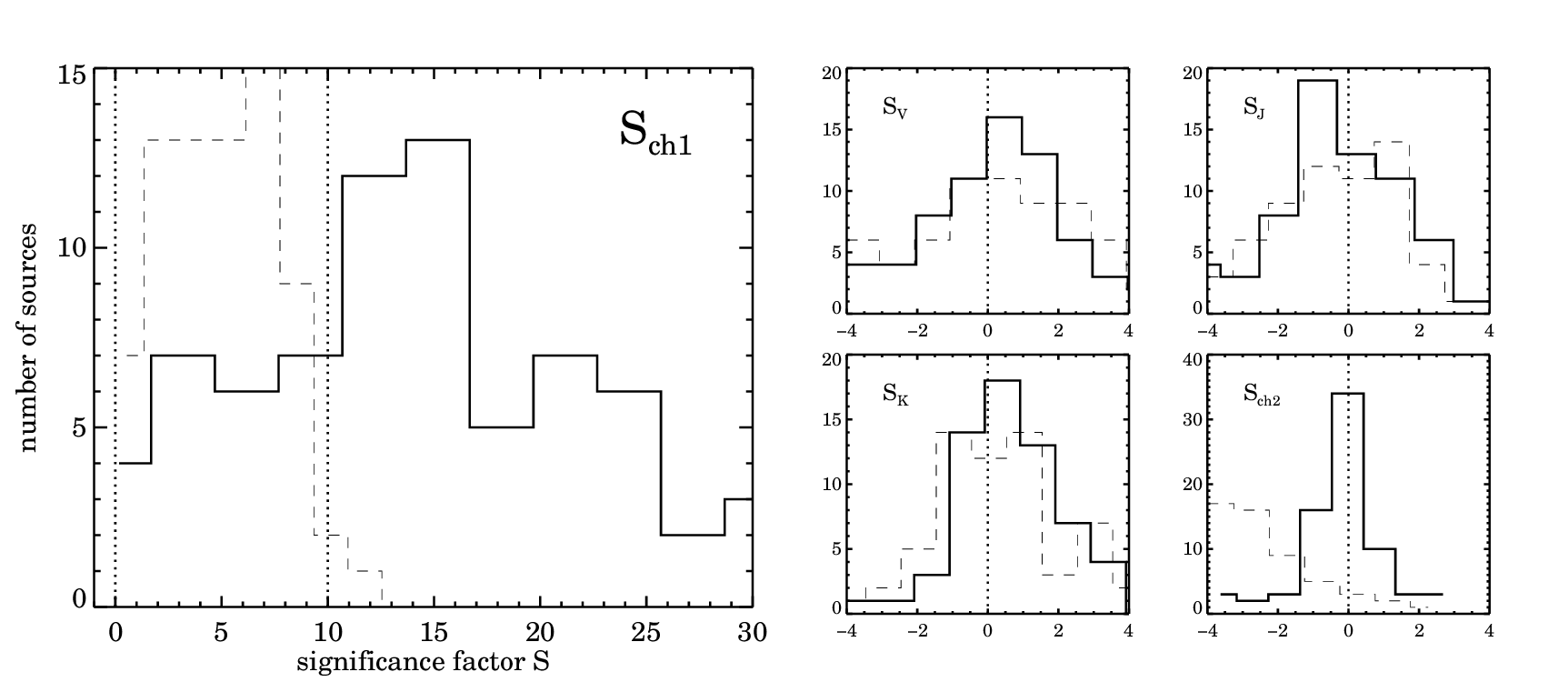}
   \caption{\label{fig:sfactor_hist}
   The distribution of ``significance factor'' $S$ in different filters
   -- IRAC ch1 (3.6\,$\mu$m), V, J, K, and IRAC ch2 (4.5\,$\mu$m).
The definition of $S$ is given in Section 3.2, Equation (2). 
    The distributions are presented for two different cases: 
   the \textit{solid} lines show SED fitting without 3.6\,$\mu$m data and
   the \textit{dashed} lines show SED fitting including 3.6\,$\mu$m data.
    Unlike the other filters ($V$, $J$, $K$, and $4.5\,\mu$m) 
which show distributions centered 
around $S=0$, the $~S_{\rm ch1}$ distribution is clearly centered 
above 0.
    Even if we try to ``fit'' the observed 3.6\,$\mu$m flux (\textit{dashed} lines),
$S_{\rm ch1}$ remains positive
   and the stellar continuum matched to 3.6\,$\mu$m over-predicts
   the 4.5\,$\mu$m flux (see the lower-right panel for $S_{\rm ch2}$ distribution).
   }
 \end{figure*}

\begin{figure*}
\epsscale{1.1}
  \plottwo{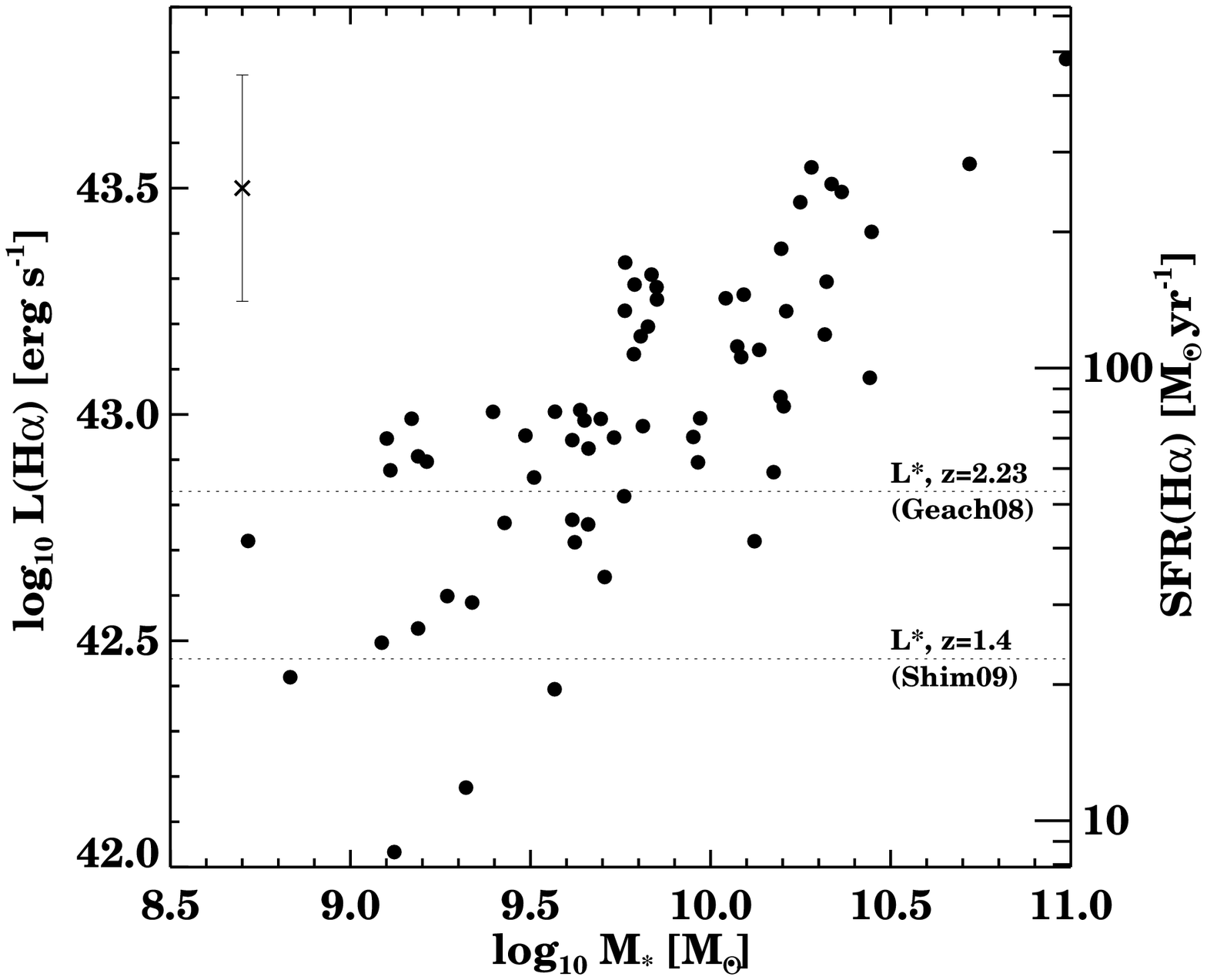}{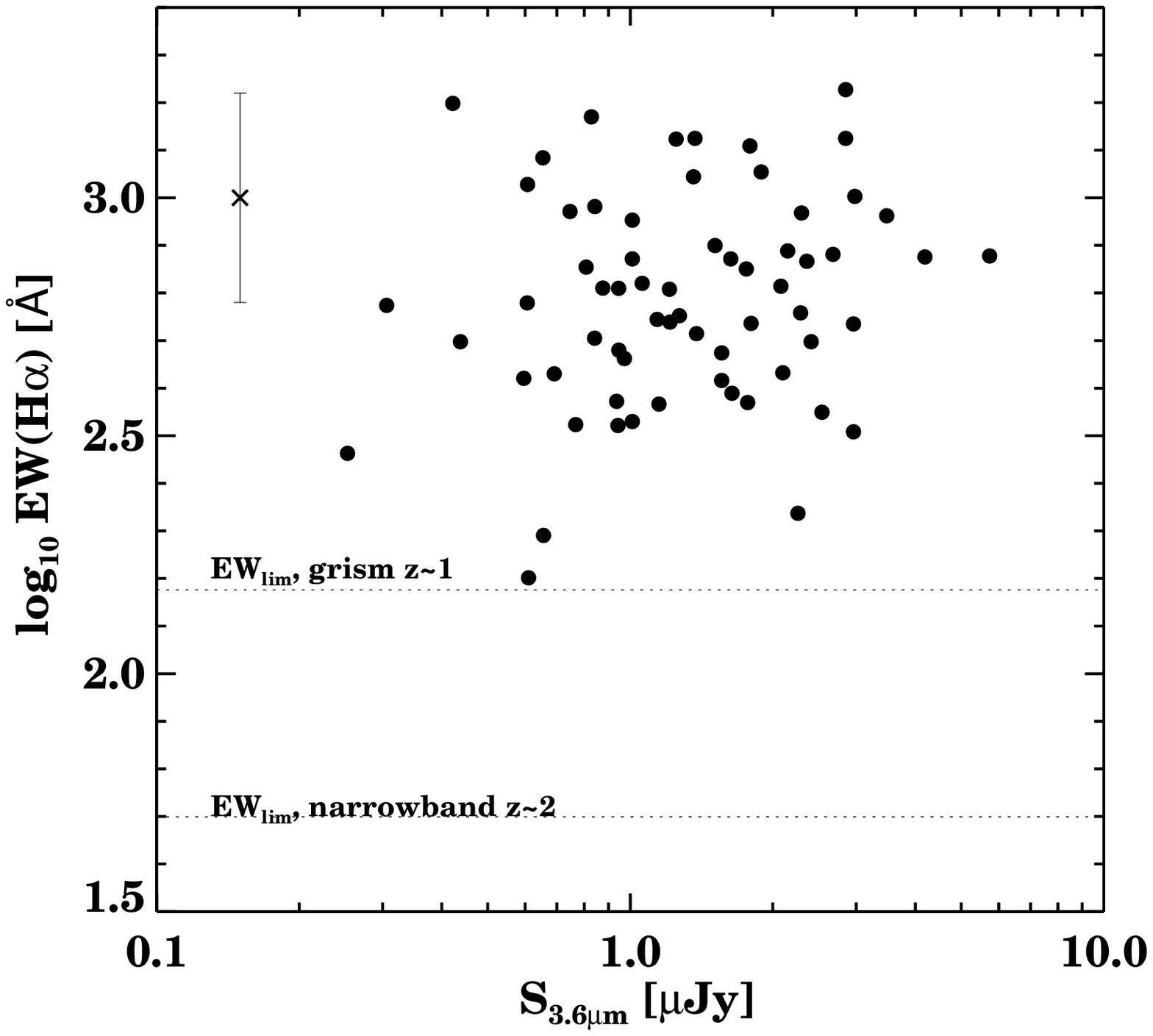}
  \caption{\label{fig:prop}
  (\textit{Left}): H$\alpha$ luminosity
  of $z\sim4$ HAE candidates as a function of stellar mass.
  The right axis shows the inferred SFR from the H$\alpha$
  luminosity. The plotted error bar represents the typical uncertainty in
H$\alpha$ luminosity.
The horizontal dotted lines indicate $L_{*, H\alpha}$ 
at $z=1.4$ (Shim et al. 2009) and $z=2.23$ (Geach et al. 2008), respectively. 
  (\textit{Right}):
H$\alpha$ equivalent width of $z\sim4$ HAE candidates 
relative to their 3.6\,$\mu$m flux density.  The error bar indicates the typical
  uncertainty in the derived H$\alpha$ equivalent width.
The horizontal dotted lines indicate the minimum equivalent width 
that could be detected in other surveys, i.e., grism survey at $z\sim1$
(Shim et al. 2009) and narrow-band imaging survey at $z\sim2$
(Geach et al. 2008).
  Clearly, the $z\sim4$ population shows stronger H$\alpha$ emission
  than star-forming galaxies at $z\sim1$ or $z\sim2$. }
 \epsscale{1.0}
\end{figure*}

\begin{figure}
  \plotone{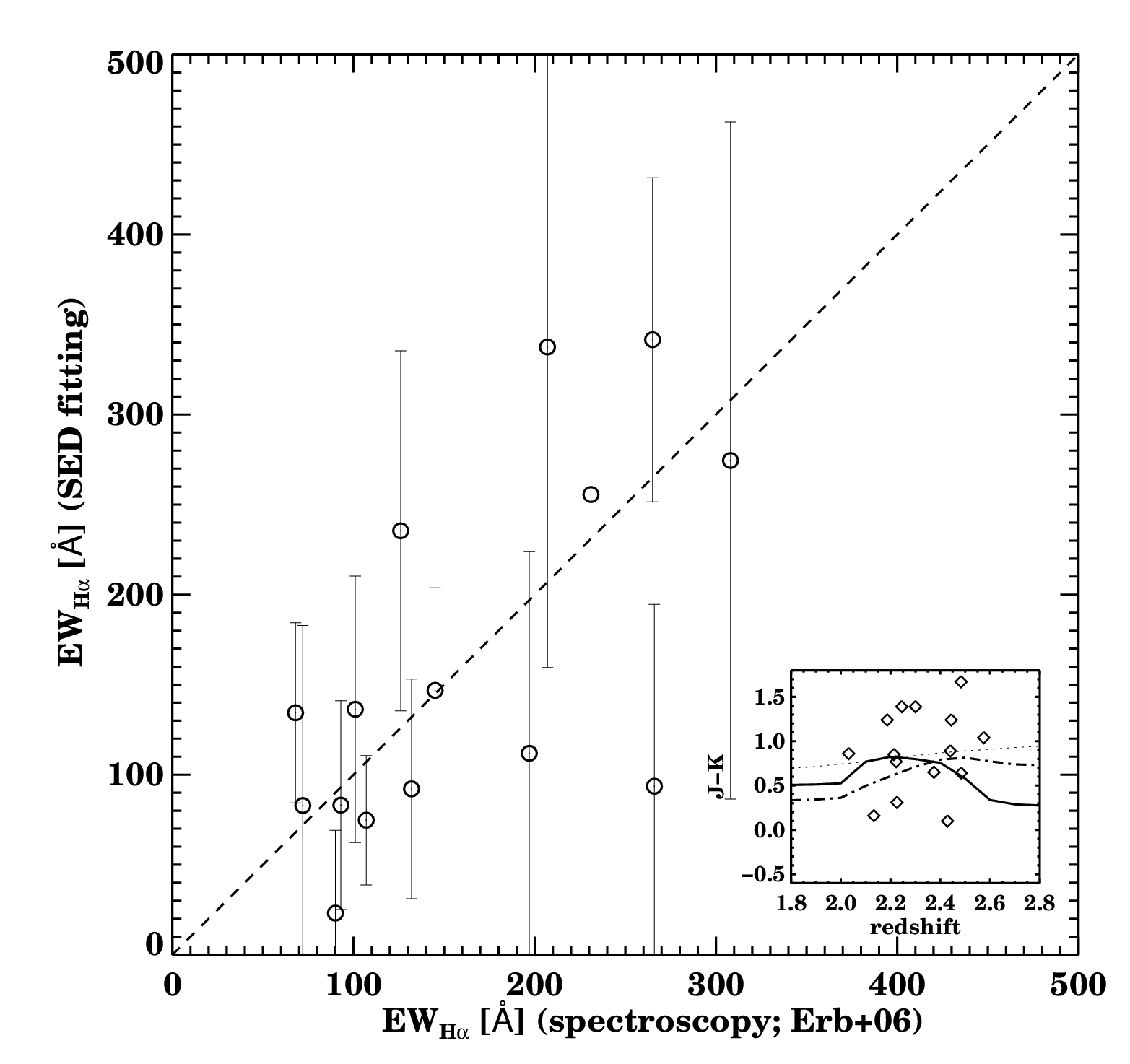}
  \caption{\label{fig:method}
   Comparison between H$\alpha$ equivalent widths derived from
broad-band photometric excess ($y$-axis)
  to equivalent widths measured from observed spectra ($x$-axis).
   We consider star-forming galaxies with
  H$\alpha$ spectroscopic observations at $2.1<z<2.5$
  (Erb et al. 2006; Reddy et al. 2006) for this test, assuming that
  the redshifted H$\alpha$ emission line is the dominant source
  of $K$-band excess.
   The inset plot shows the $K$-band excess due to the redshifted
  H$\alpha$ line as a function of redshift, where the
  \textit{solid}/\textit{dot-dashed}/\textit{dotted} line indicates
  the expected $J-K$ color for MS1512-cB58/SDSS quasar/reddest galaxy
  template. The \textit{diamonds} are
  the observed $J-K$ colors for $z\sim2$ star-forming galaxies
  (Reddy et al. 2006). The correlation between the spectroscopically measured equivalent
  widths and our photometrically derived values imply that our photometric method is robust.
  }
 \end{figure}

\begin{figure*}
\epsscale{1.1}
  \plottwo{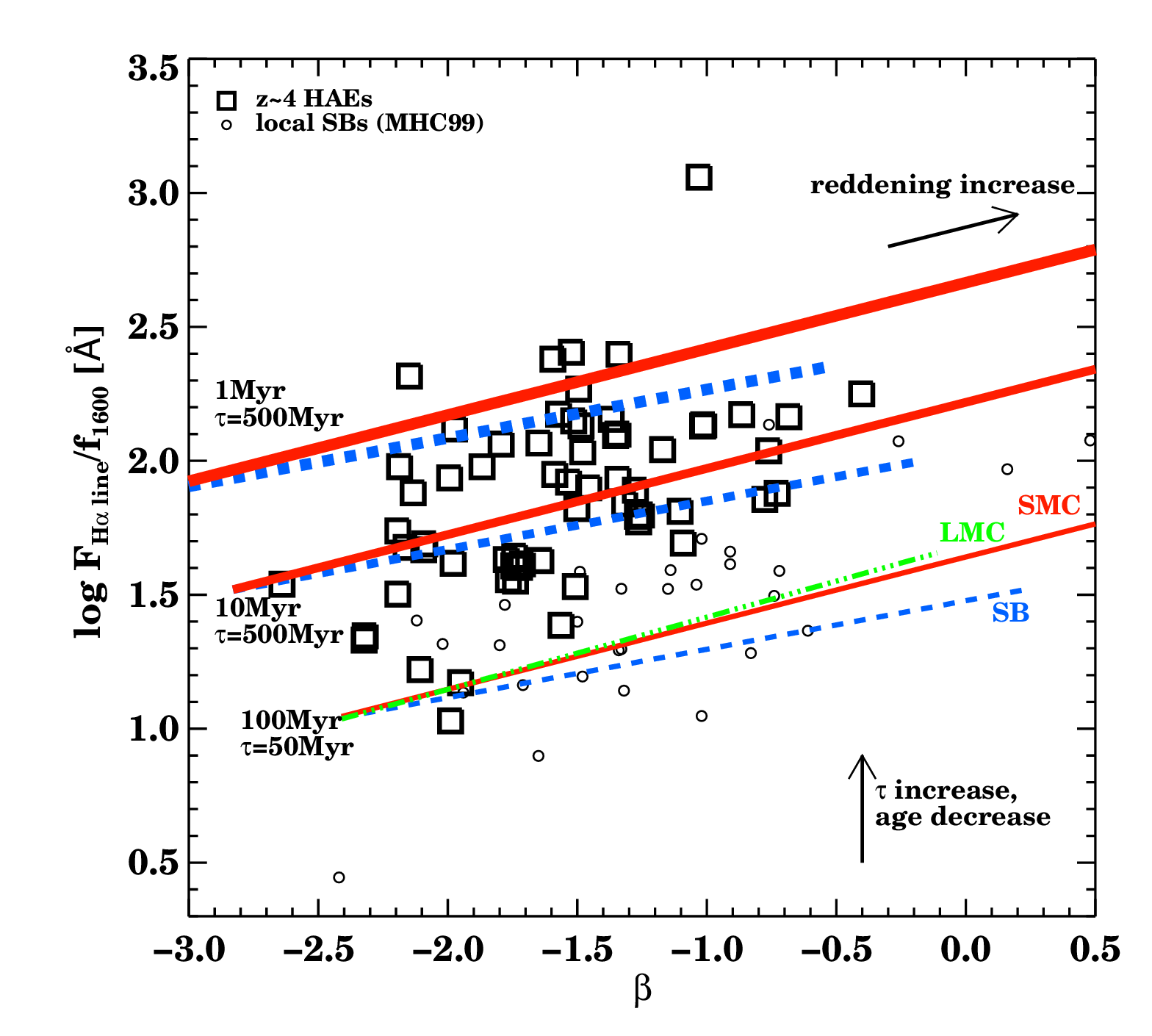}{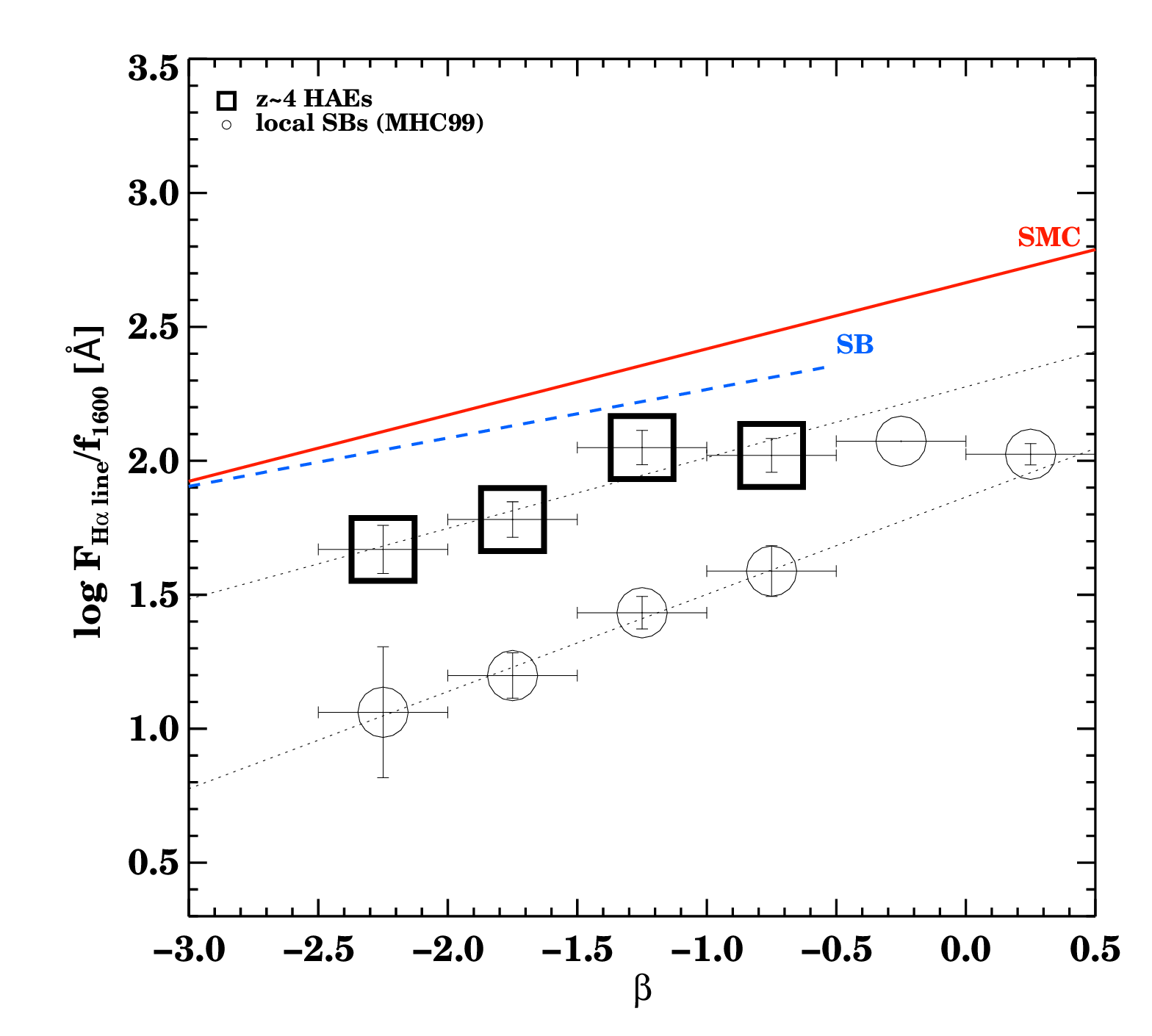}
  \caption{\label{fig:lumratio}
   (\textit{Left}): Comparison between two different extinction measures,
  UV spectral slope $\beta$ and the H$\alpha$ line flux to
  UV flux ratio. All (64) HAE candidates are plotted (\textit{squares}).
  The \textit{circles} indicate local
  starburst galaxies (data originally published by Storchi-Bergman
  et al. 1995 and McQuade et al. 1995; taken from Meurer, Heckman, \&
  Calzetti 1999). The overplotted lines show the relation
  between $\beta$ and $f_{H\alpha}/f_{1600}$ 
for different extinction laws and stellar population synthesis models.
  The \textit{solid}, \textit{dashed}, and \textit{dot-dashed} line
show the SMC (Pr\'evot 1984), SB (Calzetti et al. 2000), and
  LMC (Fitzpatrick 1986) extinction laws
 without the $2175\,\mbox{\AA}$ graphite 
feature,
  respectively. The relation moves upward when the galaxy is younger
  and the star formation history is more extended.
    (\textit{Right}): Comparison between $\beta$ vs. line-to-continuum
  ratio for local starbursts, HAE candidates, and SB/SMC extinction laws.
  The symbols represent mean $f_{H\alpha}/f_{1600}$ values in bins of 0.5 in $\beta$
  (\textit{squares} for HAE candidates, \textit{circles} for local starbursts).
  The error bars are derived through bootstrapping, and the lines are
  the best-fitted linear fits to the points. The slope of the lines
  are 0.18, 0.24, and 0.26 for SB, SMC, and LMC extinction laws.
The slope of the $\beta$ vs. line-to-continuum ratio for $z\sim4$ HAEs
  is $0.27\pm0.07$. Thus, although the HAEs appear to prefer the SMC or LMC extinction law,
the derived slope is only different by 1.5\,$\sigma$ compared to the SB extinction law.
  }
\epsscale{1.0}
\end{figure*}

\begin{figure*}
  \plotone{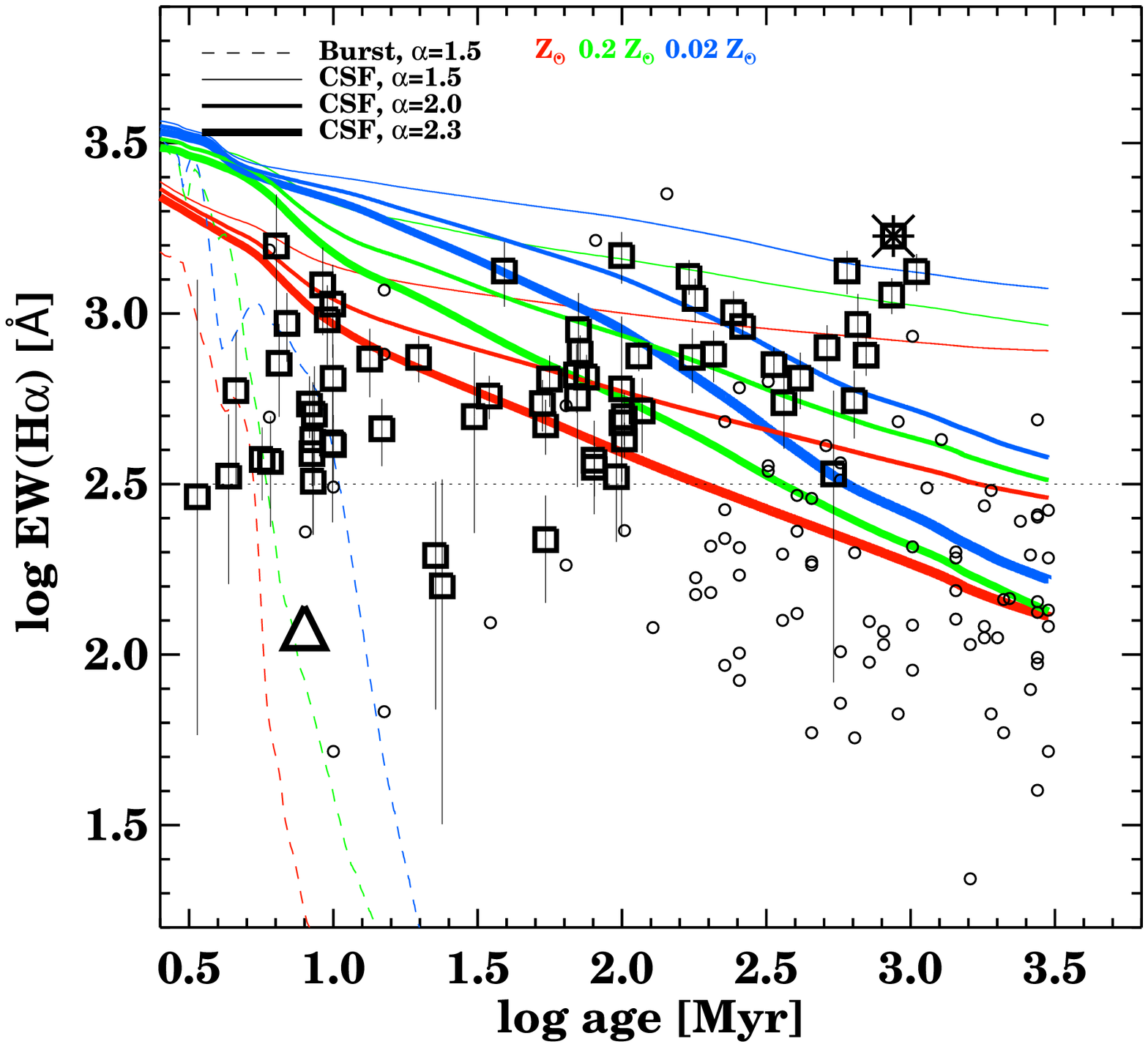}
  \caption{\label{fig:ew_age}
   Relation between the estimated EW(H$\alpha$) and 
  the stellar population age from the SED fitting.
   All (64) HAE candidates are plotted as \textit{squares},
  the \textit{asterisk} is N23308, an object with an X-ray counterpart,
probably an AGN.
   Overplotted are $z\sim2$ star-forming galaxies 
  selected in the UV (\textit{circles}; Erb et al. 2006),
  MS1512-cB58, a gravitationally lensed Lyman break galaxy at $z=2.72$
  (\textit{triangle}; Teplitz, Malkan, \& McLean 2004; Siana et al. 2008).
   The horizontal dotted line at log EW(H$\alpha$)$\sim2.5$ 
is the EW limit for galaxies
that could be identified as HAEs using 
  photometric excess in the 3.6\,$\mu$m-band (see text for details).
   We overplot the expected EW(H$\alpha$) vs. age tracks of
  models with different star formation history, metallicity, and
  IMF using STARBURST99 (Leitherer et al. 1999).
   The \textit{dashed} lines indicate galaxies that passively evolve
  after a single burst of star formation, and the \textit{solid} lines 
  indicate galaxies that constantly produce stars at a rate of 30\,$M_{\odot}$ yr$^{-1}$.
  The color of the line indicates the metallicity ($Z_{\odot}$, $0.2\,Z_{\odot}$, 
  and $0.02\,Z_{\odot}$ for \textit{red}, \textit{green}, and \textit{blue}
  lines). The thickness of the line indicates the slope of 
 the stellar initial mass function $\alpha$, i.e., $n(M)\propto M^{-\alpha}$.  
The majority of sources are consistent with an extended star-formation timescale.
Some of the highest EWs can be explained with a top-heavy IMF, 
low metallicity model.
}
 \end{figure*}

 \begin{figure*}
  \plotone{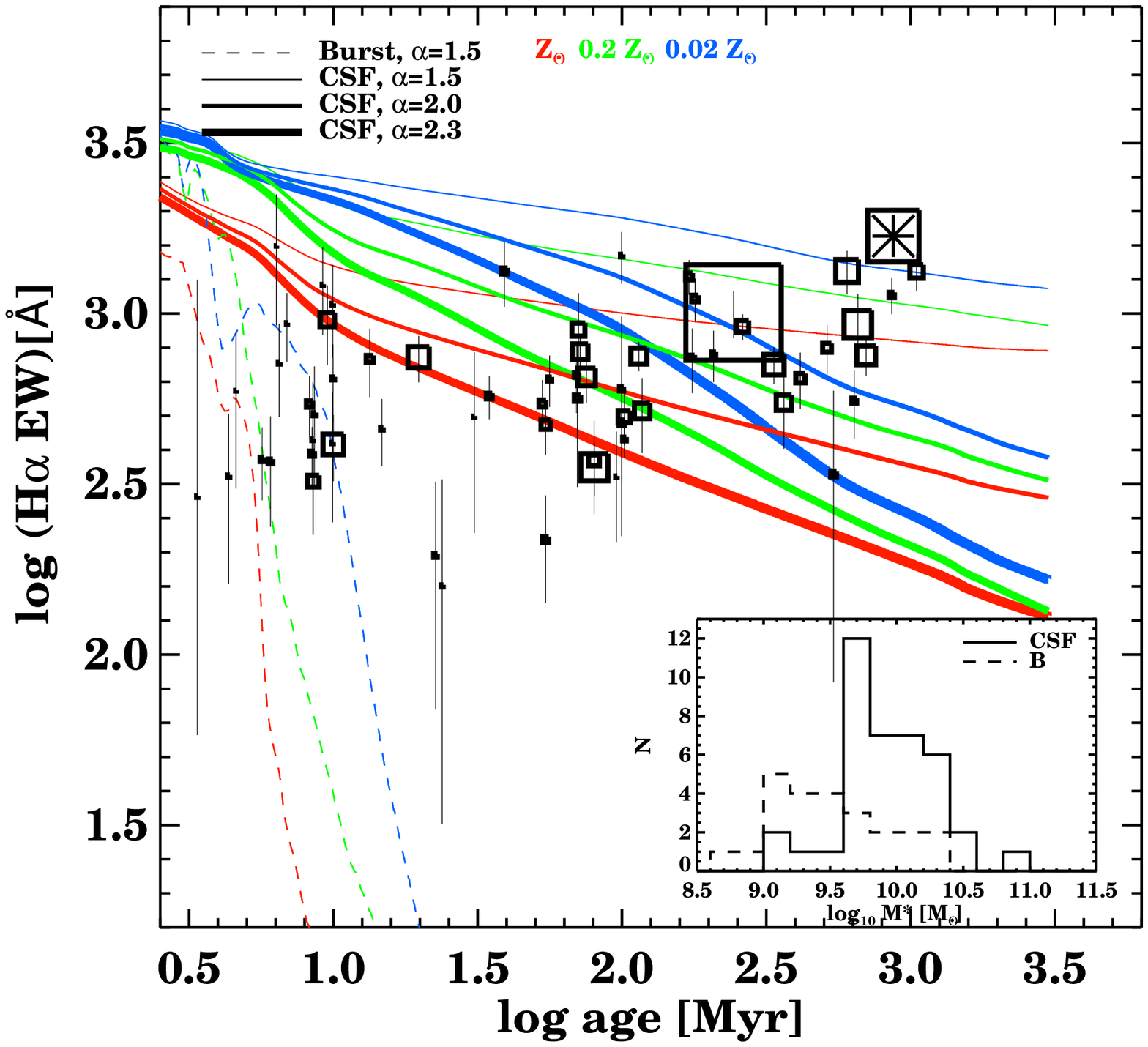}
  \caption{\label{fig:ew_age_smass}
   Relation between the estimated EW(H$\alpha$) and
  the stellar population age as a function of galaxy stellar mass. 
   Model tracks are the same as in Figure \ref{fig:ew_age}, 
  and the \textit{asterisk} again indicates N23308 as noted in Figure \ref{fig:ew_age}.
   The symbol size is proportional to the stellar mass. 
Galaxies older than 30 Myr are classified as those that prefer 
continuous star formation due to their large EW(H$\alpha$) for their evolved ages. 
The inset 
  plot shows the stellar mass distribution of galaxies that have 
  different star formation histories. 
   The \textit{solid} histogram is 
  the stellar mass distribution of the 39 galaxies that prefer 
  `continuous star formation'. 
   The \textit{dashed} histogram is the stellar mass distribution of
the 24 galaxies that prefer `instantaneous bursts'; i.e., 
are younger than 30\,Myr. 
   The median stellar masses are 
  $\langle M_* \rangle = 7.1\times10^9\,M_{\odot}$
  and $\langle M_* \rangle = 3.1\times10^9\,M_{\odot}$
  for continuous star formation and instantaneous burst, respectively.
Lower mass galaxies appear to show bursty star formation histories.
  }
 \end{figure*}

\begin{figure*}
  \plotone{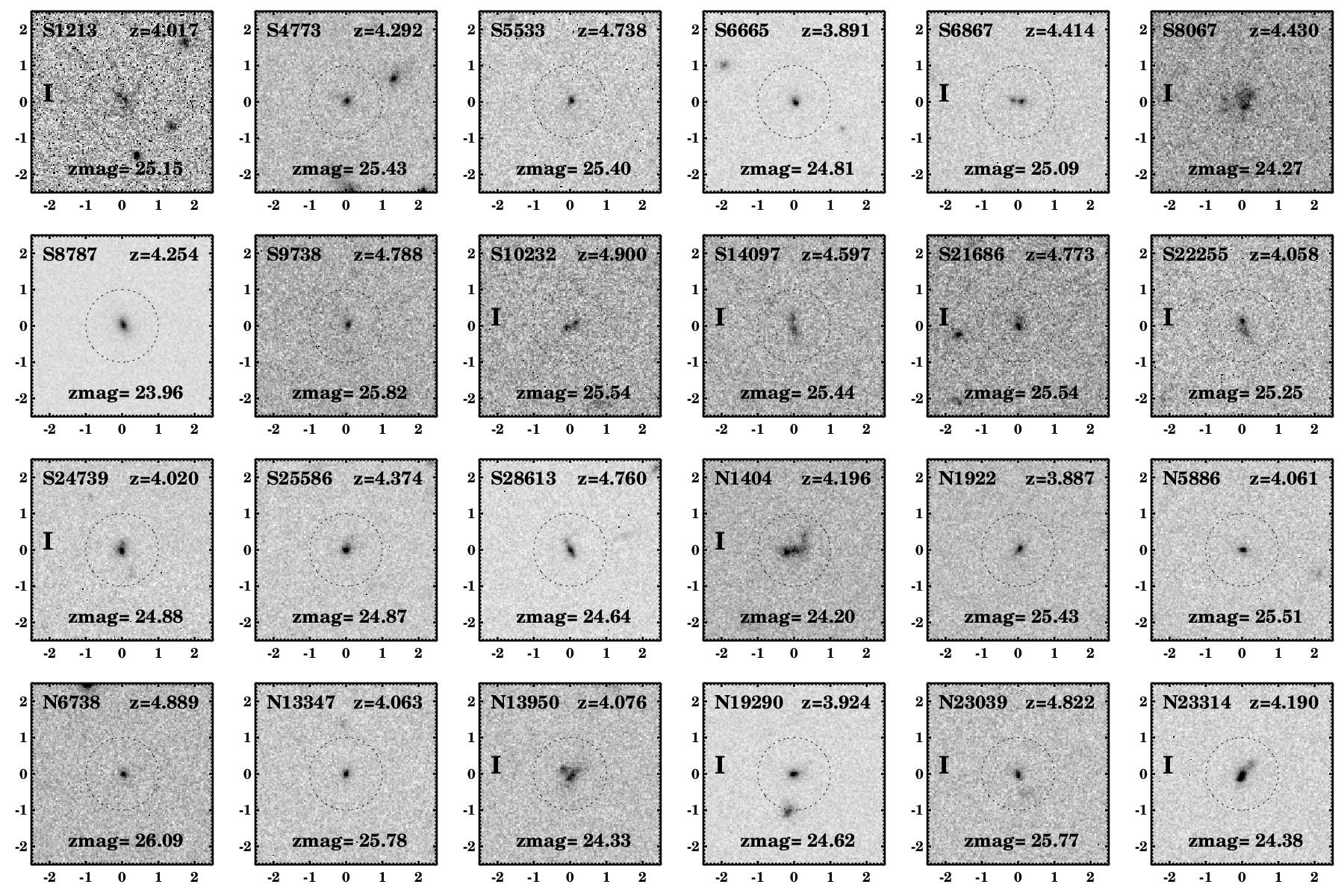}
  \caption{\label{fig:stamp_burst}
    Morphologies of the $z\sim4$ HAE candidates that have a star
   formation history corresponding to the instantaneous burst model.
    The star formation mode classification (`instantaneous' or
   `continuous') is based on the relationship between the
   H$\alpha$ EW and the stellar age (Figure \ref{fig:ew_age}): 
   we have classified galaxies older than 30 Myr as 
   galaxies with `continuous' star formation, and others as
   galaxies with `instantaneous' star formation.
    The postage stamp images are $5\arcsec\times5\arcsec$
   cutouts of \textit{HST}/ACS $z$-band images,
e.g., these are rest-frame
   FUV images at $z\sim4$. The galaxy ID, spectroscopic redshift,
   and $z$-band magnitude are indicated.
    We mark clearly merging/interacting systems with an `I' in the left of
   each postage image. Among 24 galaxies, 13 galaxies are classified as
   clear merging/interacting systems.
   }
 \end{figure*}

 \begin{figure*}
  \plotone{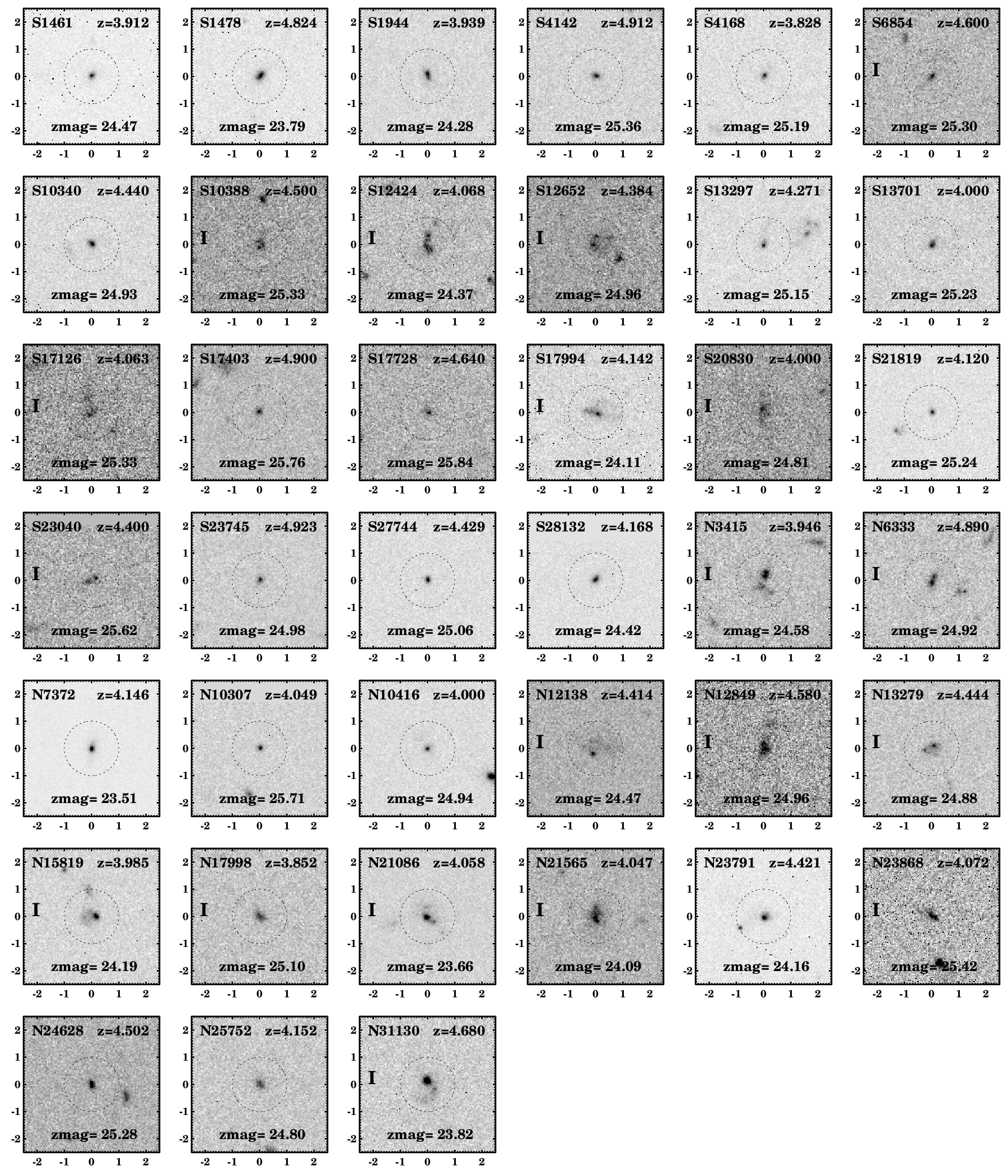}
   \caption{\label{fig:stamp_cont}
    Morphologies of the $z\sim4$ HAE candidates that have a star
   formation history corresponding to the continuous star formation model.
    The images are $5\arcsec\times5\arcsec$ cutouts of \textit{HST}/ACS
   $z$-band images as in Figure \ref{fig:stamp_burst}.
    The galaxy ID, spectroscopic redshift, $z$-band magnitude, and
   whether the object is merging/interacting system or not are marked
as in Figure \ref{fig:stamp_burst}. Among 39 galaxies, 19 galaxies
   are classified as clearly merging/interacting systems.
   }
 \end{figure*}

\begin{figure}
  \plotone{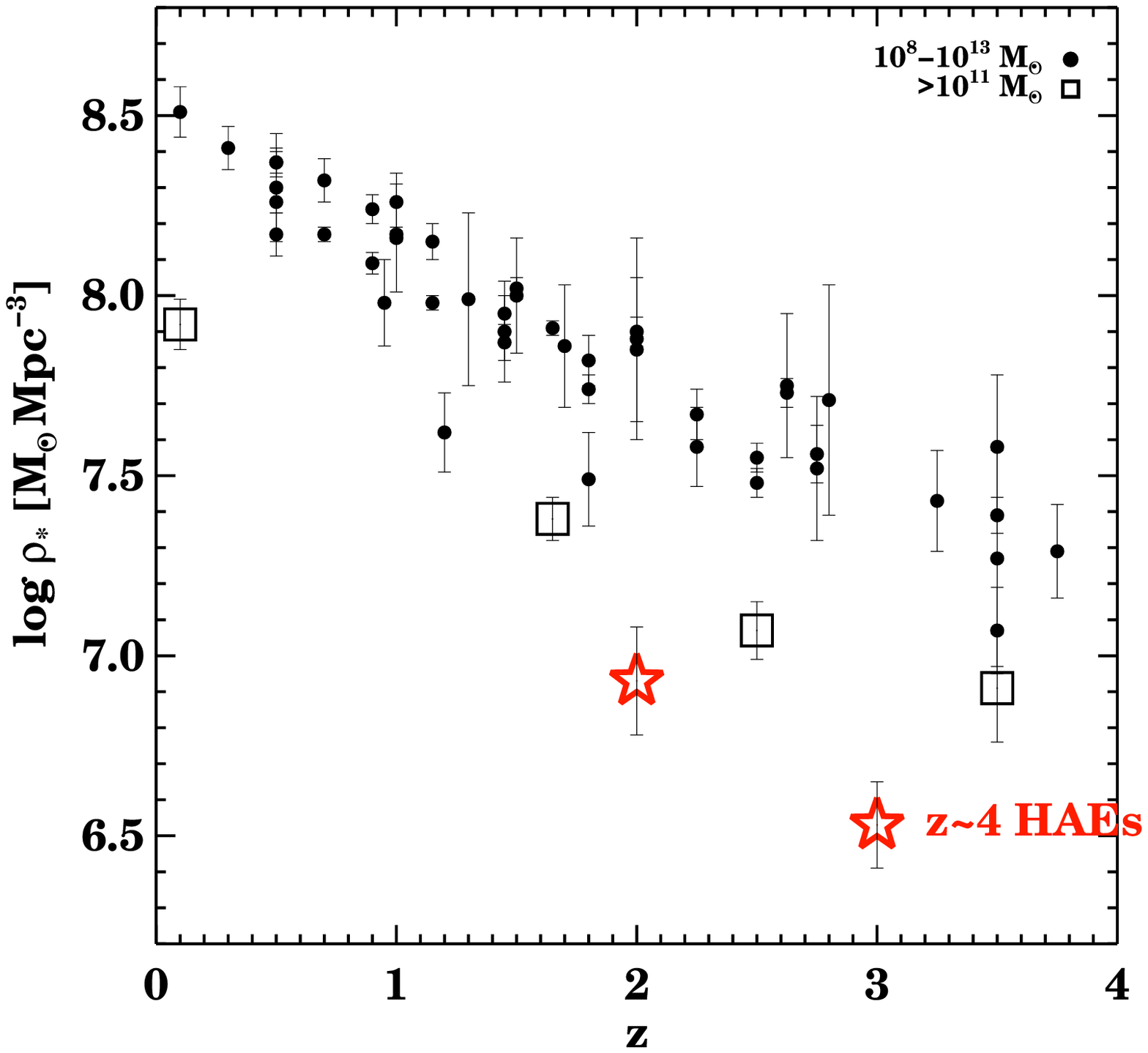}
  \caption{\label{fig:mass_z}
    The evolution of stellar mass density as a function of redshift. The
   points are from the compilation of Marchesini et al. (2009):
   \textit{filled circles} indicate the mass density of galaxies with
   stellar mass of $10^{8}-10^{13}\,M_{\odot}$, \textit{open squares}
   indicate the contribution to the mass density from
   massive ($>10^{11}\,M_{\odot}$) galaxies.
    The \textit{stars} are the ``expected'' mass density that $z\sim4$ HAEs
can harbor
 at lower redshifts if the HAEs continue to form stars at 
the SFRs they are observed to exhibit at $z\sim4$.
   }
 \end{figure}

\begin{figure}
 \plotone{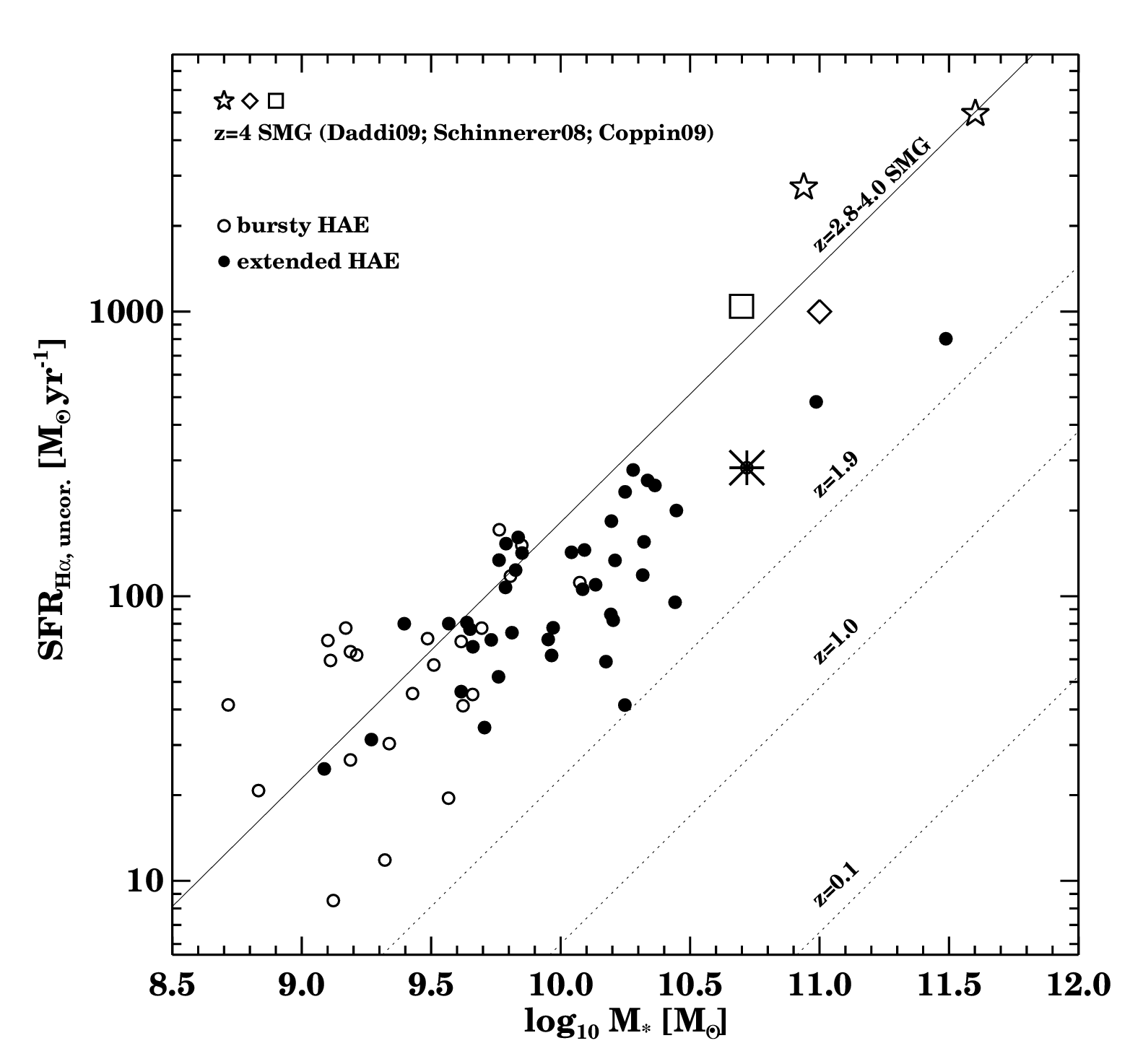}
  \caption{\label{fig:sfr_smass}
SFR vs. stellar mass correlation for HAE candidates
   (\textit{circles}). Galaxies that prefer continuous star
   formation models are marked as \textit{filled circles},
   and galaxies that prefer instantaneous burst models are
   marked as \textit{open circles}.
The \textit{asterisk} indicates N23308, an X-ray detected AGN.
  The SFR is derived using
   the derived H$\alpha$ line luminosity, and no correction
   for dust extinction is applied.
   Also plotted with different
   symbols are SFR and stellar mass of $z\sim4$ SMGs
   (\textit{stars} from Daddi et al. 2009;
   \textit{diamond} from Schinnerer et al. 2008;
   \textit{square} from Coppin et al. 2009).
The SFR and the stellar mass compared here are 
based on the assumption of Salpeter initial mass function (Salpeter 1955).
   The \textit{dotted} lines are the SFR vs. stellar mass
   correlation for star-forming galaxies at lower redshifts,
   $z=0.1$ for Elbaz et al. (2007), $z=1.0$ for Noeske et al. (2007),
   and $z=1.9$ for Daddi et al. (2007).
    The \textit{solid} line indicates the SFR vs. stellar mass 
   correlation of SMGs at $z\sim2-4$ (Daddi et al. 2009).
$z\sim4$ HAEs show SFR efficiencies similar to merger-driven SMGs, which is surprising,
because the HAEs appear to show extended star formation timescales.
This suggests that
HAEs harbor similar gas surface densities as SMGs.
   }
\end{figure}

 \begin{figure*}
  \plotone{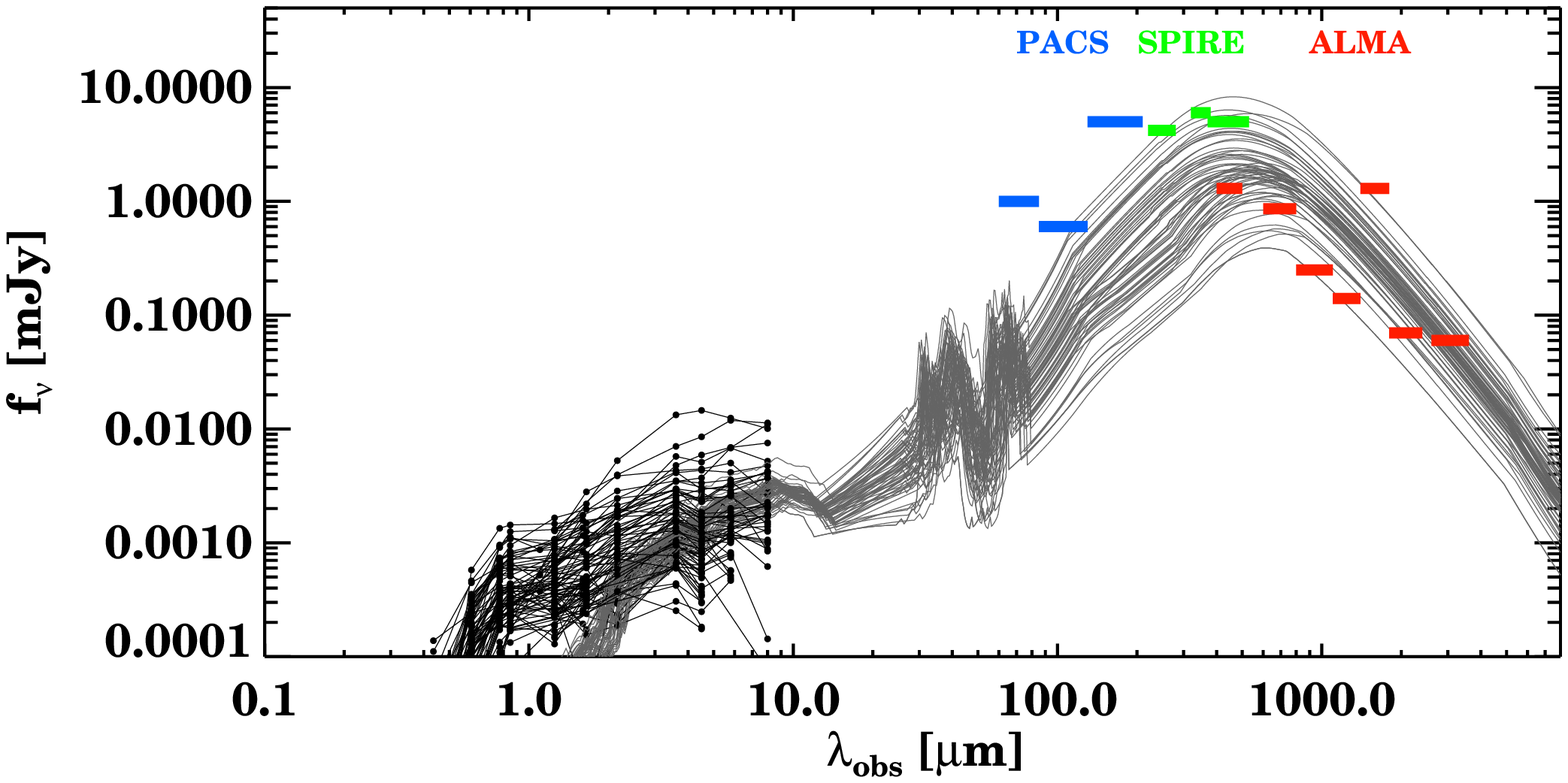}
  \caption{\label{fig:sed_ir}
   Infrared galaxy templates that represent $L_{IR}$ for each object
  at its corresponding redshift. The templates are drawn from Chary \& Pope (2010).
  The observed photometry
  data points from $B$-band to IRAC $8.0\,\mu$m is overplotted with
  small \textit{filled circles} and the connected lines.
   The thick horizontal color bars are limits for current and future surveys
  at longer wavelengths. The \textit{blue} bar represent \textit{Herschel}/PACS,
  the \textit{green} bar shows \textit{Herschel}/SPIRE limits in the GOODS-Herschel observations, 
  and the \textit{red} bar shows the expected ALMA limits.
ALMA limits indicate the expected continuum sensitivities obtained with 60 seconds exposure, 
at 12-m array configuration.
   The majority of these objects are beyond the sensitivity of 
  current far-infrared instrumentation.
  ALMA will be required to assess if the observed strong H$\alpha$ emission 
  relative to the UV continuum
  is a result of dust obscuration.
  }
 \end{figure*}

\clearpage

    \begin{landscape}

\begin{deluxetable}{c rr c cccc cc ccc cccc}
\tabletypesize{\tiny}
 \tablecaption{ \label{tab:phot} Photometry of Galaxies at $3.8<z<5.0$ }
 \tablewidth{0pt}
 \setlength{\tabcolsep}{0.05in}
 \tablehead{
   \colhead{ID} &  \colhead{R.A.} &  \colhead{Decl.}  &  \colhead{$z_{spec}$}  &
   \colhead{B}  & \colhead{V} & \colhead{i}  &  \colhead{z}  & \colhead{F110W}  &  \colhead{F160W }  &
   \colhead{J } & \colhead{H} & \colhead{K}  &
   \colhead{3.6$\mu$m} & \colhead{4.5$\mu$m} & \colhead{5.8$\mu$m} & \colhead{8.0$\mu$m}   }
\startdata
S1213  & 3 32 04.06 & -27 43 22.77 & 4.017 & $>28.9$        & 26.12$\pm$0.15 & 25.43$\pm$0.13 & 25.15$\pm$0.12 & $\ldots$       & $\ldots$       & 24.85$\pm$0.13 & 24.71$\pm$0.15 & 23.91$\pm$0.08 & 23.75$\pm$0.07 & 23.95$\pm$0.10 & 24.51$\pm$0.86 & $>22.54 $      \\
S1461  & 3 32 05.02 & -27 46 12.65 & 3.912 & 27.67$\pm$0.29 & 25.16$\pm$0.06 & 24.54$\pm$0.06 & 24.47$\pm$0.06 & $\ldots$       & $\ldots$       & 24.56$\pm$0.07 & 24.64$\pm$0.16 & 24.46$\pm$0.14 & 23.27$\pm$0.06 & 23.67$\pm$0.09 & 22.81$\pm$0.20 & 24.08$\pm$0.68 \\
S1478  & 3 32 05.08 & -27 46 56.52 & 4.825 & 30.21$\pm$3.67 & 26.22$\pm$0.13 & 23.97$\pm$0.06 & 23.79$\pm$0.06 & $\ldots$       & 24.09$\pm$0.11 & $\ldots$       & $\ldots$       & $\ldots$       & 22.00$\pm$0.06 & 22.13$\pm$0.06 & 21.82$\pm$0.10 & 22.11$\pm$0.14 \\
S1944  & 3 32 06.61 & -27 47 47.70 & 3.939 & 28.53$\pm$0.71 & 25.08$\pm$0.07 & 24.40$\pm$0.06 & 24.28$\pm$0.06 & $\ldots$       & $\ldots$       & $\ldots$       & $\ldots$       & $\ldots$       & 22.71$\pm$0.06 & 22.79$\pm$0.06 & 23.16$\pm$0.25 & 22.60$\pm$0.17 \\
S4142  & 3 32 11.71 & -27 41 49.59 & 4.912 & 28.90$\pm$0.70 & 27.27$\pm$0.16 & 25.55$\pm$0.07 & 25.36$\pm$0.07 & $\ldots$       & $\ldots$       & 25.35$\pm$0.08 & 25.28$\pm$0.10 & 24.94$\pm$0.12 & 24.44$\pm$0.09 & 24.85$\pm$0.18 & $>22.39$       & $>22.54$       \\
S4168  & 3 32 11.78 & -27 51 08.25 & 3.828 & 29.14$\pm$1.00 & 26.01$\pm$0.07 & 25.38$\pm$0.07 & 25.19$\pm$0.07 & $\ldots$       & $\ldots$       & 25.16$\pm$0.12 & 25.15$\pm$0.22 & 24.70$\pm$0.14 & 23.56$\pm$0.06 & 23.71$\pm$0.08 & $>22.39$       & $>22.54$       \\
S4773  & 3 32 12.99 & -27 48 33.75 & 4.292 & $>28.9$        & 26.74$\pm$0.12 & 25.62$\pm$0.09 & 25.43$\pm$0.08 & $\ldots$       & $\ldots$       & 26.12$\pm$0.55 & 25.06$\pm$0.34 & 24.83$\pm$0.29 & 23.97$\pm$0.03 & 24.18$\pm$0.07 & $>22.39$       & $>22.54$       \\
S5533  & 3 32 14.50 & -27 49 32.68 & 4.738 & 30.65$\pm$5.50 & 26.39$\pm$0.11 & 25.51$\pm$0.09 & 25.40$\pm$0.09 & $\ldots$       & 25.68$\pm$0.21 & 25.16$\pm$0.10 & 25.92$\pm$0.36 & 25.71$\pm$0.30 & 25.19$\pm$0.11 & 25.41$\pm$0.21 & 24.73$\pm$0.68 & $>22.54$       \\
S6665  & 3 32 16.64 & -27 42 53.35 & 3.891 & 27.03$\pm$0.18 & 25.27$\pm$0.06 & 24.86$\pm$0.06 & 24.81$\pm$0.06 & $\ldots$       & $\ldots$       & 24.60$\pm$0.05 & 24.45$\pm$0.05 & 25.55$\pm$0.23 & 24.44$\pm$0.09 & 24.46$\pm$0.13 & 23.86$\pm$0.43 & $>22.54$       \\
S6854  & 3 32 16.98 & -27 51 23.17 & 4.600 & $>28.9$        & 27.27$\pm$0.24 & 25.29$\pm$0.08 & 25.30$\pm$0.09 & $\ldots$       & 24.77$\pm$0.15 & 25.30$\pm$0.15 & 24.37$\pm$0.10 & 24.37$\pm$0.11 & 23.84$\pm$0.07 & 24.27$\pm$0.11 & $>22.39$       & $>22.54$       \\
S6867  & 3 32 17.00 & -27 41 13.72 & 4.414 & 28.90$\pm$0.90 & 26.66$\pm$0.13 & 25.30$\pm$0.07 & 25.09$\pm$0.07 & $\ldots$       & $\ldots$       & 24.73$\pm$0.07 & 24.43$\pm$0.07 & 24.31$\pm$0.10 & 23.37$\pm$0.06 & 23.83$\pm$0.09 & 23.56$\pm$0.34 & 23.44$\pm$0.32 \\
S8067  & 3 32 19.02 & -27 52 38.15 & 4.431 & 27.82$\pm$0.74 & 25.51$\pm$0.10 & 24.38$\pm$0.07 & 24.27$\pm$0.07 & $\ldots$       & $\ldots$       & 24.17$\pm$0.05 & 24.30$\pm$0.09 & 23.64$\pm$0.05 & 23.36$\pm$0.06 & 23.66$\pm$0.07 & 23.72$\pm$0.36 & 23.84$\pm$0.43 \\
S8787  & 3 32 20.29 & -27 47 18.18 & 4.254 & 27.54$\pm$0.26 & 25.70$\pm$0.07 & 24.29$\pm$0.06 & 23.96$\pm$0.05 & $\ldots$       & $\ldots$       & 23.35$\pm$0.02 & 23.16$\pm$0.02 & 23.08$\pm$0.03 & 22.72$\pm$0.06 & 23.24$\pm$0.06 & $>22.39$       & $>22.54$       \\
S9738  & 3 32 21.93 & -27 45 33.08 & 4.788 & $>28.9$        & 28.15$\pm$0.45 & 26.59$\pm$0.19 & 25.82$\pm$0.12 & $\ldots$       & 25.44$\pm$0.20 & 25.99$\pm$0.22 & 25.01$\pm$0.14 & 24.59$\pm$0.08 & 24.09$\pm$0.07 & 24.67$\pm$0.15 & $>22.39$       & $>22.54$       \\
S10232 & 3 32 22.71 & -27 51 54.38 & 4.900 & 30.65$\pm$6.50 & 27.54$\pm$0.37 & 25.81$\pm$0.13 & 25.54$\pm$0.12 & $\ldots$       & $\ldots$       & 25.25$\pm$0.11 & 27.27$\pm$1.21 & 24.83$\pm$0.13 & 24.09$\pm$0.08 & 24.42$\pm$0.12 & 24.16$\pm$0.53 & $>22.54$       \\
S10340 & 3 32 22.88 & -27 47 27.57 & 4.440 & $>28.9$        & 26.68$\pm$0.12 & 25.03$\pm$0.07 & 24.93$\pm$0.06 & $\ldots$       & $\ldots$       & 24.66$\pm$0.06 & 24.72$\pm$0.09 & 24.53$\pm$0.36 & 23.69$\pm$0.06 & 24.14$\pm$0.09 & 23.73$\pm$0.28 & 24.42$\pm$0.59 \\
S10388 & 3 32 22.97 & -27 46 29.09 & 4.500 & 28.62$\pm$1.08 & 27.45$\pm$0.34 & 25.66$\pm$0.12 & 25.33$\pm$0.10 & $\ldots$       & $\ldots$       & 24.57$\pm$0.07 & 24.34$\pm$0.09 & 23.86$\pm$0.31 & 23.10$\pm$0.06 & 23.48$\pm$0.07 & 23.40$\pm$0.24 & 23.65$\pm$0.35 \\
S12424 & 3 32 26.18 & -27 52 11.28 & 4.068 & 29.01$\pm$1.67 & 25.04$\pm$0.07 & 24.45$\pm$0.06 & 24.37$\pm$0.07 & $\ldots$       & $\ldots$       & 24.21$\pm$0.17 & 23.66$\pm$0.05 & 23.76$\pm$0.05 & 23.01$\pm$0.06 & 22.88$\pm$0.06 & 22.91$\pm$0.18 & 22.74$\pm$0.16 \\
S12652 & 3 32 26.49 & -27 41 23.97 & 4.384 & 27.29$\pm$0.32 & 25.51$\pm$0.07 & 24.93$\pm$0.07 & 24.96$\pm$0.08 & $\ldots$       & $\ldots$       & 25.44$\pm$0.16 & 24.87$\pm$0.15 & 24.04$\pm$0.13 & 23.89$\pm$0.07 & 23.68$\pm$0.08 & 24.62$\pm$0.86 & $>22.54$       \\
S13025 & 3 32 27.01 & -27 41 28.02 & 4.333 & $>28.9$        & 27.42$\pm$0.26 & 26.44$\pm$0.19 & 25.68$\pm$0.11 & $\ldots$       & $\ldots$       & 23.91$\pm$0.06 & 23.05$\pm$0.04 & 22.10$\pm$0.03 & 21.09$\pm$0.01 & 20.99$\pm$0.02 & 21.16$\pm$0.14 & 21.39$\pm$0.23 \\
S13297 & 3 32 27.37 & -27 55 27.37 & 4.271 & 29.29$\pm$1.57 & 26.43$\pm$0.11 & 25.32$\pm$0.08 & 25.15$\pm$0.08 & $\ldots$       & 25.05$\pm$0.17 & 25.55$\pm$0.18 & 24.69$\pm$0.08 & 25.18$\pm$0.17 & 23.29$\pm$0.03 & 23.21$\pm$0.04 & 22.81$\pm$0.20 & 22.84$\pm$0.24 \\
S13701 & 3 32 27.94 & -27 46 18.56 & 4.000 & $>28.9$        & 26.31$\pm$0.08 & 25.28$\pm$0.07 & 25.23$\pm$0.07 & $\ldots$       & $\ldots$       & 24.76$\pm$0.08 & 24.84$\pm$0.12 & 24.18$\pm$0.05 & 23.96$\pm$0.06 & 24.14$\pm$0.09 & 23.77$\pm$0.32 & 23.92$\pm$0.42 \\
S14097 & 3 32 28.56 & -27 40 55.72 & 4.597 & $>28.9$        & 27.32$\pm$0.21 & 25.48$\pm$0.08 & 25.44$\pm$0.08 & $\ldots$       & $\ldots$       & 25.33$\pm$0.15 & $\ldots$       & 25.69$\pm$0.40 & 24.36$\pm$0.08 & 24.39$\pm$0.13 & 24.11$\pm$0.56 & 26.01$\pm$3.43 \\
S14602 & 3 32 29.29 & -27 56 19.46 & 4.762 & $>28.9$        & 26.83$\pm$0.10 & 25.21$\pm$0.06 & 25.05$\pm$0.06 & $\ldots$       & $\ldots$       & $\ldots$       & $\ldots$       & $\ldots$       & 22.56$\pm$0.06 & 22.47$\pm$0.06 & 21.81$\pm$0.09 & 21.30$\pm$0.07 \\
S15920 & 3 32 31.19 & -27 54 29.33 & 4.005 & 28.32$\pm$0.59 & 27.03$\pm$0.18 & 26.42$\pm$0.16 & 25.82$\pm$0.11 & $\ldots$       & 23.29$\pm$0.10 & 23.82$\pm$0.03 & 22.78$\pm$0.01 & 22.43$\pm$0.01 & 22.31$\pm$0.06 & 22.28$\pm$0.06 & 22.15$\pm$0.10 & 22.81$\pm$0.18 \\
S17126 & 3 32 33.03 & -27 47 59.63 & 4.063 & 27.37$\pm$0.41 & 26.22$\pm$0.14 & 25.66$\pm$0.14 & 25.33$\pm$0.11 & $\ldots$       & $\ldots$       & 24.45$\pm$0.18 & 23.88$\pm$0.18 & 23.19$\pm$0.09 & 22.89$\pm$0.06 & 22.73$\pm$0.06 & 22.66$\pm$0.11 & 22.48$\pm$0.10 \\
S17403 & 3 32 33.47 & -27 50 30.00 & 4.900 & $>28.9$        & 28.20$\pm$0.37 & 26.47$\pm$0.14 & 25.76$\pm$0.09 & $\ldots$       & $\ldots$       & 25.51$\pm$0.09 & 25.20$\pm$0.12 & 24.79$\pm$0.07 & 23.69$\pm$0.04 & 23.78$\pm$0.06 & 23.72$\pm$0.38 & 23.88$\pm$0.56 \\
S17579 & 3 32 33.77 & -27 52 23.70 & 4.724 & $>28.9$        & 26.19$\pm$0.11 & 25.34$\pm$0.09 & 25.25$\pm$0.09 & $\ldots$       & $\ldots$       & 24.78$\pm$0.06 & 23.79$\pm$0.05 & 23.30$\pm$0.03 & 22.20$\pm$0.01 & 21.97$\pm$0.01 & 21.80$\pm$0.08 & 21.71$\pm$0.08 \\
S17728 & 3 32 33.98 & -27 48 02.03 & 4.640 & $>28.9$        & 27.29$\pm$0.18 & 26.08$\pm$0.11 & 25.84$\pm$0.10 & $\ldots$       & $\ldots$       & 25.70$\pm$0.13 & 25.46$\pm$0.18 & 24.22$\pm$0.19 & 24.04$\pm$0.06 & 24.15$\pm$0.08 & $>22.39$       & $>22.54$       \\
S17994 & 3 32 34.35 & -27 48 55.79 & 4.142 & 28.62$\pm$1.54 & 25.17$\pm$0.08 & 24.27$\pm$0.07 & 24.11$\pm$0.07 & $\ldots$       & $\ldots$       & 24.05$\pm$0.03 & 23.90$\pm$0.05 & 23.33$\pm$0.03 & 22.83$\pm$0.06 & 23.00$\pm$0.06 & 22.89$\pm$0.13 & $>22.54$       \\
S20830 & 3 32 38.73 & -27 44 13.34 & 4.000 & $>28.9$        & 26.03$\pm$0.14 & 25.04$\pm$0.10 & 24.81$\pm$0.09 & $\ldots$       & $\ldots$       & 24.26$\pm$0.09 & 24.15$\pm$0.17 & 23.32$\pm$0.08 & 23.28$\pm$0.06 & 23.44$\pm$0.07 & 23.46$\pm$0.28 & 23.29$\pm$0.27 \\
S21686 & 3 32 40.12 & -27 45 35.49 & 4.773 & 29.65$\pm$1.80 & 27.34$\pm$0.21 & 25.63$\pm$0.09 & 25.54$\pm$0.09 & $\ldots$       & $\ldots$       & 25.03$\pm$0.06 & 25.09$\pm$0.13 & 25.48$\pm$0.20 & 23.96$\pm$0.07 & 24.34$\pm$0.11 & $>22.39$       & $>22.54$       \\
S21819 & 3 32 40.38 & -27 44 31.01 & 4.120 & 29.01$\pm$0.89 & 25.74$\pm$0.07 & 25.24$\pm$0.07 & 25.24$\pm$0.07 & $\ldots$       & $\ldots$       & 25.38$\pm$0.06 & 25.06$\pm$0.09 & 24.91$\pm$0.09 & 24.10$\pm$0.07 & 24.91$\pm$0.18 & 24.52$\pm$0.75 & $>22.54$       \\
S22255 & 3 32 41.16 & -27 51 01.46 & 4.058 & $>28.9$        & 25.92$\pm$0.09 & 25.31$\pm$0.09 & 25.25$\pm$0.09 & $\ldots$       & $\ldots$       & 24.98$\pm$0.09 & 25.80$\pm$0.92 & 24.71$\pm$0.44 & 24.30$\pm$0.08 & 24.58$\pm$0.13 & $>22.39$       & $>22.54$       \\
S22746 & 3 32 42.05 & -27 47 40.58 & 4.049 & 27.64$\pm$0.38 & 25.86$\pm$0.08 & 25.47$\pm$0.09 & 25.47$\pm$0.10 & $\ldots$       & $\ldots$       & 24.96$\pm$0.07 & 23.93$\pm$0.05 & 23.60$\pm$0.04 & 23.40$\pm$0.06 & 23.44$\pm$0.06 & 23.56$\pm$0.24 & 23.28$\pm$0.20 \\
S23040 & 3 32 42.62 & -27 54 28.95 & 4.400 & 29.89$\pm$2.75 & 27.67$\pm$0.32 & 25.75$\pm$0.10 & 25.62$\pm$0.10 & $\ldots$       & $\ldots$       & 25.46$\pm$0.14 & 25.24$\pm$0.10 & 24.50$\pm$0.09 & 23.89$\pm$0.07 & 24.44$\pm$0.12 & $>22.39$       & $>22.54$       \\
S23745 & 3 32 44.07 & -27 42 27.43 & 4.923 & $>28.9$        & 27.11$\pm$0.25 & 25.01$\pm$0.08 & 24.98$\pm$0.08 & $\ldots$       & 24.90$\pm$0.16 & 25.02$\pm$0.08 & 25.02$\pm$0.19 & 24.79$\pm$0.15 & 23.76$\pm$0.07 & 23.88$\pm$0.09 & $>22.39$       & $>22.54$       \\
S23763 & 3 32 44.11 & -27 54 52.53 & 4.931 & 26.29$\pm$0.20 & 24.78$\pm$0.07 & 24.41$\pm$0.07 & 24.50$\pm$0.08 & $\ldots$       & $\ldots$       & 24.50$\pm$0.08 & 23.47$\pm$0.03 & 23.19$\pm$0.04 & 22.36$\pm$0.02 & 22.31$\pm$0.02 & 22.35$\pm$0.16 & 22.21$\pm$0.16 \\
S24739 & 3 32 46.25 & -27 48 46.98 & 4.020 & $>28.9$        & 25.65$\pm$0.06 & 24.89$\pm$0.06 & 24.88$\pm$0.06 & $\ldots$       & $\ldots$       & 24.79$\pm$0.06 & 24.96$\pm$0.14 & 24.31$\pm$0.08 & 23.93$\pm$0.06 & 24.25$\pm$0.09 & 23.62$\pm$0.26 & 24.04$\pm$0.43 \\
S25586 & 3 32 48.24 & -27 51 36.90 & 4.374 & $>28.9$        & 25.86$\pm$0.08 & 25.01$\pm$0.07 & 24.87$\pm$0.07 & $\ldots$       & $\ldots$       & 24.73$\pm$0.12 & 25.26$\pm$0.27 & 25.24$\pm$0.30 & 24.22$\pm$0.05 & 24.98$\pm$0.16 & $>22.39$       & $>22.54$       \\
S27744 & 3 32 54.04 & -27 50 00.79 & 4.430 & $>28.9$        & 25.93$\pm$0.06 & 24.95$\pm$0.06 & 25.06$\pm$0.06 & $\ldots$       & $\ldots$       & 25.33$\pm$0.09 & 25.29$\pm$0.13 & 24.98$\pm$0.09 & 24.80$\pm$0.12 & 25.20$\pm$0.25 & $>22.39$       & $>22.54$       \\
S28132 & 3 32 55.25 & -27 50 22.46 & 4.169 & 29.14$\pm$1.50 & 25.27$\pm$0.06 & 24.49$\pm$0.06 & 24.42$\pm$0.06 & $\ldots$       & $\ldots$       & 24.29$\pm$0.04 & 24.19$\pm$0.05 & 23.86$\pm$0.04 & 23.55$\pm$0.06 & 23.83$\pm$0.09 & 23.74$\pm$0.43 & 23.29$\pm$0.32 \\
S28613 & 3 32 57.17 & -27 51 45.01 & 4.760 & 28.32$\pm$0.71 & 26.02$\pm$0.09 & 24.60$\pm$0.06 & 24.64$\pm$0.06 & $\ldots$       & $\ldots$       & 24.98$\pm$0.07 & 24.78$\pm$0.07 & 24.45$\pm$0.07 & 24.13$\pm$0.06 & 24.54$\pm$0.14 & $>22.39$       & 26.70$\pm$8.13 \\
N1404  & 12 35 57.43 & 62 14 08.82 & 4.196 & $>28.9$        & 25.43$\pm$0.08 & 24.45$\pm$0.07 & 24.20$\pm$0.06 & $\ldots$       & $\ldots$       & 23.98$\pm$0.15 & $\ldots$       & 23.22$\pm$0.17 & 22.72$\pm$0.06 & 22.96$\pm$0.10 & 22.56$\pm$0.15 & 23.14$\pm$0.27 \\
N1922  & 12 36 01.25 & 62 13 19.21 & 3.887 & 28.00$\pm$0.39 & 26.07$\pm$0.07 & 25.52$\pm$0.08 & 25.43$\pm$0.08 & $\ldots$       & $\ldots$       & 25.96$\pm$0.48 & $\ldots$       & 24.96$\pm$0.42 & 25.39$\pm$0.17 & 25.79$\pm$0.38 & $>22.39$       & $>22.54$       \\
N3415  & 12 36 09.70 & 62 11 11.00 & 3.946 & 27.64$\pm$0.41 & 25.03$\pm$0.06 & 24.59$\pm$0.06 & 24.58$\pm$0.06 & $\ldots$       & $\ldots$       & 25.52$\pm$0.57 & $\ldots$       & $>24.27$       & 23.89$\pm$0.07 & 24.04$\pm$0.09 & 24.68$\pm$0.92 & $>22.54$       \\
N5886  & 12 36 20.27 & 62 12 11.51 & 4.061 & 29.14$\pm$1.00 & 26.03$\pm$0.08 & 25.48$\pm$0.07 & 25.51$\pm$0.08 & $\ldots$       & $\ldots$       & 24.03$\pm$0.19 & $\ldots$       & $>24.27$       & 24.46$\pm$0.09 & 24.67$\pm$0.14 & $>22.39$       & $>22.54$       \\
N6333  & 12 36 21.94 & 62 15 17.12 & 4.890 & $>28.9$        & 26.67$\pm$0.15 & 25.11$\pm$0.08 & 24.92$\pm$0.07 & $\ldots$       & $\ldots$       & 24.82$\pm$0.28 & $\ldots$       & 24.12$\pm$0.25 & 23.45$\pm$0.06 & 23.70$\pm$0.07 & 24.18$\pm$0.82 & $>22.54$       \\
N6660  & 12 36 23.23 & 62 12 35.80 & 4.338 & 28.09$\pm$0.38 & 25.93$\pm$0.07 & 25.51$\pm$0.07 & 25.42$\pm$0.07 & $\ldots$       & $\ldots$       & $>24.57$       & $\ldots$       & $>24.27$       & 24.45$\pm$0.08 & 25.22$\pm$0.22 & $>22.39$       & $>22.54$       \\
N6738  & 12 36 23.56 & 62 15 20.30 & 4.889 & $>28.9$        & 28.15$\pm$0.25 & 26.19$\pm$0.08 & 26.09$\pm$0.08 & $\ldots$       & $\ldots$       & $>24.57$       & $\ldots$       & $>24.27$       & 24.44$\pm$0.08 & 25.07$\pm$0.18 & $>22.39$       & $>22.54$       \\
N7372  & 12 36 25.92 & 62 09 03.95 & 4.146 & $>28.9$        & 24.50$\pm$0.05 & 23.58$\pm$0.05 & 23.51$\pm$0.05 & $\ldots$       & $\ldots$       & 23.48$\pm$0.07 & $\ldots$       & 23.33$\pm$0.12 & 23.00$\pm$0.06 & 23.40$\pm$0.07 & $>22.39$       & $>22.54$       \\
N10307 & 12 36 36.44 & 62 16 20.31 & 4.049 & $>28.9$        & 26.12$\pm$0.06 & 25.73$\pm$0.06 & 25.71$\pm$0.07 & $\ldots$       & $\ldots$       & $>24.57$       & $\ldots$       & $>24.27$       & 23.66$\pm$0.06 & 23.90$\pm$0.07 & $>22.39$       & 23.26$\pm$0.20 \\
N10416 & 12 36 36.82 & 62 12 04.30 & 4.000 & 27.34$\pm$0.33 & 25.74$\pm$0.08 & 25.05$\pm$0.07 & 24.94$\pm$0.07 & $\ldots$       & $\ldots$       & 24.35$\pm$0.31 & $\ldots$       & 22.93$\pm$0.17 & 22.76$\pm$0.06 & 22.95$\pm$0.06 & 22.87$\pm$0.16 & 23.49$\pm$0.32 \\
N12074 & 12 36 42.05 & 62 13 31.74 & 4.424 & 27.03$\pm$0.25 & 26.43$\pm$0.15 & 25.54$\pm$0.11 & 25.18$\pm$0.09 & 24.60$\pm$0.03 & 23.41$\pm$0.01 & 23.62$\pm$0.10 & $\ldots$       & 22.41$\pm$0.07 & 21.78$\pm$0.05 & 21.57$\pm$0.05 & 21.21$\pm$0.06 & 21.27$\pm$0.06 \\
N12138 & 12 36 42.25 & 62 15 23.25 & 4.414 & 29.14$\pm$2.62 & 26.42$\pm$0.18 & 24.66$\pm$0.08 & 24.47$\pm$0.07 & $\ldots$       & $\ldots$       & 24.40$\pm$0.17 & $\ldots$       & 23.51$\pm$0.15 & 23.09$\pm$0.06 & 23.36$\pm$0.06 & 23.30$\pm$0.19 & 23.32$\pm$0.21 \\
N12849 & 12 36 44.68 & 62 11 50.62 & 4.580 & 28.53$\pm$1.14 & 26.82$\pm$0.26 & 25.27$\pm$0.10 & 24.96$\pm$0.08 & 24.97$\pm$0.03 & 24.69$\pm$0.03 & 25.51$\pm$0.52 & $\ldots$       & 24.27$\pm$0.24 & 23.00$\pm$0.06 & 23.23$\pm$0.06 & 22.84$\pm$0.16 & $>22.54$       \\
N13279 & 12 36 46.16 & 62 07 01.83 & 4.444 & 28.80$\pm$1.09 & 26.07$\pm$0.09 & 24.89$\pm$0.06 & 24.88$\pm$0.07 & $\ldots$       & $\ldots$       & 25.05$\pm$0.16 & $\ldots$       & 24.48$\pm$0.22 & 23.96$\pm$0.07 & 24.25$\pm$0.13 & $>22.39$       & $>22.54$       \\
N13347 & 12 36 46.38 & 62 15 13.75 & 4.063 & 28.32$\pm$0.35 & 26.17$\pm$0.07 & 25.69$\pm$0.07 & 25.78$\pm$0.07 & $\ldots$       & 25.38$\pm$0.19 & $>24.57$       & $\ldots$       & $>24.27$       & 24.84$\pm$0.08 & 25.75$\pm$0.39 & $>22.39$       & $>22.54$       \\
N13950 & 12 36 48.41 & 62 16 43.29 & 4.076 & 28.70$\pm$1.50 & 25.22$\pm$0.07 & 24.42$\pm$0.06 & 24.33$\pm$0.06 & $\ldots$       & $\ldots$       & 24.95$\pm$0.41 & $\ldots$       & 23.77$\pm$0.24 & 23.42$\pm$0.06 & 23.67$\pm$0.07 & 23.85$\pm$0.31 & 23.62$\pm$0.27 \\
N15819 & 12 36 54.04 & 62 08 55.46 & 3.985 & 27.74$\pm$0.52 & 25.33$\pm$0.07 & 24.27$\pm$0.06 & 24.19$\pm$0.06 & $\ldots$       & $\ldots$       & 24.32$\pm$0.12 & $\ldots$       & 23.70$\pm$0.15 & 23.07$\pm$0.06 & 23.50$\pm$0.07 & 23.14$\pm$0.22 & 23.11$\pm$0.25 \\
N16443 & 12 36 55.94 & 62 14 12.76 & 4.085 & 26.85$\pm$0.24 & 24.74$\pm$0.06 & 23.95$\pm$0.06 & 23.78$\pm$0.05 & 24.05$\pm$0.02 & 23.57$\pm$0.01 & 23.64$\pm$0.13 & $\ldots$       & 22.76$\pm$0.12 & 22.54$\pm$0.06 & 22.59$\pm$0.06 & 22.52$\pm$0.10 & 22.38$\pm$0.10 \\
N17998 & 12 37 00.66 & 62 17 56.36 & 3.852 & 29.45$\pm$1.83 & 26.29$\pm$0.11 & 25.23$\pm$0.08 & 25.10$\pm$0.07 & $\ldots$       & $\ldots$       & 24.78$\pm$0.27 & $\ldots$       & 23.73$\pm$0.19 & 23.64$\pm$0.06 & 23.84$\pm$0.08 & 23.90$\pm$0.42 & $>22.54$       \\
N19290 & 12 37 05.01 & 62 17 31.34 & 3.924 & 28.62$\pm$0.77 & 25.39$\pm$0.06 & 24.74$\pm$0.06 & 24.62$\pm$0.06 & $\ldots$       & $\ldots$       & 24.20$\pm$0.15 & $\ldots$       & 23.57$\pm$0.18 & 22.97$\pm$0.06 & 23.26$\pm$0.07 & 23.23$\pm$0.23 & $>22.54$       \\
N20633 & 12 37 09.77 & 62 14 00.67 & 3.910 & 26.05$\pm$0.12 & 25.44$\pm$0.07 & 25.27$\pm$0.09 & 25.20$\pm$0.09 & $\ldots$       & $\ldots$       & 25.50$\pm$0.50 & $\ldots$       & 24.60$\pm$0.50 & 23.86$\pm$0.06 & 24.05$\pm$0.08 & $>22.39$       & $>22.54$       \\
N21086 & 12 37 11.48 & 62 21 55.83 & 4.058 & 28.90$\pm$1.70 & 25.04$\pm$0.07 & 23.99$\pm$0.06 & 23.66$\pm$0.05 & $\ldots$       & $\ldots$       & 23.61$\pm$0.09 & $\ldots$       & 22.92$\pm$0.10 & 22.34$\pm$0.06 & 22.73$\pm$0.06 & 22.64$\pm$0.17 & 22.33$\pm$0.13 \\
N21565 & 12 37 13.04 & 62 21 11.49 & 4.047 & 28.09$\pm$0.90 & 25.35$\pm$0.09 & 24.32$\pm$0.06 & 24.09$\pm$0.06 & $\ldots$       & 23.74$\pm$0.10 & 24.39$\pm$0.16 & $\ldots$       & 23.27$\pm$0.12 & 22.55$\pm$0.06 & 22.68$\pm$0.06 & 22.76$\pm$0.17 & 22.47$\pm$0.14 \\
N23039 & 12 37 18.07 & 62 16 41.72 & 4.822 & $>28.9$        & 27.42$\pm$0.18 & 25.74$\pm$0.08 & 25.77$\pm$0.08 & $\ldots$       & $\ldots$       & 25.84$\pm$0.39 & $\ldots$       & $>24.27$       & 24.36$\pm$0.08 & 25.04$\pm$0.19 & $>22.39$       & $>22.54$       \\
N23308 & 12 37 18.97 & 62 10 26.21 & 4.134 & $>28.9$        & 27.54$\pm$0.14 & 26.31$\pm$0.08 & 26.68$\pm$0.10 & $\ldots$       & $\ldots$       & 25.73$\pm$0.64 & $\ldots$       & 23.79$\pm$0.23 & 22.76$\pm$0.06 & 23.00$\pm$0.06 & 22.59$\pm$0.13 & 22.34$\pm$0.11 \\
N23314 & 12 37 19.00 & 62 19 53.84 & 4.190 & 29.89$\pm$2.75 & 25.12$\pm$0.06 & 24.45$\pm$0.06 & 24.38$\pm$0.06 & $\ldots$       & $\ldots$       & 24.51$\pm$0.13 & $\ldots$       & 24.13$\pm$0.20 & 24.19$\pm$0.08 & 24.56$\pm$0.14 & 24.22$\pm$0.60 & $>22.54$       \\
N23791 & 12 37 20.58 & 62 11 06.11 & 4.421 & $>28.9$        & 25.44$\pm$0.07 & 24.23$\pm$0.06 & 24.16$\pm$0.05 & $\ldots$       & $\ldots$       & 23.82$\pm$0.16 & $\ldots$       & 23.18$\pm$0.19 & 23.26$\pm$0.06 & 23.68$\pm$0.07 & 23.58$\pm$0.26 & 23.54$\pm$0.27 \\
N23868 & 12 37 20.84 & 62 18 43.52 & 4.072 & 28.15$\pm$0.50 & 26.26$\pm$0.10 & 25.48$\pm$0.08 & 25.42$\pm$0.08 & $\ldots$       & $\ldots$       & $>24.57$       & $\ldots$       & $>24.27$       & 23.21$\pm$0.06 & 23.34$\pm$0.07 & 23.32$\pm$0.26 & $>22.54$       \\
N24628 & 12 37 23.57 & 62 20 38.72 & 4.502 & 30.21$\pm$2.67 & 26.58$\pm$0.11 & 25.46$\pm$0.07 & 25.28$\pm$0.07 & $\ldots$       & $\ldots$       & 24.49$\pm$0.39 & $\ldots$       & $>24.27$       & 23.57$\pm$0.06 & 24.08$\pm$0.10 & 23.75$\pm$0.39 & $>22.54$       \\
N25752 & 12 37 28.03 & 62 19 54.01 & 4.152 & 28.09$\pm$0.71 & 26.08$\pm$0.11 & 24.98$\pm$0.08 & 24.80$\pm$0.08 & $\ldots$       & $\ldots$       & 25.26$\pm$0.22 & $\ldots$       & 24.15$\pm$0.19 & 23.42$\pm$0.06 & 23.56$\pm$0.08 & 23.31$\pm$0.26 & 23.01$\pm$0.22 \\
N31130 & 12 37 57.51 & 62 17 19.10 & 4.680 & $>28.9$        & 25.70$\pm$0.10 & 24.00$\pm$0.06 & 23.82$\pm$0.06 & $\ldots$       & $\ldots$       & 23.85$\pm$0.09 & $\ldots$       & 23.65$\pm$0.18 & 22.94$\pm$0.06 & 23.33$\pm$0.07 & 23.36$\pm$0.29 & 23.07$\pm$0.23 \\
\enddata
  \tablecomments{ The observed flux is given as a unit of AB magnitude
  and magnitude error for each filter. The $B, V, i,$ and $z$-band magnitudes
  indicate the \textit{HST}/ACS F435W, F606W, F775W, and F850LP magnitudes.
  The $F110W$ and $F160W$-band magnitudes are from the \textit{HST}/NICMOS.
  The $J, H,$ and $K$-band magnitudes are the \textit{VLT}/ISAAC magnitudes
  for galaxies in CDF-South,
  while $J$ and $K$-band magnitudes are from \textit{CFHT}/WIRCAM for galaxies in
  HDF-North.
Two of the objects, S15920 and N12074, appear to
show multiwavelength SED that are inconsistent with their quoted redshift and likely have erroneous
spectroscopic redshifts.   } 
\end{deluxetable}

\clearpage

     \end{landscape}

    \begin{landscape}

\begin{deluxetable}{l c r rrrr c r r c r r r}
 \tabletypesize{\scriptsize}
 \tablecaption{ \label{tab:params} Spectral Energy Distribution Fitting Results }
 \tablewidth{0pt}
 \setlength{\tabcolsep}{0.05in}
 \tablehead{
   \colhead{ID}  & \colhead{$z_{spec}$}  & \colhead{$\chi^2$}  & \colhead{log$_{10} M_*$ } &
   \colhead{$E(B-V)$}  &  \colhead{ age }  &
   \colhead{$\tau$ }  &  \colhead{ $\beta$ }  &
   \colhead{ ext.law }  & \colhead{ log$_{10} L_{H\alpha}$ }  &
   \colhead{ $EW_{H\alpha}$}      &
   \colhead{SFR$_{H\alpha}$}  & \colhead{SFR$_{UV}$} &
   \colhead{$S_{\rm ch1}$}
   \\
   \colhead{}    & \colhead{}            & \colhead{}          & \colhead{ ($M_{\odot}$)  } &
   \colhead{}          &  \colhead{ (Myr) } &
   \colhead{ (Myr) }  &  \colhead{}           &
   \colhead{}           & \colhead{ (erg s$^{-1}$) }          &
   \colhead{ ($\mbox{\AA}$) }          & 
   \colhead{ ($M_{\odot}$ yr$^{-1}$) }  & \colhead{ ($M_{\odot}$ yr$^{-1}$) } &
   \colhead{}
   }
\startdata
   S1213  &    4.017 &  0.48 &  9.66 &  0.37 &     6 &   100 &  -1.25  & SB & 43.25$\pm$0.15 &  362$\pm$93 & 141.9 &  10.6 & 11.80 \\
$\ldots$  & $\ldots$ &  0.66 &  9.76 &  0.17 &    91 &   800 &  -0.98  &SMC & 42.98$\pm$0.35 &  449$\pm$100 &  76.1 &  10.7 & 12.69 \\
   S1461  &    3.912 &  2.02 &  9.85 &  0.18 &    69 &  1000 &  -1.80  & SB & 43.46$\pm$0.02 &  896$\pm$72 & 228.6 &  20.1 & 13.76 \\
$\ldots$  & $\ldots$ &  1.89 &  9.84 &  0.07 &   170 &   900 &  -1.79  &SMC & 43.38$\pm$0.02 & 1285$\pm$106 & 188.7 &  20.6 & 16.74 \\
   S1478  &    4.825 &  3.08 & 10.99 &  0.04 &   701 &   400 &  -1.97  & SB & 43.84$\pm$0.07 &  743$\pm$69 & 548.8 &  54.2 & 10.16 \\
$\ldots$  & $\ldots$ &  3.20 & 11.01 &  0.01 &   822 &   400 &  -1.98  &SMC & 43.78$\pm$0.08 &  705$\pm$68 & 474.0 &  54.3 &  9.29 \\
   S1944  &    3.939 &  0.89 & 10.05 &  0.34 &    14 &   200 &  -1.29  & SB & 43.66$\pm$0.03 &  423$\pm$62 & 361.9 &  24.1 &  9.44 \\
$\ldots$  & $\ldots$ &  0.62 & 10.25 &  0.09 &   243 &   300 &  -1.51  &SMC & 43.56$\pm$0.04 & 1006$\pm$112 & 284.8 &  24.4 & 20.17 \\
   S4142  &    4.912 &  0.80 &  9.62 &  0.09 &    99 &   100 &  -2.09  & SB & 42.89$\pm$0.25 &  594$\pm$269 &  61.4 &  14.7 & 11.49 \\
$\ldots$  & $\ldots$ &  3.00 &  8.86 &  0.11 &     3 &  1000 &  -1.99  &SMC & 43.15$\pm$0.12 & 2557$\pm$825 & 111.1 &  16.3 & 12.70 \\
   S4168  &    3.828 &  0.90 &  9.79 &  0.30 &    39 &   700 &  -1.34  & SB & 43.69$\pm$0.08 & 1314$\pm$203 & 382.8 &   9.1 & 12.98 \\
$\ldots$  & $\ldots$ &  2.82 &  9.82 &  0.17 &   121 &   900 &  -0.89  &SMC & 43.47$\pm$0.14 & 1341$\pm$201 & 231.1 &   9.0 &  7.06 \\
   S4773  &    4.292 &  0.26 &  9.62 &  0.37 &     5 &   100 &  -1.27  & SB & 43.22$\pm$0.10 &  367$\pm$64 & 129.6 &   9.8 & 24.26 \\
$\ldots$  & $\ldots$ &  0.64 &  9.66 &  0.17 &    70 &   800 &  -0.97  &SMC & 42.91$\pm$0.24 &  406$\pm$66 &  64.2 &  10.2 & 18.06 \\
   S5533  &    4.738 &  3.09 &  8.83 &  0.15 &     4 &   800 &  -2.32  & SB & 42.62$\pm$0.14 &  584$\pm$204 &  32.8 &  13.8 &  5.16 \\
$\ldots$  & $\ldots$ &  4.43 &  8.63 &  0.06 &     1 &   900 &  -2.58  &SMC & 42.81$\pm$0.09 & 3644$\pm$1100 &  50.6 &  14.2 & 10.61 \\
   S6665  &    3.891 &  3.46 &  9.32 &  0.17 &    23 &   900 &  -1.99  & SB & 42.40$\pm$0.65 &  156$\pm$118 &  19.6 &  16.4 &  0.64 \\
$\ldots$  & $\ldots$ &  5.76 &  8.93 &  0.13 &     4 &  1000 &  -1.77  &SMC & 42.88$\pm$0.19 &  876$\pm$150 &  60.2 &  17.0 &  5.92 \\
   S6854  &    4.600 &  0.70 &  9.52 &  0.29 &     9 &   500 &  -1.55  & SB & 43.34$\pm$0.04 &  587$\pm$102 & 174.7 &  14.1 & 17.17 \\
$\ldots$  & $\ldots$ &  0.61 &  9.65 &  0.14 &    69 &   800 &  -1.28  &SMC & 43.12$\pm$0.07 &  660$\pm$106 & 103.5 &  14.6 & 19.87 \\
   S6867  &    4.414 &  0.69 &  9.81 &  0.31 &    19 &   300 &  -1.35  & SB & 43.59$\pm$0.04 &  737$\pm$82 & 307.9 &  14.0 & 24.39 \\
$\ldots$  & $\ldots$ &  0.69 &  9.81 &  0.17 &    71 &  1000 &  -1.00  &SMC & 43.35$\pm$0.07 &  792$\pm$80 & 175.0 &  14.8 & 25.40 \\
   S8067  &    4.431 &  0.64 &  9.69 &  0.25 &     8 &   200 &  -1.74  & SB & 43.32$\pm$0.07 &  382$\pm$54 & 166.9 &  32.1 & 15.14 \\
$\ldots$  & $\ldots$ &  0.82 &  9.92 &  0.09 &   101 &  1000 &  -1.66  &SMC & 43.12$\pm$0.12 &  444$\pm$71 & 104.7 &  32.2 & 15.21 \\
   S8787  &    4.254 & 23.00 & 10.08 &  0.36 &     5 &    60 &  -1.35  & SB & 43.73$\pm$0.04 &  416$\pm$66 & 421.1 &  33.6 &  2.43 \\
$\ldots$  & $\ldots$ &  2.65 &  9.76 &  0.22 &     8 &   500 &  -0.78  &SMC & 43.54$\pm$0.07 &  543$\pm$79 & 273.7 &  35.2 &  9.23 \\
   S9738  &    4.788 &  2.34 &  9.92 &  0.10 &    95 &    10 &  -1.60  & SB & 43.07$\pm$0.10 &  731$\pm$155 &  93.6 &   7.3 & 10.53 \\
$\ldots$  & $\ldots$ &  1.67 &  9.17 &  0.22 &     9 &   700 &  -0.68  &SMC & 43.17$\pm$0.08 &  825$\pm$178 & 117.8 &   7.8 & 14.42 \\
  S10232  &    4.900 &  0.25 &  9.51 &  0.30 &     8 &    50 &  -1.50  & SB & 43.26$\pm$0.13 &  499$\pm$137 & 144.8 &  12.8 & 12.46 \\
$\ldots$  & $\ldots$ &  0.38 &  9.67 &  0.14 &    80 &   700 &  -1.27  &SMC & 42.99$\pm$0.30 &  516$\pm$120 &  77.2 &  13.5 &  9.88 \\
  S10340  &    4.440 &  1.15 &  9.60 &  0.31 &     6 &   300 &  -1.51  & SB & 43.38$\pm$0.06 &  538$\pm$68 & 189.1 &  16.5 & 14.62 \\
$\ldots$  & $\ldots$ &  0.55 &  9.64 &  0.14 &    56 &   900 &  -1.32  &SMC & 43.14$\pm$0.11 &  642$\pm$78 & 108.8 &  17.3 & 24.23 \\
  S10388  &    4.500 &  0.54 & 10.03 &  0.46 &     5 &   200 &  -0.91  & SB & 43.87$\pm$0.01 &  650$\pm$74 & 580.3 &  11.6 & 27.37 \\
$\ldots$  & $\ldots$ &  0.27 & 10.04 &  0.23 &    75 &   700 &  -0.40  &SMC & 43.47$\pm$0.02 &  647$\pm$74 & 232.9 &  11.9 & 35.23 \\
  S12424  &    4.068 &  2.03 & 10.18 &  0.25 &    54 &   400 &  -1.51  & SB & 43.21$\pm$0.17 &  214$\pm$54 & 128.1 &  25.7 &  3.23 \\
$\ldots$  & $\ldots$ &  2.11 & 10.28 &  0.06 &   364 &   400 &  -1.82  &SMC & 43.26$\pm$0.15 &  590$\pm$88 & 143.6 &  26.1 &  7.45 \\
  S12652  &    4.384 &  5.52 &  9.90 &  0.18 &    80 &  1000 &  -1.79  & SB & 42.66$\pm$0.43 &  139$\pm$67 &  36.1 &  19.8 &  0.95 \\
$\ldots$  & $\ldots$ &  3.20 & 10.12 &  0.00 &   541 &   400 &  -2.64  &SMC & 42.66$\pm$0.43 &  338$\pm$182 &  36.3 &  17.7 &  3.16 \\
  S13025  &    4.333 &  7.02 & 11.51 &  0.00 &   845 &     1 &   0.76  & SB & 43.87$\pm$0.09 &  583$\pm$60 & 590.9 &   4.4 & 26.40 \\
$\ldots$  & $\ldots$ &  7.70 & 11.43 &  0.42 &    71 &     1 &   2.00  &SMC & 44.10$\pm$0.05 &  226$\pm$40 & 985.2 &   5.7 & 10.53 \\
  S13297  &    4.271 &  2.36 & 10.32 &  0.19 &   335 &   800 &  -1.57  & SB & 43.43$\pm$0.04 &  698$\pm$62 & 211.9 &  11.8 & 22.84 \\
$\ldots$  & $\ldots$ &  3.69 & 10.31 &  0.10 &   571 &  1000 &  -1.35  &SMC & 43.26$\pm$0.06 &  667$\pm$62 & 143.5 &  12.2 & 17.07 \\
  S13701  &    4.000 &  1.38 &  9.60 &  0.30 &    19 &   600 &  -1.40  & SB & 42.99$\pm$0.10 &  289$\pm$83 &  77.4 &   9.9 &  5.63 \\
$\ldots$  & $\ldots$ &  0.15 &  9.71 &  0.15 &    95 &   700 &  -1.09  &SMC & 42.79$\pm$0.17 &  332$\pm$84 &  48.3 &  10.4 & 20.90 \\ 
  S14097  &    4.597 &  0.45 &  9.57 &  0.26 &    22 &   600 &  -1.56  & SB & 42.74$\pm$0.40 &  192$\pm$90 &  43.5 &  11.9 &  4.02 \\
$\ldots$  & $\ldots$ &  1.90 &  9.19 &  0.18 &     9 &   800 &  -1.10  &SMC & 42.65$\pm$0.61 &  256$\pm$99 &  34.9 &  13.2 &  3.45 \\
  S14602\tablenotemark{a}\tablenotemark{b}  &    4.762 &  9.93 & 10.77 &  0.41 &    50 &    60 &  -0.77  & SB & 43.56$\pm$0.16 &  128$\pm$28 & 285.7 &  19.5 &  0.68 \\
$\ldots$  & $\ldots$ & 22.50 & 10.68 &  0.28 &    83 &    90 &   0.13  &SMC & $\ldots$       & $\ldots$  & $\ldots$ &  21.8 &  -2.10 \\
  S15920  &    4.005 & 72.89 & 11.03 &  0.00 &   699 &    60 &  -0.36  & SB & 42.74$\pm$0.55 &  107$\pm$25 &  43.5 &   5.9 &  0.32 \\
$\ldots$  & $\ldots$ & 10.09 & 10.41 &  0.56 &     5 &    30 &   2.43  &SMC & $\ldots$       & $\ldots$  & $\ldots$ &   3.8 &  -2.75 \\
  S17126  &    4.063 &  1.91 & 10.44 &  0.39 &    80 &   600 &  -0.87  & SB & 43.60$\pm$0.08 &  349$\pm$45 & 313.4 &   9.5 &  6.90 \\
$\ldots$  & $\ldots$ &  3.35 & 10.32 &  0.31 &    76 &   300 &   0.42  &SMC & 42.91$\pm$0.67 &   95$\pm$30 &  63.9 &   8.2 &  0.17 \\
  S17403  &    4.900 &  0.84 &  9.90 &  0.44 &     8 &    90 &  -0.91  & SB & 43.51$\pm$0.05 &  354$\pm$95 & 254.2 &   8.9 & 19.44 \\
$\ldots$  & $\ldots$ &  0.72 & 10.19 &  0.14 &   363 &   800 &  -1.02  &SMC & 43.17$\pm$0.11 &  548$\pm$104 & 117.8 &   9.5 & 27.08 \\
  S17579  &    4.724 &  8.58 & 11.14 &  0.26 &   115 &     1 &  -0.77  & SB & 43.52$\pm$0.02 &  147$\pm$3 & 263.2 &  17.3 &  5.12 \\
$\ldots$  & $\ldots$ &  8.60 & 10.98 &  0.10 &   378 &    80 &  -0.88  &SMC & 43.55$\pm$0.02 &  305$\pm$40 & 281.9 &  18.3 & 16.02 \\
  S17728  &    4.640 &  1.19 &  9.98 &  0.27 &    99 &   600 &  -1.36  & SB & 43.12$\pm$0.10 &  384$\pm$76 & 104.3 &   8.1 &  9.20 \\
$\ldots$  & $\ldots$ &  1.06 &  9.97 &  0.09 &   414 &   800 &  -1.48  &SMC & 42.98$\pm$0.14 &  645$\pm$86 &  75.8 &   9.5 & 13.47 \\
  S17994  &    4.142 &  0.70 & 10.10 &  0.26 &    34 &   400 &  -1.52  & SB & 43.58$\pm$0.05 &  465$\pm$55 & 297.7 &  29.6 & 16.51 \\
$\ldots$  & $\ldots$ &  0.46 & 10.20 &  0.09 &   207 &   600 &  -1.59  &SMC & 43.45$\pm$0.07 &  760$\pm$93 & 223.7 &  30.5 & 26.52 \\
  S20830  &    4.000 &  1.71 & 10.06 &  0.26 &    77 &   500 &  -1.43  & SB & 43.40$\pm$0.08 &  526$\pm$86 & 197.7 &  13.5 & 10.63 \\
$\ldots$  & $\ldots$ &  0.26 &  9.95 &  0.19 &    80 &   900 &  -0.73  &SMC & 43.13$\pm$0.15 &  371$\pm$80 & 107.4 &  13.8 & 19.51 \\
  S21686  &    4.773 &  0.83 &  9.49 &  0.29 &     9 &   300 &  -1.53  & SB & 43.34$\pm$0.06 &  635$\pm$118 & 173.4 &  12.4 & 16.11 \\
$\ldots$  & $\ldots$ &  1.40 &  9.63 &  0.13 &    80 &   800 &  -1.33  &SMC & 43.11$\pm$0.12 &  754$\pm$118 & 102.3 &  13.2 & 13.65 \\
  S21819  &    4.120 &  0.50 &  9.40 &  0.09 &    99 &   500 &  -2.19  & SB & 43.12$\pm$0.04 & 1459$\pm$181 & 103.8 &  12.4 & 31.55 \\
$\ldots$  & $\ldots$ &  4.87 &  8.76 &  0.09 &     1 &   900 &  -2.29  &SMC & 43.29$\pm$0.03 & 8470$\pm$855 & 155.5 &  13.2 & 17.02 \\
  S22255  &    4.058 &  0.27 &  9.27 &  0.26 &     8 &   500 &  -1.72  & SB & 42.94$\pm$0.18 &  420$\pm$144 &  68.6 &  11.6 & 15.55 \\
$\ldots$  & $\ldots$ &  4.45 &  8.98 &  0.17 &     5 &   800 &  -1.37  &SMC & 43.00$\pm$0.15 & 1011$\pm$201 &  79.8 &  12.1 &  8.62 \\
  S22746  &    4.049 &  6.16 & 10.29 &  0.08 &   739 &   700 &  -1.92  & SB & 43.26$\pm$0.10 &  880$\pm$144 & 142.6 &  10.8 &  7.12 \\
$\ldots$  & $\ldots$ &  5.79 & 10.12 &  0.09 &   469 &   900 &  -1.51  &SMC & 43.24$\pm$0.10 &  885$\pm$146 & 137.9 &  11.3 &  7.59 \\
  S23040  &    4.400 &  1.50 &  9.72 &  0.25 &    65 &   500 &  -1.52  & SB & 43.35$\pm$0.04 &  932$\pm$193 & 175.2 &   8.2 & 15.11 \\
$\ldots$  & $\ldots$ &  0.56 &  9.57 &  0.17 &    70 &   700 &  -1.01  &SMC & 43.16$\pm$0.07 &  897$\pm$178 & 114.8 &   8.7 & 23.97 \\
  S23745  &    4.923 &  0.89 &  9.93 &  0.19 &    71 &   800 &  -1.78  & SB & 43.22$\pm$0.10 &  440$\pm$81 & 130.2 &  22.2 & 11.77 \\
$\ldots$  & $\ldots$ &  0.10 & 10.20 &  0.01 &   637 &   900 &  -2.19  &SMC & 43.03$\pm$0.16 &  555$\pm$89 &  84.8 &  22.2 & 36.11 \\
  S23763\tablenotemark{a}  &    4.931 & 24.24 & 10.80 &  0.28 &    91 &    90 &  -1.29  & SB & 43.74$\pm$0.05 &  261$\pm$19 & 433.6 &  42.9 &  6.17 \\
$\ldots$  & $\ldots$ & 24.87 & 10.72 &  0.06 &   189 &    50 &  -1.50  &SMC & 43.60$\pm$0.07 &  478$\pm$47 & 311.9 &  44.9 &  8.84 \\
  S24739  &    4.020 &  0.68 &  9.43 &  0.23 &    14 &   700 &  -1.74  & SB & 43.07$\pm$0.07 &  453$\pm$73 &  93.0 &  15.3 & 13.64 \\
$\ldots$  & $\ldots$ &  5.71 &  9.11 &  0.17 &     5 &   800 &  -1.36  &SMC & 43.16$\pm$0.06 & 1060$\pm$180 & 113.0 &  16.1 &  9.52 \\
  S25586  &    4.374 &  1.17 &  9.11 &  0.18 &     6 &  1000 &  -2.09  & SB & 43.12$\pm$0.05 &  920$\pm$152 & 103.3 &  17.9 & 23.97 \\
$\ldots$  & $\ldots$ &  1.70 &  8.90 &  0.11 &     4 &   600 &  -1.92  &SMC & 43.06$\pm$0.05 & 1393$\pm$398 &  90.9 &  19.2 & 24.85 \\
  S27744  &    4.430 &  0.35 &  9.09 &  0.08 &    30 &   900 &  -2.32  & SB & 42.61$\pm$0.19 &  493$\pm$193 &  32.0 &  17.1 & 10.49 \\
$\ldots$  & $\ldots$ &  2.36 &  8.80 &  0.07 &     1 &   800 &  -2.47  &SMC & 42.94$\pm$0.08 & 3408$\pm$566 &  68.5 &  17.9 & 11.36 \\
  S28132  &    4.169 &  0.59 &  9.57 &  0.24 &     9 &   700 &  -1.76  & SB & 43.18$\pm$0.08 &  357$\pm$70 & 119.9 &  25.3 & 11.81 \\
$\ldots$  & $\ldots$ &  0.40 &  9.81 &  0.08 &   117 &  1000 &  -1.77  &SMC & 43.05$\pm$0.11 &  518$\pm$91 &  88.6 &  25.7 & 19.48 \\
  S28613  &    4.760 &  2.96 &  9.21 &  0.16 &     6 &   900 &  -2.19  & SB & 43.11$\pm$0.08 &  703$\pm$154 & 101.0 &  28.9 & 11.99 \\
$\ldots$  & $\ldots$ &  4.71 &  8.88 &  0.07 &     5 &   900 &  -2.34  &SMC & 43.14$\pm$0.07 & 1745$\pm$303 & 108.6 &  30.0 & 15.68 \\
   N1404  &    4.196 &  1.13 & 10.07 &  0.36 &     8 &    70 &  -1.26  & SB & 43.63$\pm$0.13 &  322$\pm$69 & 333.4 &  27.9 &  8.03 \\
$\ldots$  & $\ldots$ &  1.64 & 10.21 &  0.16 &   107 &   500 &  -1.02  &SMC & 43.44$\pm$0.22 &  498$\pm$88 & 219.1 &  27.6 & 10.54 \\ 
   N1922  &    3.887 &  0.97 &  9.12 &  0.26 &     3 &   500 &  -1.95  & SB & 42.38$\pm$0.78 &  290$\pm$250 &  19.0 &   8.5 &  2.05 \\
$\ldots$  & $\ldots$ &  1.45 &  8.71 &  0.12 &     1 &   900 &  -2.01  &SMC & 42.70$\pm$0.57 & 2346$\pm$390 &  39.7 &   8.7 &  9.08 \\
   N3415  &    3.946 &  1.37 &  9.68 &  0.12 &    82 &   700 &  -2.06  & SB & 42.93$\pm$0.14 &  451$\pm$90 &  67.6 &  20.5 &  6.98 \\
$\ldots$  & $\ldots$ &  1.22 &  9.66 &  0.04 &   175 &   800 &  -2.16  &SMC & 42.96$\pm$0.13 &  743$\pm$113 &  71.6 &  20.7 & 10.17 \\
   N5886  &    4.061 &  3.03 &  9.19 &  0.25 &     9 &   700 &  -1.74  & SB & 42.85$\pm$0.17 &  417$\pm$113 &  56.5 &   9.8 &  3.27 \\
$\ldots$  & $\ldots$ &  3.98 &  9.46 &  0.10 &   105 &   800 &  -1.58  &SMC & 42.63$\pm$0.34 &  423$\pm$105 &  33.7 &   9.8 &  2.85 \\
   N6333  &    4.890 &  1.86 &  9.77 &  0.29 &    10 &    90 &  -1.52  & SB & 43.48$\pm$0.03 &  465$\pm$60 & 238.9 &  23.5 &  8.66 \\
$\ldots$  & $\ldots$ &  1.32 & 10.21 &  0.04 &   513 &  1000 &  -1.99  &SMC & 43.26$\pm$0.05 &  793$\pm$93 & 145.0 &  22.9 & 14.53 \\
   N6660  &    4.338 &  5.34 &  8.76 &  0.12 &    10 &  1000 &  -2.30  & SB & 43.08$\pm$0.03 & 1901$\pm$545 &  94.1 &  11.9 &  7.87 \\
$\ldots$  & $\ldots$ &  9.02 &  8.68 &  0.08 &     1 &  1000 &  -2.38  &SMC & 43.13$\pm$0.02 & 1901$\pm$522 & 105.8 &  12.2 &  8.47 \\
   N6738  &    4.889 &  0.32 &  9.19 &  0.27 &     9 &   800 &  -1.65  & SB & 43.26$\pm$0.06 & 1066$\pm$230 & 144.2 &   8.1 & 21.52 \\
$\ldots$  & $\ldots$ &  2.94 &  8.98 &  0.19 &     5 &   600 &  -1.15  &SMC & 43.11$\pm$0.09 & 1301$\pm$255 & 101.2 &   9.2 &  7.54 \\
   N7372  &    4.146 &  0.93 & 10.16 &  0.32 &     3 &    30 &  -1.73  & SB & 43.60$\pm$0.04 &  488$\pm$48 & 316.7 &  55.2 & 11.68 \\
$\ldots$  & $\ldots$ &  0.85 &  9.76 &  0.10 &    34 &  1000 &  -1.77  &SMC & 43.32$\pm$0.08 &  573$\pm$59 & 164.7 &  55.0 & 14.06 \\
  N10307  &    4.049 &  2.01 &  9.94 &  0.08 &   394 &   300 &  -1.94  & SB & 43.27$\pm$0.03 & 1477$\pm$125 & 147.4 &   8.0 & 12.99 \\
$\ldots$  & $\ldots$ &  1.65 & 10.14 &  0.00 &  1048 &   600 &  -2.15  &SMC & 43.14$\pm$0.04 & 1329$\pm$118 & 109.2 &   7.8 & 12.75 \\
  N10416  &    4.000 &  2.82 & 10.39 &  0.09 &   120 &     1 &  -1.51  & SB & 43.50$\pm$0.05 &  819$\pm$101 & 248.7 &  14.6 & 11.16 \\
$\ldots$  & $\ldots$ &  2.80 & 10.36 &  0.08 &   602 &   600 &  -1.52  &SMC & 43.57$\pm$0.04 & 1333$\pm$138 & 291.9 &  14.2 & 14.71 \\
  N12074\tablenotemark{a}  &    4.424 & 21.30 & 11.03 &  0.66 &     5 &    10 &   0.06  & SB & $\ldots$       & $\ldots$  & $\ldots$ &  16.1 & -2.37     \\
$\ldots$  & $\ldots$ & 10.50 & 11.01 &  0.36 &   178 &   300 &   1.01  &SMC & 43.88$\pm$0.15 &  304$\pm$14 & 601.2 &   9.6 &  2.69 \\
  N12138  &    4.414 &  1.35 &  9.98 &  0.37 &     5 &    80 &  -1.30  & SB & 43.57$\pm$0.02 &  367$\pm$28 & 292.6 &  23.7 &  9.43 \\
$\ldots$  & $\ldots$ &  0.89 & 10.09 &  0.15 &   101 &   700 &  -1.10  &SMC & 43.27$\pm$0.05 &  428$\pm$35 & 147.1 &  24.1 & 13.55 \\
  N12849  &    4.580 &  1.14 & 10.40 &  0.23 &   126 &    80 &  -1.41  & SB & 43.67$\pm$0.04 &  760$\pm$146 & 368.3 &  15.4 & 20.47 \\
$\ldots$  & $\ldots$ &  1.08 & 10.45 &  0.08 &   655 &   800 &  -1.50  &SMC & 43.48$\pm$0.07 &  929$\pm$150 & 240.0 &  16.0 & 22.89\\
  N13279  &    4.444 &  0.75 &  9.76 &  0.13 &    99 &   600 &  -1.98  & SB & 42.99$\pm$0.12 &  478$\pm$111 &  78.0 &  18.6 & 11.83 \\
$\ldots$  & $\ldots$ &  5.34 &  8.99 &  0.12 &     5 &   600 &  -1.79  &SMC & 43.22$\pm$0.07 & 1632$\pm$312 & 130.0 &  20.2 & 10.53 \\
  N13347  &    4.063 &  1.25 &  8.72 &  0.17 &     6 &   800 &  -2.13  & SB & 42.95$\pm$0.05 & 1578$\pm$465 &  70.4 &   8.1 & 15.34 \\
$\ldots$  & $\ldots$ &  3.18 &  8.66 &  0.11 &     1 &  1000 &  -2.09  &SMC & 42.98$\pm$0.04 & 5122$\pm$885 &  75.9 &   8.4 & 14.40 \\
  N13950  &    4.076 &  1.14 &  9.62 &  0.25 &     9 &   300 &  -1.71  & SB & 43.28$\pm$0.08 &  413$\pm$65 & 149.9 &  24.9 & 11.17 \\
$\ldots$  & $\ldots$ &  1.34 &  9.77 &  0.11 &    70 &  1000 &  -1.53  &SMC & 43.12$\pm$0.13 &  510$\pm$75 & 104.0 &  25.6 & 12.71 \\
  N15819  &    3.985 &  3.77 &  9.74 &  0.29 &     8 &   400 &  -1.55  & SB & 43.58$\pm$0.02 &  620$\pm$50 & 299.9 &  23.6 &  8.51 \\
$\ldots$  & $\ldots$ &  2.39 &  9.85 &  0.13 &    71 &  1000 &  -1.34  &SMC & 43.38$\pm$0.03 &  773$\pm$63 & 188.6 &  24.6 & 12.95 \\
  N16443  &    4.085 & 15.89 & 10.19 &  0.36 &    11 &    60 &  -1.19  & SB & 43.54$\pm$0.31 &  214$\pm$84 & 276.8 &  29.3 &  1.28 \\
$\ldots$  & $\ldots$ & 19.65 & 10.41 &  0.13 &   226 &   500 &  -1.24  &SMC & 43.50$\pm$0.35 &  554$\pm$180 & 251.3 &  30.2 &  2.96 \\
  N17998  &    3.852 &  1.59 &  9.72 &  0.40 &     5 &    30 &  -1.14  & SB & 43.39$\pm$0.09 &  438$\pm$150 & 194.2 &   9.4 &  5.38 \\
$\ldots$  & $\ldots$ &  0.26 &  9.73 &  0.19 &    70 &   800 &  -0.76  &SMC & 43.13$\pm$0.17 &  564$\pm$181 & 106.5 &   9.5 & 18.45 \\
  N19290  &    3.924 &  0.29 &  9.85 &  0.33 &    13 &   200 &  -1.34  & SB & 43.71$\pm$0.07 &  735$\pm$119 & 408.6 &  17.7 & 28.83 \\
$\ldots$  & $\ldots$ &  2.32 & 10.03 &  0.16 &   105 &   800 &  -0.98  &SMC & 43.45$\pm$0.13 &  743$\pm$108 & 221.6 &  18.0 & 10.53 \\
  N20633  &    3.910 & 10.46 &  9.42 &  0.28 &     9 &   500 &  -1.59  & SB & 43.23$\pm$0.09 &  579$\pm$111 & 134.6 &  12.2 &  3.43 \\
$\ldots$  & $\ldots$ & 13.61 &  9.54 &  0.12 &    79 &   800 &  -1.44  &SMC & 43.10$\pm$0.13 &  891$\pm$130 &  99.6 &  12.4 &  4.43 \\
  N21086  &    4.058 &  4.85 & 10.12 &  0.34 &     8 &   400 &  -1.33  & SB & 43.90$\pm$0.02 &  539$\pm$44 & 630.6 &  36.3 &  7.17 \\
$\ldots$  & $\ldots$ &  2.02 & 10.28 &  0.14 &   114 &   700 &  -1.17  &SMC & 43.68$\pm$0.04 &  750$\pm$52 & 378.7 &  37.3 & 14.39 \\
  N21565  &    4.047 &  2.52 & 10.13 &  0.35 &    13 &    80 &  -1.21  & SB & 43.78$\pm$0.01 &  451$\pm$38 & 472.3 &  25.1 &  8.22 \\
$\ldots$  & $\ldots$ &  1.96 & 10.34 &  0.11 &   261 &   600 &  -1.37  &SMC & 43.61$\pm$0.02 &  916$\pm$56 & 324.6 &  26.4 & 14.72 \\ 
  N23039  &    4.822 &  0.94 &  9.10 &  0.22 &     9 &   800 &  -1.87  & SB & 43.24$\pm$0.05 & 1213$\pm$248 & 136.6 &  10.9 & 15.94 \\
$\ldots$  & $\ldots$ &  3.36 &  8.64 &  0.10 &     4 &  1000 &  -1.98  &SMC & 43.19$\pm$0.06 & 3367$\pm$600 & 121.8 &  11.4 & 12.35 \\
  N23308\tablenotemark{a}  &    4.134 &  2.27 & 10.72 &  0.29 &   871 &   700 &  -1.03  & SB & 43.91$\pm$0.04 & 1687$\pm$134 & 641.8 &   3.6 & 17.64 \\
$\ldots$  & $\ldots$ &  3.97 & 10.64 &  0.12 &  1222 &   500 &  -0.81  &SMC & 43.67$\pm$0.08 & 1644$\pm$120 & 369.7 &   3.6 & 12.99 \\
  N23314  &    4.190 &  0.88 &  9.34 &  0.20 &     4 &    80 &  -2.11  & SB & 42.85$\pm$0.26 &  333$\pm$123 &  56.1 &  27.0 &  6.89 \\
$\ldots$  & $\ldots$ &  2.18 &  9.02 &  0.10 &     5 &  1000 &  -2.01  &SMC & 42.94$\pm$0.20 &  795$\pm$180 &  68.1 &  28.2 &  9.73 \\
  N23791  &    4.421 &  1.17 &  9.73 &  0.26 &     6 &   600 &  -1.74  & SB & 43.41$\pm$0.04 &  440$\pm$60 & 205.4 &  36.2 & 11.90 \\
$\ldots$  & $\ldots$ &  0.64 &  9.79 &  0.10 &    52 &  1000 &  -1.64  &SMC & 43.23$\pm$0.06 &  544$\pm$66 & 134.6 &  37.4 & 19.51 \\
  N23868  &    4.072 &  1.27 &  9.86 &  0.41 &     9 &    30 &  -0.99  & SB & 43.64$\pm$0.02 &  528$\pm$66 & 343.3 &   9.5 &  9.92 \\
$\ldots$  & $\ldots$ &  0.58 & 10.32 &  0.07 &   859 &   800 &  -1.59  &SMC & 43.36$\pm$0.05 & 1132$\pm$97 & 179.5 &   9.6 & 22.32 \\
  N24628  &    4.502 &  1.48 &  9.59 &  0.30 &    11 &   300 &  -1.46  & SB & 43.54$\pm$0.05 &  840$\pm$100 & 274.3 &  13.2 & 14.27 \\
$\ldots$  & $\ldots$ &  1.43 &  9.83 &  0.11 &   179 &  1000 &  -1.49  &SMC & 43.29$\pm$0.09 & 1106$\pm$113 & 155.1 &  13.5 & 16.49 \\
  N25752  &    4.152 &  2.14 &  9.97 &  0.27 &    54 &   600 &  -1.46  & SB & 43.35$\pm$0.07 &  471$\pm$62 & 174.9 &  14.4 &  8.39 \\
$\ldots$  & $\ldots$ &  2.70 &  9.98 &  0.13 &   181 &  1000 &  -1.29  &SMC & 43.20$\pm$0.09 &  636$\pm$74 & 125.8 &  15.0 &  8.58 \\
  N31130  &    4.680 &  1.59 & 10.15 &  0.14 &    82 &  1000 &  -1.97  & SB & 43.50$\pm$0.04 &  567$\pm$68 & 247.6 &  49.2 & 13.41 \\
$\ldots$  & $\ldots$ &  1.30 & 10.09 &  0.09 &   101 &   900 &  -1.75  &SMC & 43.35$\pm$0.05 &  498$\pm$62 & 174.9 &  52.9 & 13.02
\enddata
\tablenotetext{a}{X-ray detected Active Galactic Nuclei}
\tablenotetext{b}{Sub-millimeter galaxy with $f_{850\mu m}\sim5$mJy (Coppin et al. 2009)}
\end{deluxetable}

     \end{landscape}


\begin{thebibliography}{}
\bibitem[Adelberger \& Steidel (2000)]{AS00}
  Adelberger, K. L. \& Steidel, C. C. 2000, ApJ, 544, 218
\bibitem[Alexander et al.(2003)]{Alexsander03}
  Alexander, D. M. et al. 2003, AJ, 126, 539
\bibitem[Ando et al.(2004)]{Ando04}
  Ando, M., Ohta, K., Iwata, I., Watanabe, C., Tamura, N., Akiyama, M., \& Aoki, K. 2004, ApJ, 610, 635
\bibitem[Atek et al.(2010)]{Atek10}
  Atek, H. et al. 2010, ApJ, 723, 104 
\bibitem[Bouwens et al.(2007)]{Bouwens07}
  Bouwens, R. J., Illingworth, G. D., Franx, M., \& Ford, H. 2007, ApJ, 670, 928
\bibitem[Bouwens et al.(2009)]{Bouwens09}
  Bouwens, R. J. et al. 2009, ApJ, 705, 936
\bibitem[Bouwens et al.(2010)]{Bouwens10}
  Bouwens, R. J. et al. 2010, ApJL, 708, 60
\bibitem[Brandt et al. (2004)]{Brandt04}
  Brandt, W. N., Alexander, D. M., Bauer, F. E., \& Vignali, C. 2004,
  X-ray Survey Results on Active Galaxy Physics and Evolution (arXiv:astro-ph/0403646)
\bibitem[Bruzual (2007)]{Bruzual07}
  Bruzual, G. 2007, From Stars to Galaxies: Building the Pieces to Build Up the Universe
  (ASP Conf. Ser. 374), ed. A. Vallenari et al. (San Francisco, CA: ASP), 303 
\bibitem[Calzetti (2000)]{Calzetti00}
  Calzetti, D., Armus, L., Bohlin, R. C., Kinney, A. L.,
  Koornneef, J., Storchi-Bergmann, T. 2000, ApJ, 533, 682
\bibitem[Calzetti (2001)]{Calzetti01}
  Calzetti, D. 2001, New Astronomy Reviews, 45, 601
\bibitem[Capak et al.(2008)]{Capak08}
  Capak, P., et al., 2008, ApJ, 681, L53
\bibitem[Carilli et al.(2008)]{Carilli08}
  Carilli, C. L. et al. 2008, ApJ, 689, 883
\bibitem[Chary, Stern, \& Eisenhardt (2005)]{CSE05}
  Chary, R.-R., Stern, D., Eisenhardt, P. 2005, ApJL, 635, 5
\bibitem[Chary \& Pope (2010)]{CP10}
  Chary, R.-R., \& Pope, A. 2010 (arXiv:1003.1731)
\bibitem[Conselice \& Arnold (2009)]{CA09}
  Conselice, C. J., \& Arnold, J. 2009, MNRAS, 397, 208
\bibitem[Conselice et al.(2011)]{Conselice11}
  Conselice, C. J. et al. 2011, MNRAS, 413, 80
\bibitem[Coppin et al.(2009)]{Coppin09}
  Coppin, K. E. K., 2009, MNRAS, 395, 1905
\bibitem[Daddi et al.(2007)]{Daddi07}
  Daddi, E. et al. 2007, ApJ, 670, 156
\bibitem[Daddi et al.(2009)]{Daddi09}
  Daddi, E. et al. 2009, ApJ, 694, 1517
\bibitem[Dekel et al.(2009)]{Dekel09}
  Dekel, A. et al. 2009, Nature, 457, 451
\bibitem[Dickinson (1999)]{Dickinson99}
  Dickinson, M. 1999, After the Dark Ages: When Galaxies were Young
 (the Universe at $2<z<5$) (AIP Conf. Proc. 470), ed. S. Holt \& E. Smith (Melville, NY: AIP), 122
\bibitem[Dickinson et al.(2003)]{Dickinson03}
  Dickinson, M., Papovich, C., Ferguson, H. C., \& Budav\'ari, T. 2003, ApJ, 587, 25
\bibitem[Elbaz et al.(2007)]{Elbaz07}
  Elbaz, D. et al. 2007, A\&A, 468, 33
\bibitem[Erb et al.(2006)]{Erb06}
  Erb, D. K., Steidel, C. C., Shapley, A. E., Pettini, M., Reddy, N. A., \&
  Adelberger, K. L. 2006, ApJ, 647, 128
\bibitem[Fitzpatrick (1986)]{Fitzpatrick86}
  Fitzpatrick, E. L. 1986, AJ, 92, 1068
\bibitem[Gallego et al.(1997)]{Gallego97}
  Gallego, J., Zamorano, J., Rego, M., \& Vitores, A. G. 1997, ApJ, 475, 502
\bibitem[Geach et al.(2008)]{Geach08}
  Geach, J. E., Smail, I., Best, P. N., Kurk, J., Casali, M., Ivison, R. J., \&
  Coppin, K. 2008, MNRAS, 388, 1473
\bibitem[Giavalisco et al.(2004)]{Giavalisco04}
  Giavalisco, M. et al. 2004, ApJL, 600, 93
\bibitem[Hayes et al.(2010)]{Hayes10}
  Hayes, M. et al. 2010, Nature, 464, 562
\bibitem[Kaspi, Brandt, \& Schneider (2000)]{Kaspi00}
  Kaspi, S., Brandt, W. N., \& Schneider, D. P. 2000, AJ, 119, 2031
\bibitem[Kennicutt (1998)a]{Kennicutt98a}
  Kennicutt, R. C. 1998, ARA\&A, 36, 189
\bibitem[Kennicutt (1998)b]{Kennicutt98b}
  Kennicutt, R. C. 1998, ApJ, 498, 541 
\bibitem[Laidler et al.(2007)]{Laidler07}
  Laidler, V. G. et al. 2007, PASP, 119, 1325
\bibitem[Leitherer \& Heckman (1995)]{LH95}
  Leitherer, C., \& Heckman, T. M. 1995, ApJS, 96, 9
\bibitem[Leitherer et al.(1999)]{Leiterer99}
  Leiterer, C. et al. 1999, ApJS, 123, 3
\bibitem[Luo e tal.(2008)]{Luo08}
  Luo, B. et al. 2008, ApJS, 179, 19
\bibitem[Madau (1995)]{Madau95}
  Madau, P. 1995, ApJ, 441, 18
\bibitem[Malhotra \& Rhoads (2002)]{MR02}
  Malhotra, S., Rhoads, J. E. 2002, ApJL, 565, 71
\bibitem[Marchesini et al.(2009)]{Marchesini09}
  Marchesini, D., van Dokkum, P. G., F\"orster Schreiber, N. M.,
  Franx, M., Labb\'e, I., \& Wuyts, S. 2009, ApJ, 701, 1765
\bibitem[McQuade et al.(1995)]{McQuade95}
  McQuade, K., Calzetti, D., \& Kinney, A. L. 1995, ApJS, 97, 331
\bibitem[Meurer, Heckman, \& Calzetti (1999)]{MHC99}
  Meurer, G. R., Heckman, T. M., \& Calzetti, D. 1999, ApJ, 521, 64
\bibitem[Mouhcine et al.(2005)]{Mouhcine05}
  Mouhcine, M., Lewis, I., Jones, B., Lamareille, F., Maddox, S. J., \& Contini, T. 2005, MNRAS, 362, 1143
\bibitem[Muzzin et al.(2009)]{Muzzin09}
  Muzzin, A., van Dokkum, P., Franx, M., Marchesini, D., Kriek, M., \& Labb\'e, I. 2009, ApJL, 706, 188
\bibitem[Noeske et al.(2007)]{Noeske07}
  Noeske, K. G. et al. 2007, ApJL, 660, 47
\bibitem[Ouchi et al.(2004)]{Ouchi04}
  Ouchi, M. et al. 2004, ApJ, 611, 660
\bibitem[Papovich et al. (2006)]{Papovich06}
  Papovich, C. et al. 2006, ApJ, 640, 92
\bibitem[Pettini et al.(2002)]{Pettini02}
  Pettini, M., Rix, S. A., Steidel, C. C., Adelberger, K. L., Hunt, M. P., \& Shapley, A. E. 2002, ApJ, 569, 742
\bibitem[Pope \& Chary (2010)]{Pope10}
  Pope, A., \& Chary, R., 2010, ApJ, 715, L171
\bibitem[Pr\'evot et al. (1984)]{Prevot84}
  Pr\'evot, M. L., Lequeux, J., Pr\'evot, L., Maurice, E., \& Rocca-Volmerange, B. 1984, A\&A, 132, 389
\bibitem[Ravindranath et al.(2006)]{Ravindranath06}
  Ravindranath, S. et al. 2006, ApJ, 652, 963
\bibitem[Reddy et al.(2006)]{Reddy06}
  Reddy, N. A., Steidel, C. C., Fadda, D., Yan, L., Pettini, M.,
  Shapley, A. E., Erb, D. K., \& Adelberger, K. L. 2006, ApJ, 644, 792
\bibitem[Reddy et al.(2010)]{Reddy10}
  Reddy, N. A., Erb, D. K., Pettini, M., Steidel, C. C., \& Shapley, A. E. 2010, ApJ, 712, 1070
\bibitem[Retzlaff et al.(2010)]{Retzlaff10}
  Retzlaff, J., Rosati, P., Dickinson, M., Vandame, B., Rite, C., Nonino, M., Cesarsky, C.,
   \& GOODS Team. 2010, A\&A, 511, 50
\bibitem[Salpeter (1955)]{Salpeter1955}
  Salpeter, E. E. 1955, ApJ, 121, 161
\bibitem[Schinnerer et al.(2008)]{Schinnerer08}
  Schinnerer, E. et al. 2008, ApJL, 689, 5
\bibitem[Schmidt (1959)]{Schmidt59}
  Schmidt, M. 1959, ApJ, 129, 243
\bibitem[Shim et al.(2009)]{Shim09}
  Shim, H., Colbert, J., Teplitz, H., Henry, A., Malkan, M., McCarthy, P., \& Yan, L. 2009, ApJ, 696, 785
\bibitem[Shioya et al.(2009)]{Shioya09}
  Shioya, Y. et al. 2009, ApJ, 696, 546
\bibitem[Siana et al.(2008)]{Siana08}
  Siana, B., Teplitz, H. I., Chary, R.-R., Colbert, J., Frayer, D. T. 2008, ApJ, 689, 59
\bibitem[Siana et al.(2009)]{Siana09}
  Siana, B., et al. 2009, ApJ, 698, 1273
\bibitem[Steidel et al.(2003)]{Steidel03}
  Steidel, C. C., Adelberger, K. L., Shapley, A. E.,
  Pettini, M., Dickinson, M., \& Giavalisco, M. 2003, ApJ, 592, 728
\bibitem[Storchi-Bergmann et al.(1995)]{Storchi-Bergmann95}
  Storchi-Bergmann, T., Kinney, A. L., \& Challis, P. 1995, ApJS, 98, 103 
\bibitem[Taniguchi et al.(2009)]{Taniguchi09}
  Taniguchi, Y. et al. 2009, ApJ, 701, 915
\bibitem[Teplitz, Malkan, \& McLean (2004)]{Teplitz04}
  Teplitz,  H. I., Malkan, M. A., McLean, I. S. 2004, ApJ, 608, 36
\bibitem[Vanden Berk et al.(2001)]{VandenBerk01}
  Vanden Berk, D. E. et al. 2001, AJ, 122, 549
\bibitem[Vanzella et al.(2005)]{Vanzella05}
  Vanzella, E. et al. 2005, A\&A, 434, 53
\bibitem[Vanzella et al.(2006)]{Vanzella06}
  Vanzella, E. et al. 2006, A\&A, 454, 423
\bibitem[Vanzella et al.(2008)]{Vanzella08}
  Vanzella, E. et al. 2008, A\&A, 478, 83
\bibitem[Wang et al.(2010)]{Wang10}
  Wang, W.-H., Cowie, L. L., Barger, A. J., Keenan, R. C., \& Ting, H.-C. 2010, ApJS, 187, 251
\bibitem[Zackrisson, Bergvall, \& Leiter (2008)]{ZBL08}
  Zackrisson, E., Bergvall, N., \& Leiter, E. 2008, ApJL, 676, 9
\end{thebibliography}
\end{document}